\documentclass[a4paper,11pt]{article}
\pdfoutput=1
\usepackage[utf8]{inputenc}
\usepackage{amssymb}
\usepackage{amsmath}
\usepackage{amsthm}
\usepackage{amsfonts}
\usepackage[pdftex]{graphicx}
\usepackage{geometry}
\usepackage{braket}
\usepackage{enumerate}
\usepackage{comment}
\usepackage{lscape}

\usepackage{subcaption}

\allowdisplaybreaks

\usepackage[table,usenames,dvipsnames]{xcolor}

\usepackage{setspace}
\usepackage[in]{fullpage}
\usepackage{hyperref}
\usepackage{array}
\newcolumntype{P}[1]{>{\centering\arraybackslash}p{#1}}
\newcolumntype{M}[1]{>{\centering\arraybackslash}m{#1}}
\usepackage{cancel}
\usepackage{wrapfig}
\usepackage[font=small,labelfont=bf]{caption}
\usepackage{pifont}

\usepackage[dvipsnames]{xcolor}
\usepackage{mathtools}

\usepackage{color}

\usepackage{tikz}
\usetikzlibrary{arrows,shapes}
\usetikzlibrary{trees}
\usetikzlibrary{matrix,arrows} 
\usetikzlibrary{positioning}
\usetikzlibrary{calc,through}
\usetikzlibrary{decorations.pathreplacing}
\usepackage{pgffor}	

\numberwithin{equation}{section}

\usepackage[english]{babel}

\usepackage{caption}

\usepackage[nottoc,notlot,notlof]{tocbibind}
\usepackage[nosort]{cite}   
\usepackage{color}

\usepackage{parskip}
\usepackage[T1]{fontenc}

\usepackage{lmodern}
\usepackage{yfonts}
\usepackage{bbm}
\usepackage{physics}
\usepackage{slashed}
\usepackage{makeidx}
\usepackage[vcentermath]{youngtab}
\usepackage{amsthm}
\usepackage[scr=boondoxo]{mathalfa}

\hypersetup{
	colorlinks=true,
	linktoc=page,
	citecolor=blue,
	linkcolor=blue,
	urlcolor=blue} 

\usepackage{tikz}
\usetikzlibrary{arrows,shapes}
\usetikzlibrary{trees}
\usetikzlibrary{matrix,arrows} 				
\usetikzlibrary{positioning}				
\usetikzlibrary{calc,through}				
\usetikzlibrary{decorations.pathreplacing}  
\usepackage{pgffor}							

\usetikzlibrary{decorations.pathmorphing}	
\usetikzlibrary{decorations.markings}
\tikzset{
   vector/.style={decorate, decoration={snake, amplitude=1pt, segment length=6pt}, draw,double},
   vector2/.style={decorate, decoration={snake, amplitude=1pt, segment length=6pt}, draw},
	provector/.style={decorate, decoration={snake,amplitude=2.5pt}, draw},
	antivector/.style={decorate, decoration={snake,amplitude=-2.5pt}, draw},
    fermion/.style={draw=black, postaction={decorate},
        decoration={markings,mark=at position .55 with {\arrow[draw=black]{>}}}},
    fermionbar/.style={draw=black, postaction={decorate},
        decoration={markings,mark=at position .55 with {\arrow[draw=black]{<}}}},
    fermionnoarrow/.style={draw=black},
    gluon/.style={decorate, draw=black,
        decoration={coil,amplitude=4pt, segment length=5pt}},
    scalar/.style={dashed,draw=black, postaction={decorate},
        decoration={markings,mark=at position .55 with {\arrow[draw=black]{>}}}},
    scalarbar/.style={dashed,draw=black, postaction={decorate},
        decoration={markings,mark=at position .55 with {\arrow[draw=black]{<}}}},
    scalarnoarrow/.style={dashed,draw=black},
    electron/.style={draw=black, postaction={decorate},
        decoration={markings,mark=at position .55 with {\arrow[draw=black]{>}}}},
	bigvector/.style={decorate, decoration={snake,amplitude=4pt}, draw},
}
\tikzset{cross/.style={cross out, draw, 
         minimum size=2*(#1-\pgflinewidth), 
         inner sep=0pt, outer sep=0pt}}

\tikzstyle{block} = [draw, rectangle, 
    minimum height=3em, minimum width=6em]

\newcommand{\agl}[2]{\langle#1 #2 \rangle}
\newcommand{\sqr}[2]{\lbrack #1 #2 \rbrack}

\newcommand{\lu}[1]{\lambda^{#1}}
\newcommand{\ltu}[1]{\tilde{\lambda}^{\dot{#1}}}
\newcommand{\ld}[1]{\lambda_{#1}}
\newcommand{\ltd}[1]{\tilde{\lambda}_{\dot{#1}}}
\newcommand{\muu}[1]{\mu^{#1}}
\newcommand{\mtu}[1]{\tilde{\mu}^{\dot{#1}}}
\newcommand{\mud}[1]{\mu_{#1}}
\newcommand{\mtd}[1]{\tilde{\mu}_{\dot{#1}}}

\DeclareMathOperator{\RR}{\mathbb{R}}
\DeclareMathOperator{\CC}{\mathbb{C}}
\newcommand{\cA}{\mathcal{A}}
\newcommand{\cN}{\mathcal{N}}
\newcommand{\cL}{\mathcal{L}}

\usepackage{float}
\usepackage{marvosym}
\usepackage{empheq}

\usepackage{tikz}
\usetikzlibrary{shapes}
\usetikzlibrary{arrows}
\usetikzlibrary{positioning}
\usetikzlibrary{decorations.pathmorphing}
\usetikzlibrary{decorations.pathreplacing}
\usetikzlibrary{decorations.markings}

\def\cA{\mathcal{A}}
\def\uno{\mbox{1 \kern-.59em {\rm l}}}

\newcommand{\zb}{\bar{z}}

\newcommand{\yb}{\bar{y}}
\newcommand{\partdiv}[2]{\frac{\partial #1}{\partial #2}}

\DeclareMathOperator{\cJ}{\mathcal{J}}

\DeclareMathOperator{\half}{\frac{1}{2}}
\DeclareMathOperator{\slC}{\text{SL}(2,\mathbb{C})}

\newcommand{\tvec}[1]{\underline{#1}}

\DeclareMathOperator{\ep}{\varepsilon}
\DeclareMathOperator{\lam}{\lambda}
\DeclareMathOperator{\lamt}{\tilde{\lambda}}
\newcommand\ontop[2]{\genfrac{}{}{0pt}{}{#1}{#2}}

\DeclareMathOperator*{\SumInt}{%
\mathchoice%
  {\ooalign{$\displaystyle\sum$\cr\hidewidth$\displaystyle\int$\hidewidth\cr}}
  {\ooalign{\raisebox{.14\height}{\scalebox{.7}{$\textstyle\sum$}}\cr\hidewidth$\textstyle\int$\hidewidth\cr}}
  {\ooalign{\raisebox{.2\height}{\scalebox{.6}{$\scriptstyle\sum$}}\cr$\scriptstyle\int$\cr}}
  {\ooalign{\raisebox{.2\height}{\scalebox{.6}{$\scriptstyle\sum$}}\cr$\scriptstyle\int$\cr}}
}

\makeatletter
\newcommand{\oset}[3][0ex]{%
  \mathrel{\mathop{#3}\limits^{
    \vbox to#1{\kern+0.4\ex@        
    \hbox{$\scriptstyle#2$}\vss}}}}
\makeatother

\newcommand{\kibnbnormal}{\oset{\!\!\textbf{\fontsize{3.5pt}{3.5pt}\selectfont(\quad)}}{\kb_i}}

\newcommand{\kbnbnormal}[1]{\oset{\textbf{\fontsize{3.5pt}{3.5pt}\selectfont(\quad)}}{\kb}\negmedspace\negmedspace{}_{#1}}

\makeatletter
\newcommand{\osetsm}[3][0ex]{%
  \mathrel{\mathop{#3}\limits^{
    \vbox to#1{\kern+0.7\ex@         
    \hbox{$\scriptstyle#2$}\vss}}}}
\makeatother

\newcommand{\kibnbsmall}{\osetsm{\!\!\textbf{\fontsize{2.5pt}{2.5pt}\selectfont(\quad)}}{\kb_i}}

\newcommand{\kbnbsmall}[1]{\osetsm{\textbf{\fontsize{2.5pt}{2.5pt}\selectfont(\quad)}}{\kb}\!{}_{#1}}

\newcommand{\cb}{\bar{c}}
\newcommand{\db}{\bar{d}}
\newcommand{\hb}{\bar{h}}

\newcommand{\ub}{\bar{u}}
\newcommand{\pld}[1]{\partial_{#1}}
\newcommand{\pltd}[1]{\tilde{\partial}_{\dot{#1}}}

\theoremstyle{definition}

\usepackage{tikz-cd}
\newcommand{\wb}{\bar{w}}
\newcommand{\kb}{\bar{k}}
\newcommand{\tb}{\bar{t}}

\newcommand{\CCP}{\mathbb{CP}}
\newcommand{\RRP}{\mathbb{RP}}

\newcommand{\cS}{\mathcal{S}}
\newcommand{\cC}{\mathcal{C}}

\newcommand{\lambdat}{\tilde\lambda}
\newcommand{\sbar}{\bar{s}}
\newcommand{\mut}{\tilde\mu}
\newcommand{\cI}{\mathcal{I}}
\newcommand{\cR}{\mathcal{R}}
\newcommand{\sgn}{\text{sgn}}
\newcommand{\epsilonb}{\bar\epsilon}

\newcommand{\ZZ}{\mathbb{Z}}
\newcommand{\ct}{\tilde{c}}
\newcommand{\dt}{\tilde{d}}

\newcommand{\yt}{\tilde{y}}

\newcommand{\at}{\tilde{a}}

\newcommand{\Hbf}{\mathbf{H}}
\newcommand{\Hb}{\bar{H}}
\newcommand{\Htl}{\tilde{H}}
\DeclareMathOperator{\slR}{\text{SL}(2,\mathbb{R})}
\newcommand{\ut}{\tilde{u}}
\newcommand{\mt}{\tilde{m}}

\newcommand{\cK}{\mathcal{K}}
\newcommand{\kap}{\kappa}
\newcommand{\kapb}{\bar\kappa}
\newcommand{\cG}{\mathcal{G}}

\begin{document}


\begin{flushright}
	 QMUL-PH-22-37\\
\end{flushright}

\vspace{20pt} 

\begin{center}

	{\Large \bf  {Celestial Twistor Amplitudes} }  \\
	\vspace{0.3 cm}

	\vspace{25pt}

	{\mbox {\sf  \!\!\!\! Graham~R.~Brown$^{a,b}$, Joshua~Gowdy$^{a,b}$, Bill~Spence$^{a}$
	}}
	\vspace{0.5cm}

	\begin{center}
		{\small \em
			(a)~Centre for Theoretical Physics\\
			Department of Physics and Astronomy\\
			Queen Mary University of London\\
			Mile End Road, London E1 4NS, United Kingdom
		}
		
	\end{center}
		\begin{center}
		{\small \em
			(b)~Kavli Institute for Theoretical Physics\\
			University of California, Santa Barbara\\
			CA~93106, USA
		}
		
	\end{center}

	\vspace{40pt}  

	{\bf Abstract}
\end{center}

\vspace{0.3cm}

\noindent

We show how to formulate celestial twistor amplitudes in Yang-Mills (YM) and gravity. This is motivated by a refined holographic correspondence between the twistor transform and the light transform in the boundary Lorentzian CFT. The resulting amplitudes are then equivalent to light transformed correlators on the celestial torus. Using an ambidextrous basis of twistor and dual twistor variables, we derive formulae for the three and four-point YM and gravity amplitudes. The four-point amplitudes take a particularly simple form in terms of elementary functions, with a striking correspondence between the YM and gravity expressions. We derive celestial twistor BCFW recursion relations and show how these may be used to generate the four-point YM amplitude, illuminating the structure it inherits from the three-point amplitude and paving the way for the calculation of higher multiplicity light transformed correlators. Throughout our calculations we utilise the unique properties of the boundary structure of split signature, and in order to properly motivate and highlight these properties we first develop our methodology in Lorentzian signature. This also allows us to prove a holographic correspondence between Fourier transforms in Lorentzian signature and shadow transforms in the Euclidean boundary CFT.

\vfill
\hrulefill
\newline
\vspace{-1cm}
\!\!{\tt\footnotesize\{graham.brown, j.k.gowdy, w.j.spence\}@qmul.ac.uk}

\setcounter{page}{0}
\thispagestyle{empty}
\newpage


\setcounter{tocdepth}{4}
\hrule height 0.75pt
\tableofcontents
\vspace{0.8cm}
\hrule height 0.75pt
\vspace{1cm}
\setcounter{tocdepth}{2}


\section{Introduction}

Celestial holography seeks to describe the properties of fundamental theories in holographic terms from a conformal field theory perspective defined on the celestial sphere of asymptotically flat spacetimes. This utilises insights and techniques from the well-established AdS/CFT correspondence, as well as from the infrared structure of gauge and gravity theories in flat space directly.
Major progress has been made in this field in the past few years and 
there are now a number of excellent reviews of the subject - see, for example,  \cite{Strominger:2017zoo, Raclariu:2021zjz, Pasterski:2021rjz, Prema:2021sjp, Pasterski:2021raf, McLoughlin:2022ljp}.
An important part of this work is the study of the scattering amplitudes of particles on the celestial sphere, which can be interpreted as conformal correlators. Considerable work has been done on this
at tree-level \cite{Pasterski:2016qvg, Lam:2017ofc,  Stieberger:2018edy, Pasterski:2017ylz, Nandan:2019jas, Brandhuber:2021nez, Jiang:2021xzy, Ferro:2021dub,  Casali:2020uvr}, and in particular general formulae for tree-level MHV and NMHV $n$-point amplitudes  have been derived in \cite{Schreiber:2017jsr}. At loop level, results are now being obtained \cite{Albayrak:2020saa, Gonzalez:2020tpi, Gonzalez:2021dxw, Nastase:2021izh },
but in general the celestial amplitudes  found so far
take rather implicit or relatively complex forms, notably in comparison with the simplicity of some standard results for spacetime amplitudes, such as  
the $n$-point Parke-Taylor formula for tree-level MHV amplitudes in four-dimensional spacetime. A simple way to obtain celestial amplitudes is to perform a Mellin transform in the energy of each external particle, which turns momentum eigenstates into boost eigenstates \cite{Pasterski:2017ylz, Pasterski:2017kqt}.  However, massless  four-point amplitudes obtained in this way  contain a delta function enforcing momentum conservation from the bulk, which complicates their interpretation as conformal correlators. Thus, we are motivated to investigate other spaces and transformations which describe flat space amplitudes in a conformal way, in particular twistor space. 

Witten's pioneering paper on amplitudes in twistor space 
\cite{Witten:2003nn} described a geometry underlying the spinor formulations of  amplitudes, leading to many new insights. 
For $N=4$ super Yang-Mills (YM) amplitudes, for example, the \lq\lq half-Fourier'' transformation to the twistor space formulation of an amplitude was  developed in detail (see \cite{Korchemsky:2009jv} and references therein).
Somewhat later, in  \cite{Arkani-Hamed:2009hub} ambidextrous twistor amplitudes in $(2,2)$ signature spacetime were shown to take remarkably simple forms, with this work prompting a variety of new developments in amplitudes research. Much has happened since - for a recent brief overview of amplitudes research see \cite{Bern:2022jnl}, and for detailed reviews of different areas see the SAGEX papers \cite{Travaglini:2022uwo, Brandhuber:2022qbk, Bern:2022wqg, Abreu:2022mfk, Blumlein:2022zkr, Papathanasiou:2022lan, Geyer:2022cey, Herrmann:2022nkh, Heslop:2022qgf, Chicherin:2022nqq,Dorigoni:2022iem, McLoughlin:2022ljp, White:2022wbr, Bjerrum-Bohr:2022blt,Kosower:2022yvp, DelDuca:2022skz}.
The ambitwistor string  has played a key role in a number of  developments (c.f. \cite{Geyer:2022cey}). This has recently been applied to derive operator expressions for a broad class of celestial amplitudes in \cite{Casali:2020uvr} and develop celestial OPEs  in \cite{Adamo:2021zpw}.
 The half-Fourier transformation to twistor space has a closely related ambidextrous light transform analogue; this was explored in \cite{Sharma:2021gcz}, providing insights into the light-transformed basis of states, as well as a concise integral formulation for $n$-point celestial amplitudes, localising on positive energy regions of the Grassmannian. All-order formulae in the MHV sector have been derived using the twistor string very recently in \cite{Adamo:2022wjo}.

Given the insights that twistor theory has provided in amplitudes research, and recent progress in this field, it is then  natural to ask  if the study of {\it celestial} twistor amplitudes might  provide a route to further progress in the celestial programme. %
In this paper we will develop a formalism to derive these celestial twistor amplitudes, and show how this gives explicit results for Yang-Mills and gravity.

The outline of the paper is as follows. In the first half of the paper, Sections 
\ref{sec: prelude} to \ref{sec: twistortolight}, we develop the general formalism required to discuss celestial amplitudes in Lorentzian or split signature spacetime and the various bases we have available to describe them. While the second half, Sections \ref{sec: celesttwistamps} and \ref{sec: bcfw}, is devoted to celestial twistor amplitudes and the BCFW recursion relations they satisfy.

We begin in Section \ref{sec: prelude} by outlining the structure of quotient spaces and representations that will arise in subsequent sections.
Then in Section \ref{sec: 13Sig}
we will review a general framework for defining the asymptotic states used to build scattering amplitudes in asymptotically flat spacetimes with signature $(1,3)$, beginning in momentum space before shifting focus onto conformal primary states which live on the celestial sphere at the null boundary of spacetime. This includes a discussion of the extended little group and the chiral Mellin transform introduced in \cite{Brandhuber:2021nez}. In Section \ref{sec: shadowspace} we will discuss the the basis of conformal primaries which are constructed via the shadow transform, and its relation to the Fourier basis.
Following from this, in Section
\ref{sec: fourier2shad} we bring these results together, showing how the Fourier, shadow and Mellin transforms are related within a commuting diagram, and in Section \ref{sec: fourieramps} we describe the Fourier-transformed amplitudes.
Then, in Section \ref{sec: 22sig} we use the same methodology to explore how these results are modified in $(2,2)$ signature spacetime. Here the structure of the extended little group implies that celestial states have imaginary helicity and also discrete parity indices labelling even and odd representations. Once again, we discuss how the chiral Mellin transform in split signature constructs celestial states. In Section
\ref{sec: lightspace} we give the light transformed basis, and finally in Section 
\ref{sec: twistortolight} we show how the half-Fourier, light and Mellin transforms are related by a commuting diagram. 

We then turn in Section \ref{sec: celesttwistamps} to the study and derivation of celestial twistor amplitudes. With insights from the earlier analysis, we begin with analytically continued $(2,2)$ signature amplitudes, mapping these to Mellin space via a chiral Mellin transform. We use this approach to find the three-point Yang-Mills amplitudes, taking care to identify the even and odd states under the $\ZZ_2$ parity subgroups of the extended little group. We discuss the regularisation needed to evaluate the integrals, basing this on the key requirement that the Mellin transform is invertible on a suitably defined strip of definition.

Next, we derive four-point amplitudes using the same general approach. Beginning with the $(+-+-)$ helicity amplitude, we show that there are eight separate non-vanishing cases, with different parities. The integrals require careful regularisation but lead to a very simple expression based on four rational functions of the cross-ratios and weights, from which the different parity cases can be found via (anti)symmetrisation. The $(++--)$ amplitude is related to this result by a certain derivative operator, and we see that this has a simple action, leading to a similarly concise form of the amplitude.

Following this, we  apply these methods to derive the three and four-point gravity amplitudes.  
The three-point amplitudes are found to be analogues of  the YM cases, differing just by changes from sgn functions to mod functions and the doubling of the number of poles, as might be expected. The four-point amplitudes are also found to be strikingly similar to the YM case, exhibiting the same basic correspondence mentioned at three-points.

Finally, using Mellin transforms we derive a celestial analogue of the BCFW twistor recursion relations  in Section \ref{sec: bcfw} which enables an iterative construction of celestial twistor amplitudes. Furthermore, we find an equivalent but formally simpler recursion relation involving sub-amplitudes with the glued internal legs in the Mellin basis and not light transformed. To illustrate the application of these results we use them to rederive the four-point amplitudes presented earlier. This explains the origin of the remaining key features in the four point amplitude not fixed by conformal covariance, a conclusion which we expect to apply in more general cases.

After concluding remarks in Section \ref{sec: conclusions}, summarising our results and highlighting further possible extensions of our work,
we give appendices covering: the all but one leg celestial twistor amplitudes required in the four-point BCFW (Appendix \ref{app: almostceltwistamps}), details of the map between Fourier to shadow transforms and back (Appendix \ref{app: fourierequalsshadow}), formulae for the bulk conformal generators  (Appendix \ref{app: conformalgens}), proofs that the shadow and light transforms are self-inverse (Appendix \ref{app: Ssquared}), the result of all-leg shadow transformed gluon amplitudes in split signature (Appendix \ref{app: all legfouriergluon}), a discussion of transforms in complexified spacetime and slices thereof (Appendix \ref{app: complexify}), the inverse chiral Mellin transform  (Appendix \ref{sec: Mellininversion}) and our spinor conventions (Appendix \ref{app: conventions}).


\section{A Prelude on Quotients and Representations}\label{sec: prelude}

Before diving into the myriad asymptotic states and transforms of the celestial sphere/torus, we will  discuss a few guiding principles of harmonic analysis and representation theory that appear frequently (see for example \cite{LinearReps,GELFANDtextbook66} for a more detailed exposition on the ideas explained here). Repeatedly we will encounter spaces and their quotient spaces, related to each other by ``dividing out'' by some symmetry group. Our goal here will be to make this notion precise and, in particular, to explain how functions living on spaces and their quotients are related. We shall do this through two simple examples that will  appear later in this work, before stating the general framework. 

For the first example consider the space of real numbers without zero: $\RR_{*}$. This space has a $\mathbb{Z}_2$ symmetry which maps points $x\mapsto-x$. We can quotient this space using this symmetry by `gluing' points together with the equivalence relation $x\sim -x$,  giving us: $(\RR_* /\sim) \cong \RR_+$. The map that takes us from the original space to the quotient space is the projection map $p$, which in this case is simply the absolute value $p(x)=\lvert x \lvert$. Diagrammatically it is common to write these spaces vertically, with the quotient space below the original one
\begin{equation}
\begin{tikzcd}
& \RR_{*} \arrow[d, "p"] \\
& \RR_{+}
\end{tikzcd}\,.
\end{equation}
The structure we have just described is actually a  trivial principal $\mathbb{Z}_2$-bundle. In fact, all the examples of quotient spaces we will consider are principal $G$-bundles, for some Abelian group $G$, however, we will try to avoid using such language whenever possible. 

Now let us consider a function living on the original space, $f: \RR_{*}\rightarrow \RR$. A natural question is how we can encode this function in the quotient space without losing any information. The answer is that the single function $f$ on the original space becomes two functions $\tilde f_0$ and  $\tilde f_1$ on the quotient space, labelled by the trivial and fundamental representations of $\mathbb{Z}_2$. These are of course just the even ($\tilde f_0$) and odd ($\tilde f_{1}$) parts of the function $f$, but their interpretation in terms of representations generalises to more complicated examples. These even and odd pieces are constructed from the original function $f$ in the usual way, by taking even and odd linear combinations of the function, but let us write this in a compact and suggestive form
\begin{equation}
    \tilde f_{s}(x)=\frac{1}{2}\sum_{c \in \mathbb{Z}_2=\{+1,-1\}} c^{s} f(c x)\,,
\end{equation}
for $s=0,1$. In other words, we construct the $\tilde f_s$ by summing over the group elements in $\mathbb{Z}_2$ while letting these elements act on the argument of $f$, and include a phase factor depending on the representation. This is a recipe that we will use for all of the other groups in this work, be they continuous or discrete. Note here that the functions $\tilde f_s(x)$ can be considered as functions on either $\RR_{+}$ or the original space $\RR_{*}$, since their even and odd properties let us extend their definition from $x>0$ to $x<0$. This is a freedom we will often use in other examples. In particular, this lets us recover the original function on $\RR_{*}$ by simply summing over the representations
\begin{equation}
    f(x)= \tilde f_{0}(x)+\tilde f_{1}(x)\,. 
\end{equation}
Thus, even though the projection map $p$ is not invertible, we can go between functions on the full space and the quotient space at will.

As a second example consider the space of real numbers $\RR$ with the symmetry group of translations $x\mapsto x+c$, also denoted by $\RR$. If we quotient the space $\RR$ by gluing points together related under translations ($x\sim x+c$ for any $c\in\RR$) then the quotient space is simply a point, which we denote as $\{0\}\cong \RR/ \sim  $. The projection map is then trivial: $p(x)=0$ for all $x\in \RR$, and we can write the spaces diagrammatically as
\begin{equation}
\begin{tikzcd}
& \RR \arrow[d, "p"] \\
& \{0\}
\end{tikzcd}\,.
\end{equation}

Now again let us consider a function on the original space $f:\RR\rightarrow\RR$. To describe this function on the quotient space we follow the same recipe as the previous example: one function on the space $\RR$ will become a family of functions on the quotient space $\{0\}$, labelled by representations of the group of translations $\RR$.  The irreducible representations of the translations (plus the trivial representation) are labelled by a number $k\in\RR$, so we write the functions in the quotient space as $\tilde{f}(k,x)$. They are again constructed from the original function by letting the group act on $f$ and summing over the group, with a phase corresponding to the representation 
\begin{equation}\label{eq: preludeFourier}
    \tilde{f}(k,x)= \frac{1}{2\pi}\int_{-\infty}^{+\infty}dc\, e^{i k c}f(x+c)\,.
\end{equation}
As before, the $ \tilde{f}(k,x)$ can be thought of as living at the point $x=0$ (or any other point) or as a function on the  original space $\RR$. This is due to the property 
\begin{equation}
    \tilde{f}(k,x+c)=e^{ikc}\tilde{f}(k,x)\,.
\end{equation}
Thus if you are given the function $\tilde{f}(k,0)$, and the property above, you can extend it to the whole real line.  Once again this property lets us recover the original function by summing over the representations 
\begin{equation}\label{eq: preludeInvFourier}
    \int_{-\infty}^\infty dk\, \tilde{f}(k,x)=\frac{1}{2\pi}\int_{-\infty}^{+\infty} dk \int_{-\infty}^{+\infty}dc\, e^{-i k c}f(x+c)=  f(x)\,. 
\end{equation}
The integral transforms \eqref{eq: preludeFourier} and \eqref{eq: preludeInvFourier} above are, of course, Fourier transforms. However, normally one would work exclusively with $\tilde{f}(k,0)$  and consider $k$ and $x$ as conjugate variables in separate spaces. There are benefits to each one of these interpretations and we shall use both in the rest of the paper.

Let us briefly summarise the general story - for a generic space $P$ and a symmetry group $G$ that acts on points $x\in P$ as $x.g$, the quotient space $P/G$ is the set of equivalence classes under the relation $x\sim x.g$ for all $g \in G$. The projection map $p:P\rightarrow P/G$ then maps points in $P$ to their equivalence class
\begin{equation}
\begin{tikzcd}
 & P \arrow[d,"p"] \\
& P/G
\end{tikzcd}\,.
\end{equation}\
 For the spaces that we will consider in this work, a function $f$ on the original space $P$ is encoded in the quotient space by a family of functions $\tilde f_r$, labelled by the irreducible representations (and trivial representation) $r$ of $G$. These $\tilde{f}_r$ are always constructed using the same recipe:
\begin{equation}\label{eq: repDecombPrelude}
\tilde f_r(x)= \SumInt_{g\in G} K(r,g) f(x.g)\,,
\end{equation}
where $K(r,g)$ is a function (often simply a phase) of the group element $g$ and the representation $r$. This kernel is ``covariant'' under the action of $G$ in the sense that it satisfies\footnote{Note that, as previously mentioned, every group $G$ we will consider here is Abelian, so the order of multiplication does not matter.}
\begin{equation}
    K(r,gh)= K(r,g) K(r,h)\,.
\end{equation}
for all $g,h\in G$. This implies that the $\tilde{f}_r(x)$ has the following homogeneity property under the action of $g\in G$
\begin{equation}
\tilde{f}_r(x.g) = K(r,g^{-1}) \tilde{f}_r(x)    
\end{equation}
Thus, the $\tilde{f}_r$ can be thought of as living either on $P$ or the quotient space $P/G$ since the above property lets us restrict or extend their definition.
Finally, to recover the original function $f$ we simply sum over all of the $\tilde{f}_r$
\begin{equation}\label{eq: repSumPrelude}
    f(x)= \SumInt_{r} \tilde{f}_{r}(x)\,.
\end{equation}

In this work the space $P$ will be a space of spinors, either in Lorentzian or split signature. We shall then quotient this space by the (extended) little group to obtain the null cone and celestial sphere/torus. We can also quotient by translations of spinors to obtain twistor space etc. The discussion above concerned functions on these spaces, but applies equally well to the asymptotic states and amplitudes that we will be interested in. Thus to find maps between states in each space we can simply follow the general formulae \eqref{eq: repDecombPrelude} and \eqref{eq: repSumPrelude}. These maps will be the Mellin, Fourier  transforms etc. and their respective inverses. 

In the next section we will begin with an analysis of the Lorentzian story, discussing split signature in Section \ref{sec: 22sig}. One could also begin with general independent complex spinors and then take various slices; we comment on this approach in Appendix \ref{app: complexify}.

\section{Lorentzian Signature Spacetime}\label{sec: 13Sig}

In this section, we will review a general framework for defining asymptotic states in flat spacetimes. We will begin in Lorentzian momentum space ($(+,-,-,-)$ signature) before shifting focus to the celestial sphere at null infinity. We call the space of states on the celestial sphere Mellin space since it is obtained by a Mellin transform of momentum eigenstates. These Mellin space states transform as conformal primaries under $\slC$, and hence amplitudes built from these states transform as conformal correlators. There are diverse approaches taken in defining these conformal primary states; in this section and what follows, we will find it very useful to follow Banerjee's extended little group approach \cite{Banerjee:2018gce}.

 Here and throughout the rest of the paper we will 
 write everything in terms of spinor-helicity variables $\ld{\alpha}, \ltd{\alpha}$ such that massless momenta is given by $p_{\alpha \dot\alpha} = \ld{\alpha}\ltd{\alpha}$ (we will often suppress the indices on the spinors in the discussion). We always regard $\lambda, \lambdat$ as homogeneous coordinates for a particular projective space - be it the null cone of massless momenta, or the celestial sphere/torus etc. Homogeneous coordinates make our formulae considerably tidier and make manifest the links between the momentum space and Mellin space representations. From the perspective of the holographic CFT, the use of spinor-helicity variables $\lambda, \lambdat$ is simply a version of the embedding space formalism specific to a two-dimensional CFT.

\subsection{Little Group}\label{sec: 13LG}
 The little group plays a central role in defining asymptotic states.
 We begin in momentum space where asymptotic particle states are labelled by a complex spinor $\lam$ corresponding to momentum $p_{\alpha\dot{\alpha}}=\ld{\alpha}\ltd{\alpha}$. We denote the complex conjugate spinor by $\lamt$
 \begin{equation}\label{eq: 13realitycondition}
    \lamt=\epsilon\lambda^*\,,
\end{equation}
where $\epsilon = \pm 1$ for either incoming or outgoing momentum. Thus in $(1,3)$ signature the space of on-shell spinors is $(\lambda, \epsilon) \in \CC_*^2 \times \ZZ_2$. 
 
 This parameterisation of a null momentum in terms of spinors is not unique, since the following little group transformations leave the momentum invariant
\begin{equation}\label{eq: 13LittleGroup}
     \ld{\alpha} \mapsto e^{i\theta}\, \ld{\alpha}, \quad \ltd{\alpha} \mapsto e^{-i\theta} \ltd{\alpha}.
 \end{equation}
 Hence, the little group of massless momentum in (1,3) spacetime is $U(1)$, corresponding to rotations about the direction of the null momentum\footnote{Here we are ignoring, as usual, the two "continuous spin" generators of the massless little group. This is so we recover the usual notion of helicity.}. The freedom to transform spinors according to \eqref{eq: 13LittleGroup} implies the spinor variables act as homogeneous coordinates for the space of massless momenta, and the projection is exactly $(\lam,\lamt)\mapsto p_{\alpha\dot{\alpha}}=\ld{\alpha}\ltd{\alpha}$. In other words, we quotient the space of spinors by little group transformations to obtain the space of massless momenta, given by $p_{\alpha\dot{\alpha}} \in \RR_{+} \times \CCP^1 \times \ZZ_2$.
 
 Asymptotic particle states are homogeneous functions of the spinor variables and transform in representations of the little group. For the little group of null momenta, $U(1)$ representations are labelled by a half-integer helicity $J \in \ZZ/2$. Thus, we write asymptotic particle states in Lorentzian spacetime as
\begin{equation}\label{eq: 31momentumEigenstate}
    \vert \lam, \lamt ; J\rangle\, ,
\end{equation}
and under little group transformations we have
\begin{equation}\label{eq: 13LittleGrouphomogeneity}
     \boxed{\vert e^{i\theta}\lam, e^{-i\theta}\lamt ; J\rangle = \left(e^{i\theta}\right)^{-2\,J}\vert \lam, \lamt ; J\rangle\, .}
\end{equation}

The universal cover of the Lorentz group in (1,3) is $\slC$ (since $\text{SO}^+(1,3)=\slC/\mathbb{Z}_2$) and the transformation of these states under this group is encoded in the spinors through
\begin{equation}
    \ket{\ld{\alpha} , \ltd{\alpha};J}\rightarrow \ket{M_\alpha^{\,\,\beta}\ld{\beta} , \bar{M}_{\dot{\alpha}}^{\,\,\dot{\beta}}\ltd{\beta}; J }\,,
\end{equation}
where $M_\alpha^{\,\,\beta}$ is an $\slC$ matrix. 

The process we have described above is our first example of the general story laid out in Section \ref{sec: prelude}. We have a group action on the space of spinors which we then quotient out. We then define states in the quotient space as functions of spinors (regarded as projective coordinates on the space) labelled by representations of the group and transforming homogeneously under the group action.

\subsection{Banerjee's Extended Little Group}\label{sec: 13ELG}

The little group transformations described above exactly correspond to rotations about the direction of the null momentum, and so stabilise $p_{\alpha\dot{\alpha}}$. On the celestial sphere however, we are only interested in preserving the direction the momentum is pointing. Thus, one could consider an extended little group which only preserves the null direction. This is the approach presented in \cite{Banerjee:2018gce}, which extends the little group to include boosts. 

In $(1,3)$ signature these boosts simply scale the momentum by a positive real number, $b \in \RR_+$
\begin{equation}\label{eq: 13momboost}
    p_{\alpha\dot{\alpha}}\rightarrow  b\, p_{\alpha\dot{\alpha}}\,.
\end{equation}
Note that there is no $\slC$ transformation acting on the spinors which changes the sign of the energy of the momentum, i.e., there is no boost in $\text{SO}^+(1,3)$ which turns an incoming state into an outgoing one. This is a crucial difference with $(2,2)$ signature which we will discuss in Section \ref{sec: 22ELG} \footnote{One could, however, still quotient by $\mathbb{Z}_2$ in Lorentzian signature and consider even and odd combinations of ingoing and outgoing states. Such even combinations were used in \cite{Fan:2021isc}.}. We can write the new boost scalings \eqref{eq: 13momboost} as a re-scaling of the spinors
\begin{equation}\label{eq: 13boostGroup}
\begin{split}
    \lam\rightarrow \sqrt{b} \lam, \quad \lamt\rightarrow \sqrt{b}\lamt . 
\end{split}
\end{equation}
Combining the positive real boost of the spinors with the usual $\text{U}(1)$ little group gives us the extended little group $\CC_*= \text{U}(1)\times \RR_+$ in Lorentzian signature, which we write as
\begin{equation}\label{eq: 13ELG}
    \lam\rightarrow y \lam, \quad \lamt\rightarrow  \yb \lamt\,,
\end{equation}
for any $y\in \CC_*$. Again following the general recipe, we then quotient the space of on-shell spinors by the complex re-scalings \eqref{eq: 13ELG} to define the celestial sphere $\CCP^1\coloneqq \CC_*^2/\CC_*$. Next we build conformal primaries from celestial states in Lorentzian signature, with discrete helicity $J$ labeling the representation of $\text{U}(1)$, and continuous conformal dimension $\Delta$ labeling the representation of $\RR_+$
\begin{equation}\label{eq: 13CelestState}
    \vert \lam,\lamt; J,\Delta \rangle = \vert \lam,\lamt; h, \hb \rangle\, ,
\end{equation}
where we have  $h:=\frac{1}{2}(\Delta+J)$ and $\hb:=\frac{1}{2}(\Delta-J)$. These states \eqref{eq: 13CelestState} have the following homogeneity property
\begin{equation}\label{eq: 13ELGhomo}
    \boxed{\vert y\lam,\yb\lamt; h, \hb \rangle =y^{-2h} \yb^{-2\hb} \vert \lam,\lamt; h, \hb \rangle\, ,}
\end{equation}
and transform under $\slC$ like conformal primaries in a two-dimensional \emph{Euclidean} CFT
\begin{equation}\label{eq: 13HomoSL2C}
\begin{split}
    \vert\ld{\alpha} , \ltd{\alpha};h, \hb\rangle &\rightarrow \vert M_\alpha^{\,\,\beta}\ld{\beta} , \bar M_{\dot{\alpha}}^{\dot{\beta}}\ltd{\beta};h, \hb \rangle, \\
    \ket{\begin{pmatrix}
			z \\
			1  
		\end{pmatrix},\epsilon\begin{pmatrix}
			\zb \\
			1  
		\end{pmatrix}; h, \hb}&\rightarrow (cz+d)^{-2h}(\cb\zb+\db)^{-2\hb}\ket{\begin{pmatrix}
			z' \\
			1  
		\end{pmatrix},\epsilon\begin{pmatrix}
			\zb' \\
			1  
		\end{pmatrix}; h, \hb}\,,
\end{split}
\end{equation}
where the transformed coordinates $z', \zb'$ are given by a M\"{o}bius transformation. In the above we have defined the affine coordinate $z=\frac{\lam_1}{\lam_2}$ (in the patch $\lam_2\neq 0$) and used the homogeneity property which implies
\begin{equation}\label{eq: GlobalToaffine}
\ket{\begin{pmatrix}
			z \\
			1  
		\end{pmatrix},\epsilon\begin{pmatrix}
			\zb \\
			1  
		\end{pmatrix}; h, \hb} = \lam_2^{2h} \bar\lam_2^{2\hb} \vert\ld{\alpha} , \ltd{\alpha};h, \hb\rangle.
\end{equation}
It is only when we consider affine coordinates on the celestial sphere that the dependence on the incoming/outgoing parameter $\epsilon$ from \eqref{eq: 13realitycondition} is made manifest. In \eqref{eq: GlobalToaffine} we can only scale by $\bar\lam_2$ (the complex conjugate of $\lam_2$) and so $\epsilon$ appears explicitly. Note here that (extended-)little group scalings can be applied to every leg of a celestial amplitude \textit{individually}. In other words, we can choose to scale one leg using \eqref{eq: 13ELGhomo} and leave the others alone, in contrast to the $\slC$ transformations which are applied to every leg. This clean separation between $\slC$ and the little group is lost once we use the use the affine coordinates $(z,\zb)$, as can be seen in \eqref{eq: 13HomoSL2C}. 

Our job now is to find an explicit construction to go between the usual momentum eigenstates and these celestial states. There are, in fact, two distinct but interrelated bases for celestial conformal primaries in a Euclidean CFT - the Mellin basis and the shadow basis - and each has its own integral transform. In the following section we will first study the Mellin transform and only later will we study the shadow basis. We will again use homogeneous coordinates to make a number of expressions more compact. In addition we will find that the chiral Mellin transform discussed in \cite{Brandhuber:2021nez} is often more natural than the usual non-chiral Mellin transform over just the particle energy. Finally, before moving on let us summarise the structure of the spaces we have just discussed diagrammatically
\begin{equation}
\begin{tikzcd}
 &\text{Spinor space:}\CC_*^2 \times \ZZ_2\,\, \arrow[d,"/U(1)"] \\
&\text{Null cone:}\,\, \RR_{+} \times \CCP^1 \times \ZZ_2 \arrow[d,"/\RR_+"]\\
&\text{Celestial Sphere: }\,\,\CCP^1\times \ZZ_2
\end{tikzcd}\,.
\end{equation}\

\subsection{Chiral Mellin Transform}\label{sec: 13chiralMellin}

As explained in Section \ref{sec: 13ELG}, states on the celestial sphere  can be expressed in terms of homogeneous conformal primaries. Constructing these homogeneous states can be achieved with a Mellin transform and, as discussed in Section \ref{sec: prelude}, the use of integral transforms to build homogeneous functional representations of groups is very general. In a recent paper by the current authors and Brandhuber and Travaglini \cite{Brandhuber:2021nez}, a ``chiral'' form of the Mellin transform in affine coordinates was introduced (we note here that this chiral Mellin transform has appeared in the literature previously, see for example \cite{GELFANDtextbook66}). In this section we will recap this formalism, but working in homogeneous coordinates from the start and then showing how to recover the affine expressions.

We define (1,3) celestial states in terms of the following (1,3) chiral Mellin transform
\begin{equation}\label{eq: 13chiralMellin}
    \boxed{\vert \lam,\lamt; h, \hb \rangle := \frac{1}{2\pi i}\int_{\mathbb{C}_*} \frac{d\ub}{\ub}\wedge \frac{du}{u}\, u^{2h}\ub^{2\hb} \vert u\lam,\ub\lamt\rangle,}
\end{equation}
 Note that this transform is clearly in the general form of $\eqref{eq: repDecombPrelude}$. The celestial state \eqref{eq: 13chiralMellin} is homogeneous under complex rescalings of the spinors with weights $h,\hb$ and thus, using \eqref{eq: 13HomoSL2C}, is also a conformal primary with the same weights.

Since we are interested in building celestial conformal correlators for a particular theory, be it Yang-Mills or gravity, we will always Mellin transform a function which is already homogeneous under the little group with some helicity $l$.
 Thus the definition above is equivalent to the usual non-chiral Mellin transform over the particle energy \cite{Pasterski:2017kqt}, once we integrate out the little group degree of freedom. This integral over the compact $U(1)$ takes the form of a discrete Fourier transform and we recover a Kronecker delta
\begin{equation}
    \vert \lam,\lamt; h, \hb, l \rangle = \delta_{J,l}\int_0^\infty\!d\omega \,\omega^{\Delta-1} \vert \sqrt{\omega}\lam,\sqrt{\omega}\lamt; l \rangle ,
\end{equation}
which identifies the bulk helicity $l$ with the holographic helicity $J:=h-\hb$. A consequence of this is that the only freely varying weight of a celestial state is the conformal dimension $\Delta$\footnote{The chiral Mellin transform may act on a general Lorentz covariant function of a spinor and by design will build a function which transforms as a conformal primary with weights $\Delta, J$, where now these are both free parameters.}. To make our notation more compact, we will often drop the explicit label $l$ with the understanding that the Kronecker delta imposes $h-\hb=l$.      

The non-compact integral is a Mellin transform over the $\RR_+$ scale and maps to modes labelled by a continuous complex weight $\Delta$. The convergence of the Mellin transform depends on the behaviour of the function $\vert \sqrt{\omega}\lam,\sqrt{\omega}\lamt \rangle$ for large and small values of $\omega$. In general, a Mellin transform converges for some `strip of definition' \cite{Bertrand:2000ibk}. For the $\omega$ Mellin transform with $\Delta=a+i\RR$ we then have convergence for real $a$ in some interval $(a_1, a_2)$. Assuming asymptotic behaviour $ \mathcal{O}(\abs{\ld{\alpha}}^{-p})$ for large $\abs{\ld{\alpha}}\sim \abs{\ltd{\alpha}}$ and $ \mathcal{O}(\abs{\ld{\alpha}}^{-q})$ for small $\abs{\ld{\alpha}}\sim \abs{\ltd{\alpha}}$ then the strip of definition $(a_1, a_2)$ must lie within the strip $(q,p)$. Then the Mellin inversion theorem, see \cite{Bertrand:2000ibk} for example, guarantees that we can invert by integrating $\Delta$ along a contour within the strip of definition $(a_1, a_2)$; this ensures that at least $q<a<p$. Mellin inversion is demonstrated directly in Appendix \ref{sec: Mellininversion} for a celestial state in homogeneous coordinates and is only slightly different from the usual inverse Mellin transform in affine coordinates.

The generators of the conformal group of the four-dimensional bulk spacetime are naturally written in terms of spinors $\lam, \lamt$, for example generating the conformal symmetry of Yang-Mills. As such they can also be defined  in Mellin space using $\lam, \lamt$ as homogeneous coordinates. A full list of such generators can be found in Appendix \ref{app: conformalgens}. While the conformal generators in affine coordinates are discussed at length in \cite{Brandhuber:2021nez} they take a much simpler form in homogeneous coordinates, as shown below. We define the generators in Mellin space by simply commuting them with the chiral Mellin transform. As an example we can consider the spinor derivative $\partial_{\alpha}$ and act with a chiral Mellin transform which scales $\lam \rightarrow u\lam$ 
\begin{equation}\label{eq: spinorderivhomo}
  \int_{\mathbb{C}} \frac{d\ub}{\ub}\wedge \frac{du}{u} \, u^{2h} \ub^{2\hb}  \frac{1}{u} \pld{\alpha} \ket{u \lam,\ub \lamt} = \pld{\alpha}e^{-\frac{1}{2}\partial_h}  \ket{\lam, \lamt;h,\hb}\,.
\end{equation}
Thus the spinor derivative acting in Mellin space is simply $\pld{\alpha}e^{-\frac{1}{2}\partial_h}$.
To compare this with the affine coordinate version, using \eqref{eq: GlobalToaffine} we have
\begin{equation}
    \begin{split}
        \pld{\alpha}e^{-\frac{1}{2}\partial_h}  \ket{\lam , \lamt;h,\hb}
        =& \pld{\alpha}e^{-\frac{1}{2}\partial_h}\lam_{2}^{-2h}\bar\lam_{2}^{-2\hb} \ket{\begin{pmatrix}
			z \\
			1  
		\end{pmatrix},\epsilon\begin{pmatrix}
			\zb \\
			1  
		\end{pmatrix}; h, \hb}\\
        =&\frac{1}{\ld{2}}\begin{pmatrix}
            \ld{2}\partdiv{}{\ld{2}}-z\partial_{z}\\
            -\partial_{z}
        \end{pmatrix}e^{-\frac{1}{2}\partial_h}\ld{2}^{-2h}\bar\lam_{2}^{-2\hb} \ket{\begin{pmatrix}
			z \\
			1  
		\end{pmatrix},\epsilon\begin{pmatrix}
			\zb \\
			1  
		\end{pmatrix}; h, \hb}\\
        =& \ld{2}^{-2h}\bar\lam_{2}^{-2\hb} \begin{pmatrix}
            -2h+1-z\partial_{z}\\
            -\partial_{z}
        \end{pmatrix}e^{-\frac{1}{2}\partial_h}\ket{\begin{pmatrix}
			z \\
			1  
		\end{pmatrix},\epsilon\begin{pmatrix}
			\zb \\
			1  
		\end{pmatrix}; h, \hb}\,,
    \end{split}
\end{equation}
where we changed variables $\{\ld{1},\ld{2}\}\rightarrow\{z,\ld{2}\}$ with $z=\ld{1}/\ld{2}=-\lu{2}/\lu{1}$, recovering the expression for the spinor derivative found in \cite{Brandhuber:2021nez}.

\section{Shadow Transformed Basis}\label{sec: shadowspace}

We have described celestial states that are built from a Mellin transform, but this is only one of the bases for conformal primaries. In an Euclidean CFT (corresponding to a Lorentzian signature bulk spacetime) there is another basis of conformal primaries which is constructed via the shadow transform see \cite{Osborn:2012vt, Crawley:2021ivb}. On the other hand, we will see, in Section \ref{sec: lightspace}, that in a Lorentzian CFT (split signature bulk) there are also bases built by performing a light transform on either of the spinors $\lam$ or $\lamt$.

This network of conformal primary bases, and the integral transforms that build them, has been explored in $d$-dimensions for Euclidean CFTs in \cite{Karateev:2018oml} and for Lorentzian CFTs in \cite{Kravchuk:2018htv}. Importantly, the conformal primaries in these bases transform in \emph{equivalent} representations of the conformal group, meaning that there exist intertwining operators which map between them. These intertwining operators are exactly the shadow and light transforms, and their existence is controlled by the restricted Weyl group.
 The restricted Weyl group is a finite group of reflection transformations of the weights $\Delta$ and $J$ such that the eigenvalues of the Casimir operators of the conformal group are left invariant, \emph{and} continuous and discrete weights are not mixed \cite{Kravchuk:2018htv}. The quadratic Casimir operators of the two-dimensional conformal group of the celestial sphere/torus are given by those of the Lorentz group of the bulk. These are\footnote{See, for example, Appendix F of \cite{Cuomo:2017wme}.} 
\begin{equation}\label{eq: quadcasmiroperators}
\begin{split}
    &C_2^{+} := \frac{1}{4}(m_{\alpha\beta}m^{\alpha\beta}+ \mt_{\dot\alpha\dot\beta}\mt^{\dot\alpha\dot\beta}) \sim Tr(M^2),\\
    &C_2^{-} := \frac{1}{4}(m_{\alpha\beta}m^{\alpha\beta}- \mt_{\dot\alpha\dot\beta}\mt^{\dot\alpha\dot\beta}) \sim \epsilon_{\mu\nu\rho\sigma}M^{\mu\nu}M^{\rho\sigma},
\end{split}
\end{equation}
where we have employed the notation used in \cite{Brandhuber:2021nez} for the Lorentz generators and, at the level of the generators of the Lorentz algebra, there is no distinction between (1,3) or (2,2) signature. Note that the existence of the parity-odd quadratic Casimir $C_2^{-}$ is a particular aspect of the four-dimensional Lorentz algebra (and so specific to two-dimensional CFTs).

Using the explicit form of the Lorentz generators in Appendix \ref{app: conformalgens} we find that the eigenvalues of the above Casimir operators acting on a conformal primary are given by
\begin{equation}\label{eq: quadcasmirvalues}
\begin{split}
    &c_2^+=2h(1-h)+2\hb(1-\hb)=\Delta(2-\Delta)-J^2 ,\\
    &c_2^{-}=2h(1-h)-2\hb(1-\hb)=2\,J(1-\Delta).
\end{split}
\end{equation}

In the case of a Euclidean CFT the spin $J$ is a discrete, half-integer weight and so cannot mix with the continuous weight $\Delta$; as such we can only map $\Delta \rightarrow 2-\Delta$ and $J \rightarrow -J$. Thus, the restricted Weyl group is $\ZZ_2$ \cite{Kravchuk:2018htv} and is generated by the shadow transform.

The Lorentzian CFT case, that is split signature bulk, is dealt with in Section \ref{sec: lightspace} where the new feature is that the spin $J$ is continuous and so the restricted Weyl group is larger with new elements which mix $\Delta$ and $J$. This gives rise to the light transform. We will also discuss unifying the Euclidean CFT and Lorentzian CFT pictures in Appendix \ref{app: complexify} by complexifying spacetime.

For now we focus on the Euclidean CFT case, where the shadow transform in affine coordinates is given by \cite{Osborn:2012vt, Ferrara:1972uq,Simmons-Duffin:2012juh,Crawley:2021ivb}
\begin{align}\label{eq: affineShadow}
\begin{split}
    &S \bigg\{\ket{\begin{pmatrix}
			z \\
			1  
		\end{pmatrix},\epsilon\begin{pmatrix}
			\zb \\
			1  
		\end{pmatrix}; h, \hb}\bigg\}\\& \qquad = \frac{i^{2h-2\hb}\Gamma(2-2\hb)}{2\pi i\,\Gamma(2h-1)} \int_{\mathbb{C} \mathbb{P}^1} d\zb\wedge dz\, (w-z)^{2h-2} (\wb-\zb)^{2\hb-2} \ket{\begin{pmatrix}
			z \\
			1  
		\end{pmatrix},\epsilon\begin{pmatrix}
			\zb \\
			1  
		\end{pmatrix}; h, \hb}\,.
\end{split}
\end{align}
where the normalisation of the above is chosen such that $S^2 = \text{Id}$, as shown in Appendix \ref{app: Ssquared}. 

The shadow transform \eqref{eq: affineShadow} can be written in homogeneous coordinates, in which case we integrate over the complex projective spinor $\lam$.  We define a shadowed celestial state as\footnote{Note that in our spinor bracket conventions, listed in Appendix \ref{app: conventions}, $\agl{\lam}{d\lam}\wedge\sqr{\lamt}{d\lamt}= -dz\wedge d\zb = 2i\, d(\text{Re}(z)) d(\text{Im}(z))$ which explains the additional normalisation by a factor $\frac{1}{2i}$ in the definition of the shadow transform when compared to \cite{Crawley:2021ivb}.}
\begin{equation}\label{eq: 13shadow}
    \boxed{\ket{\mu, \mut;1-h, 1-\hb } = \frac{i^{2h-2\hb}\Gamma(2-2\hb)}{2\pi i\,\Gamma(2h-1)}\int_{\mathbb{CP}^1} \agl{\lam}{d\lam} \wedge \sqr{\lamt}{d\lamt}\, \agl{ \lam}{\mu}^{2h-2} \sqr{ \mut}{\lamt}^{2\hb-2} \ket{\lam, \lamt;h, \hb}.}
\end{equation}
where we have used homogeneous coordinates $\mu, \mut$ to describe the shadowed celestial conformal primary\footnote{We have also departed slightly from the historical notation conventions appearing in the literature where the conjugate of $\ld{\alpha}$ is labelled with a tilde, $\tilde\mu_{\alpha}$. These previous conventions mean that a tilde always denotes a helicity $+\frac{1}{2}$ object, with our choice this is not the case. Instead, a tilde denotes an object with index $\dot\alpha$ transforming with a complex conjugate $\slC$ matrix.}. It is now trivial to check that the integral over $\lam, \lamt$ is projectively well-defined and the shadowed state has weights $h \rightarrow 1-h,\, \hb \rightarrow 1-\hb$ or $\Delta \rightarrow 2-\Delta,\,J \rightarrow -J$ under rescalings of the $\mu, \mut$ spinors - thanks to the economy of the homogeneous coordinates this is enough to show that it transforms as an $\slC$ conformal primary. Furthermore, the conjugate $\mu,\mut$ spinors take the same incoming/outgoing prescription as the $\lam,\lamt$ and this ensures that the integration kernel and measure are independent of $\epsilon$. 

The shadow transform above can be combined with the chiral Mellin transform to give a single transform which takes momentum eigenstates and maps them to shadowed conformal primaries on the celestial sphere
\begin{equation}\label{eq: 13compactshadcelstate}
\begin{split}
    \ket{\mu, \mut;1-h, 1-\hb} =& \frac{i^{2h-2\hb}\Gamma(2-2\hb)}{2\pi i\,\Gamma(2h-1)}\int_{\mathbb{CP}^1} \agl{\lam}{d\lam} \wedge \sqr{\lamt}{d\lamt}\, \agl{ \lam}{\mu}^{2h-2} \sqr{ \mut}{\lamt}^{2\hb-2} \\
    &\qquad \qquad \qquad \qquad \times \frac{1}{2\pi i}\int_{\mathbb{C}_*} \frac{d\ub}{\ub}\wedge \frac{du}{u}\, u^{2h}\ub^{2\hb} \vert u\lam,\ub\lamt\rangle \\
    =&\frac{i^{2h-2\hb}\Gamma(2-2\hb)}{(2\pi)^2\Gamma(2h-1)}\int_{\mathbb{C}_*^2} d^2\lamt\wedge\, d^2\lam \, \agl{ \lam}{\mu}^{2h-2} \sqr{ \mut}{\lamt}^{2\hb-2} \ket{\lam, \lamt} ,
\end{split}
\end{equation}
where we have used that the measure $d^2\lamt\wedge\, d^2\lam$ is given by $-u\,\ub\, \agl{\lam}{d\lam} \wedge \sqr{\lamt}{d\lamt}\wedge d\ub \wedge du$.

The shadow transformed state is labelled by homogeneous coordinates $\mud{\alpha}, \mtd{\alpha}$ to emphasise that it can also be regarded as living on an entirely different dual space to the usual momentum space and its associated celestial sphere. We will explore this point of view in the next section where we show that the shadow transformed state can be recovered by performing a chiral Mellin transform on a Fourier transformed state. This adds a new (1,3) signature version to the existing duality between light transformed conformal primaries and twistor eigenstates found in \cite{Sharma:2021gcz}.



\section{From Fourier to Shadow} \label{sec: fourier2shad}
\begin{figure}
\[
\begin{tikzcd}
\text{Momentum Basis} \arrow[rrr, "\scalebox{1.5}{$F$}"] \arrow[dd, "\scalebox{1.5}{$C_{h,\hb}$}"'] &  &  & \text{Fourier Basis} \arrow[dd, "\scalebox{1.5}{$C_{1-h,1-\hb}$}"] \\                &  &  & \\                                   
\text{Mellin Basis} \arrow[rrr, "\scalebox{1.5}{$S$}"]  &  &  &  \text{Shadow Basis}           
\end{tikzcd}
\]
\caption{We denote the chiral Mellin transform with weights $h,\hb$ by $C_{h,\hb}$, Fourier transform by $F$ and shadow transform by $S$.}\label{fig: fouriershadow}
\end{figure}
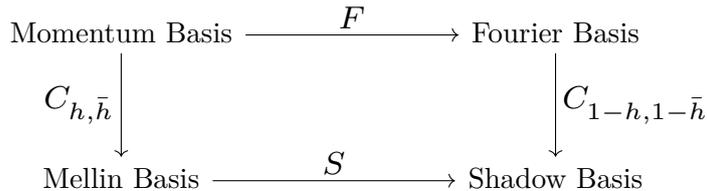

The expression for the shadow transform above \eqref{eq: 13compactshadcelstate} has many similarities with a Fourier transform acting on the spinors. In fact we can make this completely concrete in terms of a commuting diagram of integral transforms, shown in Figure \ref{fig: fouriershadow}.

We now prove the closure of this diagram by showing that a Fourier transform in spinor space followed by a chiral Mellin transform is exactly the shadow \eqref{eq: 13compactshadcelstate}. The Fourier transform we will consider is the following
\begin{equation}\label{eq: Fourier}
    \boxed{\vert \mu, \mut\rangle_ =\frac{1}{4\pi^2} \int_{\mathbb{C}^2}\!d^2\lambda\wedge d^2\tilde{\lambda}\,\, e^{i(\langle \ld{} \mu \rangle+[ \mut\lamt])}\vert\lam, \lamt\rangle\,.}
\end{equation}
Here we work in Minkowski signature so the spinors brackets are complex, and the integration kernel above can be written as $\exp(2i\Re \agl{\lam}{\mu})$\footnote{In our conventions, listed in Appendix \ref{app: conventions}, $\langle \ld{} \mu  \rangle^*=[ \mut\lamt]$.}.
As before, the conjugate $\mu,\mut$ spinors take the same incoming/outgoing prescription as the $\lam,\lamt$.

The transform \eqref{eq: Fourier} differs from the more conventional null Fourier integral
\begin{equation}\label{eq: NullFourier}
    \int d^4p \,\delta^{+}(p^2) e^{-ip.k}\,.
\end{equation}
However, such a transform is little group invariant and we have seen that the little group is crucial in controlling the holographic behaviour of scattering states. Our Fourier transform is the combination of two `twistor' or half-Fourier transforms - it has the same kernel, $i Z\cdot W$, as the full Fourier transform from twistor to dual twistor space but is integrated over an orthogonal region of the phase space $\lam,\lamt,\mu,\mut$. We can show by direct calculation that the Fourier transform \eqref{eq: Fourier} is self-inverse $F^2=\text{Id}$. This is different from the usual Fourier transform property of $F^2=P$ where $P$ is a parity or `time' reversal operator. In our case the angle/square bracket products are skew symmetric and this ensures that $F$ is exactly self-inverse.

A familiar consequence of the little group covariant Fourier transform is that the spinor $\mu$ now has helicity $+1/2$ and $\mut$ has helicity $-1/2$, which is the opposite to the spinors $\lam$ and $\lamt$. This can be seen by commuting the helicity operator through the Fourier transform
\begin{align}\label{eq: fourierhelicityoperator}
    \cJ:= &-\frac{1}{2} \lu{\alpha} \pld{\alpha} + \frac{1}{2} \ltu{\alpha} \pltd{\alpha} \sim \frac{1}{2}\muu{\alpha}\frac{\partial}{\partial \muu{\alpha}}-\frac{1}{2}\mtu{\alpha}\frac{\partial}{\partial \mtu{\alpha}}=:-\cJ_F,
\end{align}
where $\sim$ here means `conjugate to' under transform \eqref{eq: Fourier}.

The eigenvalue of the operator $\cJ$ remains unchanged, which is guaranteed by simply commuting $\cJ$ through the Fourier transform. However, we have also defined the operator $\cJ_F$, which generates a little group transformation of the $\mud{\alpha},\mtd{\alpha}$ spinors based on their $\slC$ holomorphicity. That is, \begin{equation}\label{eq: fourierJ}
    \cJ_F=\cJ\vert_{\lam\rightarrow\mu, \lamt\rightarrow \mut}\,,
\end{equation}
and so performs a little group transformation in a way that exactly mimics that of the spinors $\ld{\alpha}, \ltd{\alpha}$ 
\begin{equation}
   \vert \mu, \mut\rangle\rightarrow \vert e^{i\theta}\mu, e^{-i\theta}\mut\rangle\,.
\end{equation}
Thus if the eigenvalue of $\cJ$ is $J$ then the eigenvalue of $\cJ_F$ is $-J$. This is reminiscent of the action of the shadow transform and we can see this even more clearly with the dilatation operator
\begin{align}\label{eq: fourierdilation}
    d := \frac{1}{2}\lu{\alpha} \pld{\alpha} + \frac{1}{2} \ltu{\alpha} \pltd{\alpha} +1 \sim  -\frac{1}{2}\muu{\alpha}\frac{\partial}{\partial \muu{\alpha}}-\frac{1}{2}\mtu{\alpha}\frac{\partial}{\partial \mtu{\alpha}}-1 =: -d_F,
\end{align}
where again we have defined the operator $d_F$ which is defined in Fourier space and generates a dilatation of the $\mu, \mut$ spinors.

Now we transform to the celestial sphere by performing a chiral Mellin transform \emph{on a Fourier state}  \eqref{eq: Fourier} with weights $k,\kb$. Then relations \eqref{eq: fourierhelicityoperator} and \eqref{eq: fourierdilation} become
\begin{equation}
\begin{split}
    \cJ=h-\hb \sim -(k-\kb)=-\cJ_F,\\
    \quad d=-h-\hb+1 \sim k+\kb-1=-d_F,
\end{split}
\end{equation}
so we must identify $k=1-h$ and $\kb=1-\hb$ which are the same shifts generated by a shadow transform.

The Mellin transformed Fourier state is given by 
\begin{equation}
\begin{split}
    \vert \mu, \mut;k,\kb\rangle =& \frac{1}{2\pi i}\int_{\mathbb{C}^*} \frac{d\tb}{\tb}\wedge \frac{dt}{t} t^{2k}\tb^{2\kb}\vert t\mu, \tb \mut\rangle\\
    =&\frac{1}{2\pi i}\int_{\mathbb{C}^*} \frac{d\tb}{\tb}\wedge \frac{dt}{t} t^{2k}\tb^{2\kb}\frac{1}{4\pi^2}\int_{\mathbb{C}^2}\!d^2\lam\wedge d^2\tilde{\lam}\, e^{i t\agl{\lam}{ \mu} }e^{i\tb\sqr{\mut}{\tilde{\lam}}}\vert\lam, \tilde{\lam} \rangle\,,
\end{split}
\end{equation}
which we will now show is a shadowed conformal primary. First we change variables $\sigma=t\agl{\lam}{\mu}$, $\bar{\sigma}= \tb \sqr{\mut}{\tilde{\lam}}$ to obtain 
 \begin{equation}
      \vert \mu, \mut;k,\kb\rangle = \frac{1}{2\pi i}\int_{\mathbb{C}^*}  \frac{d\bar{\sigma}}{\bar{\sigma}}\wedge\frac{d\sigma}{\sigma} \sigma^{2k}\bar{\sigma}^{2\kb}e^{i(\sigma+\bar{\sigma})}\int_{\mathbb{C}^2}\!d^2\lam\wedge d^2\tilde{\lam}\, \agl{\lam}{ \mu}^{-2k} \sqr{\mut}{\tilde{\lam}}^{-2\kb}\vert \lam, \tilde{\lam}\rangle\,.
 \end{equation}
We can write this in a form similar to the shadow transform if we encode the expected shifts of the weights as $k=1-h$, $\kb=1-\hb$
\begin{equation}\label{eq: chiralFourier2}
     \vert \mu, \mut;1-h,1-\hb\rangle =\frac{\cI}{(2\pi i)(4\pi^2)} \int_{\mathbb{C}^2}\!d^2\lam\wedge d^2\tilde{\lam}\, \agl{\lam}{ \mu}^{2h-2} \sqr{\mut}{\tilde{\lam}}^{2\hb-2}\vert \lam, \tilde{\lam} \rangle\,,
\end{equation}
where we have defined
\begin{equation}\label{eq: Iintegralfouriershad}
    \cI :=\int_{\mathbb{C}^*}  d\bar\sigma \wedge d\sigma\, \sigma^{1-2h}\bar{\sigma}^{1-2\hb}e^{i(\sigma+\bar{\sigma})}\,.
\end{equation}

The final, although nontrivial, step is to evaluate the integral $\cI$ and compare equation \eqref{eq: chiralFourier2} with the shadow transform \eqref{eq: 13compactshadcelstate}. The evaluation of $\cI$ is contained in Appendix \ref{app: fourierequalsshadow} and we find that
\begin{equation}
    \cI = 2\pi i\, i^{-2J}\frac{\Gamma(2-2h)}{\Gamma(2\hb-1)}= 2\pi i\, i^{2J}\frac{\Gamma(2-2\hb)}{\Gamma(2h-1)} .
\end{equation}

To compare to the shadow transform we use the second form above. Plugging this expression for $\cI$ back in to \eqref{eq: chiralFourier2} we find
\begin{equation}
    \begin{split}
        \vert \mu, \mut;1-h,1-\hb\rangle =\frac{i^{2h-2\hb}\Gamma(2-2\hb)}{4\pi^2\Gamma(2h-1)} \int_{\mathbb{C}^2}\!d^2\lam\wedge d^2\tilde{\lam}\, \agl{\lam}{ \mu}^{2h-2} \sqr{\mut}{\tilde{\lam}}^{2\hb-2}\vert\lam, \tilde{\lam} \rangle\,,
    \end{split}
\end{equation}
which agrees exactly with the shadow transform \eqref{eq: 13compactshadcelstate} hence proving that the  diagram of integral transforms commutes. In Appendix \ref{app: fourierequalsshadow}, we also prove this equivalence in the other direction, by performing an inverse chiral Mellin transform on a shadowed conformal primary to recover a Fourier transformed momentum eigenstate.


\section{Fourier Amplitudes}\label{sec: fourieramps}

Having established the commuting diagram in Figure \ref{fig: fouriershadow}, we now have a new method of computing shadow transformed celestial amplitudes. One can take an amplitude in momentum space and Fourier transform any number of the external legs with \eqref{eq: Fourier}. Then, after a chiral Mellin transform, these legs become shadow transformed operators in a conformal correlator. We can choose how many legs to shadow transform, for example in the literature single leg shadowed gluon amplitudes have been considered in \cite{Fan:2021isc, Fan:2021pbp} while in \cite{Chang:2022jut} both single leg and all-leg shadow transforms were computed for amplitudes involving massless and massive scalars.

To illustrate the logic above, we examine a simple amplitude: the tree-level four-point amplitude from massless $\phi^4$ theory
\begin{equation}
    A_4^{\phi^4}= g \, \delta^{(4)}\bigg(\sum_{i=1}^4 \lam_{i\alpha}\lamt_{i\dot{\alpha}}\bigg)\, .
\end{equation}
where $g$ is the dimensionless coupling. This theory has been considered before in the celestial context  at loop level in \cite{Banerjee:2017jeg}. If we simply Mellin transform this amplitude we will run into the usual delta function constraint $\delta(z-\bar{z})$ of four-point kinematics, where $z$ is the conformal cross-ratio on the celestial sphere \cite{Pasterski:2017ylz}. This constraint demands that the external legs lie on a great circle, which is formed from the intersection of the scattering plane and the celestial sphere. To remove this constraint one might hope that performing various shadow transforms will ``smear out'' this singularity. As demonstrated above, we can do this by first performing Fourier transforms on, say, the first $m$ legs
\begin{equation}\label{eq: kLgegFourier}
    \begin{split}
         A_4^{m,\phi^4} \coloneqq \frac{1}{(4\pi^2)^{m}}\int\prod_{i=1}^m \left(  d^2{\lam_i}\wedge d^2\!\lamt_{i}\, e^{i(\langle \ld{} \mu \rangle+[ \mut\lamt])}\right) A_4^{\phi^4}\,,
    \end{split}
\end{equation}
 and then performing Mellin transforms on all legs.
We can rewrite the momentum conserving delta function appearing in the amplitude as an integral
\begin{equation}
    \delta^{(4)}\bigg(\sum_{j=1}^4 \lam_{j}\lamt_{j}\bigg) = \frac{1}{(2\pi)^4}\int d^4x\, e^{i\sum_{j=1}^4 [\lamt_j\lvert x\lvert\lam_j\rangle}\, .
\end{equation} 
Substituting this into  \eqref{eq: kLgegFourier} and exchanging the order of integration we have
\begin{equation}
    A_4^{m,\phi^4} = \frac{g}{(2\pi)^{2m+4}}\int d^4x \int\prod_{i=1}^m d^2{\lam_i}\wedge d^2\!\lamt_{i}\, \exp(i\sum_{j=1}^4 [\lamt_j\lvert x\lvert\lam_j\rangle + \sum_{i=1}^m i(\langle \ld{} \mu \rangle+[ \mut\lamt])) \, .
\end{equation}
The $\lam_i$ integrals are just complex Gaussian integrals which we can perform by completing the square 
\begin{equation}
     A_4^{m,\phi^4} =\frac{ g }{(2\pi)^{4}} \int d^4x \frac{1}{(x^2)^{m}} \exp(-i\sum_{i=1}^m  [\tilde{\mu}_{i}\vert x^{-1}\vert\mu_{i}\rangle + i\sum_{j=m+1}^4 [\lamt_{j}\vert x\vert\lam_{j}\rangle)\,.
\end{equation}
Generically performing this final $x$ integral is difficult, however, if we choose to Fourier transform all legs  we find
\begin{equation}\label{eq: n-Fourier}
    A_4^{4,\phi^4} = \frac{g(-1)^4}{(2\pi)^{4}} \int d^4(x^{-1})\,  \exp(-i\sum_{i=1}^n  [\tilde{\mu}_{i}\vert x^{-1}\vert\mu_{i}\rangle)=   g\,\delta^{(4)}\bigg(\sum_{j=1}^4 \mu_{j}\tilde{\mu}_{j}\bigg) .
\end{equation}
The Fourier transformed amplitude we have obtained above is identical in form to the original amplitude. However, the delta function now imposes ``special conformal conservation'' since  the generator of special conformal transformations $k_{\alpha\dot{\alpha}}$ in the Fourier basis acts multiplicatively. Indeed, the momentum and special conformal generators swap roles under the action of the Fourier transform
\begin{equation}
   k_{\alpha\dot{\alpha}}=\partdiv{}{\lam^\alpha}\partdiv{}{\lamt^{\dot{\alpha}}}\xrightarrow[]{F}  -\mu_{\alpha}\tilde{\mu}_{\dot{\alpha}}\,, \quad p_{\alpha\dot{\alpha}}= \ld{\alpha}\ltd{\alpha}\xrightarrow[]{F} - \partdiv{}{\mu^\alpha}\partdiv{}{\tilde{\mu}^{\dot{\alpha}}}
\end{equation}
This can be seen by simply pulling the usual momentum space generator through the Fourier transform, see Appendix \ref{app: conformalgens} for more details. In fact, we could have predicted the delta function in \eqref{eq: n-Fourier} since tree-level amplitudes in $\phi^4$ theory are conformally invariant. Note that the all-leg Fourier amplitude \eqref{eq: n-Fourier} has mass dimension +4 which is the opposite of the original amplitude and this trend continues at $n$-points. This flipping of the mass dimension, which is due to the mass dimension of the measure of \eqref{eq: Fourier}, is reminiscent of the action of conformal inversion. In Appendix \ref{app: conformalgens} we argue that the variable $ k_{\alpha\dot{\alpha}}= -\mu_{\alpha}\tilde{\mu}_{\dot{\alpha}}$ is the dual momentum associated to conformally inverted position space, and so the action of Fourier transforming induces an active conformal inversion transformation in position space \footnote{That Fourier conjugation is directly related to bulk conformal inversion is not too surprising, given that conformal inversion on the boundary variables $z, \zb$ is related to the non-identity element of the Weyl group, that is the shadow transform \cite{Koller:1974ut}.}. That an active conformal inversion should be associated with exchanging $\lam$ and $\mu$ spinors was also noted in \cite{Korchemsky:2009jv}.

If we now perform a Mellin transform on every leg of the Fourier transformed amplitude \eqref{eq: n-Fourier} we will obtain the all-leg shadowed celestial amplitude. However, this will simply give us the original celestial amplitude but with $\lam\rightarrow\mu$ and $\lamt\rightarrow \tilde{\mu}$ and, in particular, there will still be a delta function of the conformal cross-ratios! Thus although the shadow transform can `smear out' delta function singularities this is clearly not always the case - the all-leg shadow can re-introduce delta functions of its own.

One would now like to study Fourier amplitudes of gravitons and gluons, however the integrals required become significantly more involved than the above example, even at low points. Thus, here we will simply make some comments on Fourier transformed pure gluon amplitudes and leave the further study of these ideas to future work.
As discussed already, and explored more in Appendix \ref{app: conformalgens}, under the Fourier transform \eqref{eq: Fourier} the conformal generators are related according to an automorphism of the algebra. The generators $d, J, m, \bar m$ are sent to minus their equivalents in $\mu, \mut$ while the momentum and special conformal transformations are exchanged. In short, a Fourier transformed gluon amplitude inherits conformal symmetry and is annihilated by identical differential operators, but now written in terms of $\mu, \mut$. It should then have significant similarities with the momentum space amplitude. For example, special conformal symmetry is generated by the multiplicative operator $\mud{\alpha}\mtd{\alpha}$ (summed over all legs) and hence the gluon amplitude with all legs Fourier transformed should have a delta function which enforces special conformal invariance.

The difficulty of actually performing Fourier transforms \eqref{eq: Fourier} on amplitudes is in part due to the fact that the spinors $\lam,\lamt$ are complex conjugates in Lorentzian signature. The integrals above become more tractable if instead we work in split signature spacetime where the spinors $\lam,\lamt$ are real and independent. Indeed, this will allow us to show in Appendix \ref{app: all legfouriergluon} that the analogous all-leg Fourier transformed tree-level gluon amplitude in split signature takes the same form as the original amplitude. This extends the result shown here in $\phi^4$ theory to pure Yang-Mills, but only in this particular spacetime signature.

Thus, in the following section and for the rest of the paper we will work in split signature. We begin by first repeating the analysis of the (extended-)little group for null momentum which is qualitatively different in (2,2) signature. Our main purpose in using split signature is that this will allow us to construct ``half'' versions of the the Fourier and Mellin transforms which only act on $\lam$ or $\lamt$ and, as was shown in \cite{Sharma:2021gcz,Sharma:2022talk}, these are related to the ``half'' version of the shadow transform: the light transform. Using these new transformations, and the relations between them, we will see in Section \ref{sec: celesttwistamps} that certain light transformed celestial amplitudes take a particularly simple form.


\section{Split Signature Spacetime}\label{sec: 22sig}

We now shift our focus away from Lorentzian spacetime and consider split signature spacetime instead. As we shall see this has a number of benefits including allowing connections to twistor space, see for example \cite{Witten:2003nn, Korchemsky:2009jv, Mason:2009sa}. Our aim is to study celestial twistor amplitudes in (2,2) signature and link these to light transformed correlators. Celestial twistor amplitudes are defined, as one might expect, through a chiral Mellin transform of an amplitude in twistor space. In order to land on the correct form of the chiral Mellin transform we follow an analogous path to that of Section \ref{sec: 13Sig} - we start by examining the little group in (2,2) signature. 

\subsection{Little Group}\label{sec: 22LG}
 
In split signature spacetime, $(+,-,+,-)$, we require that null momenta $p_{\alpha\dot{\alpha}}=\ld{\alpha}\ltd{\alpha}$ be real. This is achieved by making $\lam$ and $\lamt$ independent and \emph{real} two component spinors
\begin{equation}\label{eq: 22realitycondition} 
    \lam=\lam^*,\quad \lamt = \lamt^*\,.
\end{equation}
In $(2,2)$ spacetime the null boundary has only one component, thus asymptotic states are not labelled by an incoming/outgoing parameter. In fact, whether momentum is future or past pointing is not a Lorentz invariant notion in $(2,2)$ since there exist $\text{SO}^+(2,2)$ transformations which take $p_{\alpha\dot{\alpha}}$ to $-p_{\alpha\dot{\alpha}}$, as has been noted before in \cite{Atanasov:2021oyu, Hu:2022syq, Sharma:2022talk} for example. The space of spinors in split signature is thus $(\lambda,\lambdat) \in \RR_*^2 \times\RR_*^2$.

In $(2,2)$ the little group is $\RR_* =\RR_{+} \times \ZZ_2$ and its action on the spinors is given by
\begin{equation}\label{eq: 22LG}
     \ld{\alpha} \mapsto c\, \ld{\alpha}, \quad \ltd{\alpha} \mapsto c^{-1} \ltd{\alpha},
 \end{equation}
 for any non-zero real number $c$. In order to land on the space of massless momenta in $(2,2)$ signature we then quotient by $\RR_*$ to give locally $p_{\alpha\dot{\alpha}} \in \RR_*^2 \times \RRP^1$.

The representations of the little group $\RR_* =\RR_{+} \times \ZZ_2$ are labelled by two numbers: 
\begin{enumerate}
    \item A continuous imaginary helicity $J \in i\RR$ which labels a representation  $\RR_+$,
    \item A discrete `helicity' $s_J \in {0,1}$ labeling the two representations of $\ZZ_2$.
\end{enumerate}
The asymptotic particle states are then denoted as
\begin{equation}\label{eq: 22momentumEigenstate}
    \vert \lam, \lamt ; J, s_J\rangle\, ,
\end{equation}
 and transform homogeneously under the little group 
\begin{equation}\label{eq: 22LittleGroup}
     \boxed{\vert c\lam, c^{-1}\lamt ; J, s_J\rangle = \abs{c}^{-2\,J}\sgn(c)^{-s_J}\vert \lam, \lamt , J, s_J\rangle\, .}
\end{equation}
The appearance of absolute values and $\sgn$ functions is a completely general feature of split signature spacetime objects and is a consequence of the real disconnected little group $\RR_*=\RR_+\times \ZZ_2$.

The `helicity' $s_J$ denotes states which are even $ s_J=0$ and odd $ s_J=1$ under the little group transformation which flips the sign of both spinors
\begin{equation}\label{eq: Z2LG}
     \vert {-}\lam, {-}\lamt ; J,  s_J\rangle = ({-}1)^{- s_J}\vert \lam, \lamt , J, s_J\rangle\,.
\end{equation}
The fact that $s_J$ controls even and odd symmetry in (2,2) suggests we should associate $s_J=0$ with bosons and $s_J=1$ with fermions\footnote{In Appendix \ref{app: complexify} we demonstrate, by considering complexified spacetime, that $s_J$ is directly associated to the (1,3) helicity $J_{1,3}$, giving $s_J=0$  for integer $J_{1,3}$ and $s_J=1$ for half-integer $J_{1,3}$.}. Interestingly, this ``Bose-Fermi symmetry'' is completely independent of the  continuous helicity $J$ of the particles. Split signature is generally discussed as an analytic continuation of theories in $(1,3)$ signature with half-integer helicity, and we shall follow this approach here as well. Bootstrapping theories exclusively in $(2,2)$ signature using its unique little group is an interesting problem in its own right. In the context of CFTs, continuous helicity has appeared multiple times, see for example \cite{Kravchuk:2018htv, Caron-Huot:2017vep, Homrich:2022mmd}.

Finally, in $(2,2)$  signature the universal cover of the Lorentz group is $\slR\times \slR$ and it acts on states through  
\begin{equation} 
    \ket{\ld{\alpha} , \ltd{\alpha};J, s_J}\rightarrow \ket{M_\alpha^{\,\,\beta}\ld{\beta} , \tilde{M}_{\dot{\alpha}}^{\,\,\dot{\beta}}\ltd{\beta}; J, s_J }\,,
\end{equation}
where $M_\alpha^{\,\,\beta}$ and $\tilde{M}_{\dot{\alpha}}^{\,\,\dot{\beta}}$ are independent $\slR$ matrices.


\subsection{Banerjee's Extended Little Group}\label{sec: 22ELG}

The extended little group in (2,2) is given by independent rescalings of either spinor by a non-zero real number. It is the group $\RR_* \times \RR_* = \RR_+ \times \RR_+ \times \ZZ_2 \times \ZZ_2$ and acts according to the transformation 
\begin{equation}\label{eq: 22ELG}
    \lam\rightarrow y \lam, \quad \lamt\rightarrow  \yb \lamt\,,
\end{equation}
where $y$ and $\yb$ are non-zero real numbers. On top of the usual little group \eqref{eq: 22LG} this corresponds to an extension which includes boost rescalings of the momentum by any non-zero real number
\begin{equation}\label{eq: 22momboosts}
p \rightarrow b p,
\end{equation}
where $b$ may be negative. The existence of Lorentz transformations which flip the sign of the momenta is a special feature of (2,2) signature. As mentioned previously, there is a single null asymptotic boundary in split signature and all scattering states are defined on this space. This leads to the novel situation of a scattering vector which combines with a vector of asymptotic particle states to compute the amplitude for a particle `process' to occur in the bulk \cite{Atanasov:2021oyu}.

We conclude that in (2,2) signature the spinors $\lam, \lamt$ serve as real homogeneous coordinates for the projective space $\RRP^1 \times \RRP^1= \RR_*^2\times \RR_*^2/(\RR_*\times \RR_*)$, which is the celestial torus. Conformal primaries on the  celestial torus are built from (2,2) signature celestial states with imaginary helicity $J$ and a discrete helicity $s_J$ as well as two new weights:
\begin{enumerate}
    \item a complex conformal dimension $\Delta \in 1+i\RR$ labelling representations of the positive real boosts,
    \item a discrete weight $s_{h} \in \{0,1\}$ which labels the $\ZZ_2$ symmetry under $\lam \rightarrow -\lam$. Correspondingly, we have $s_{\hb}\in \{0,1\}$ which labels the $\ZZ_2$ symmetry under $\lamt \rightarrow -\lamt$.
\end{enumerate} 
Hence, (2,2) signature celestial states are denoted as
\begin{equation}\label{eq: 22CelestState}
    \vert \lam,\lamt; J, s_J, \Delta, s_{h} \rangle \equiv \vert \lam,\lamt; h, s_h, \hb, s_{\hb} \rangle\, ,
\end{equation}
where we have also defined $h:=\frac{1}{2}(\Delta+J)$\footnote{We use $h, \hb$ for the weights in both (1,3) and (2,2) signature despite them being distinct quantities. Which of the two is intended should be clear from context.} and $\hb:=\frac{1}{2}(\Delta-J)$. We have taken a slightly altered route in defining (2,2) signature states in this way. 
    
The transformation law of the states \eqref{eq: 22CelestState} under the extended little group re-scalings is
\begin{equation}\label{eq: 22ELGTransformation}
    \boxed{\vert y\lam,\yb\lamt; h, s_h, \hb, s_{\hb} \rangle = \abs{y}^{-2h} \sgn(y)^{- s_h} \abs{\yb}^{-2\hb} \sgn(\yb)^{-s_{\hb}}\vert \lam,\lamt; h, s_h, \hb, s_{\hb} \rangle \,,}
\end{equation}
where once again we have appearance of absolute values and $\sgn$ functions which is due to the unique topological properties of $\RR^{2,2}$ \footnote{An excellent explanation of these properties of split signature is given throughout \cite{Mason:2009sa}.}. Note that only positive numbers, like $\abs{y}$, are raised to a complex power, while $\sgn(y)$ is raised to either the power zero or one\footnote{Similarly, in (1,3) signature wherever we have a phase it is also always raised to integer powers $2J$ while the modulus of a complex number is raised to a complex power $\Delta$.}. This has the benefit of removing branch cut ambiguities arising from raising a negative number to a complex power.

The (2,2) signature celestial states transform under $\slR \times \slR$ like conformal primaries in a two-dimensional \emph{Lorentzian} CFT
\begin{equation}\label{eq: 22HomoSL2R}
\begin{split}
    &\vert\ld{\alpha} , \ltd{\alpha};h, s_h, \hb, s_{\hb}\rangle \rightarrow \vert M_\alpha^\beta\ld{\beta} , \tilde{M}_{\dot{\alpha}}^{\dot{\beta}}\ltd{\beta};h, s_h, \hb, s_{\hb} \rangle\,
    \\
    &\ket{\begin{pmatrix}
			z \\
			1  
		\end{pmatrix},\begin{pmatrix}
			\zb \\
			1  
		\end{pmatrix}; h, s_h, \hb, s_{\hb}}\\
		&\rightarrow \abs{cz+d}^{-2h}\sgn(cz+d)^{-s_{h}} \abs{\ct\zb+\dt}^{-2\hb}\sgn(\ct\zb+\dt)^{s_{\hb}}\ket{\begin{pmatrix}
			z' \\
			1  
		\end{pmatrix},\begin{pmatrix}
			\zb' \\
			1  
		\end{pmatrix}; h, s_h, \hb, s_{\hb}}\, .
\end{split}
\end{equation}

\begin{figure}
\includegraphics[width=\linewidth]{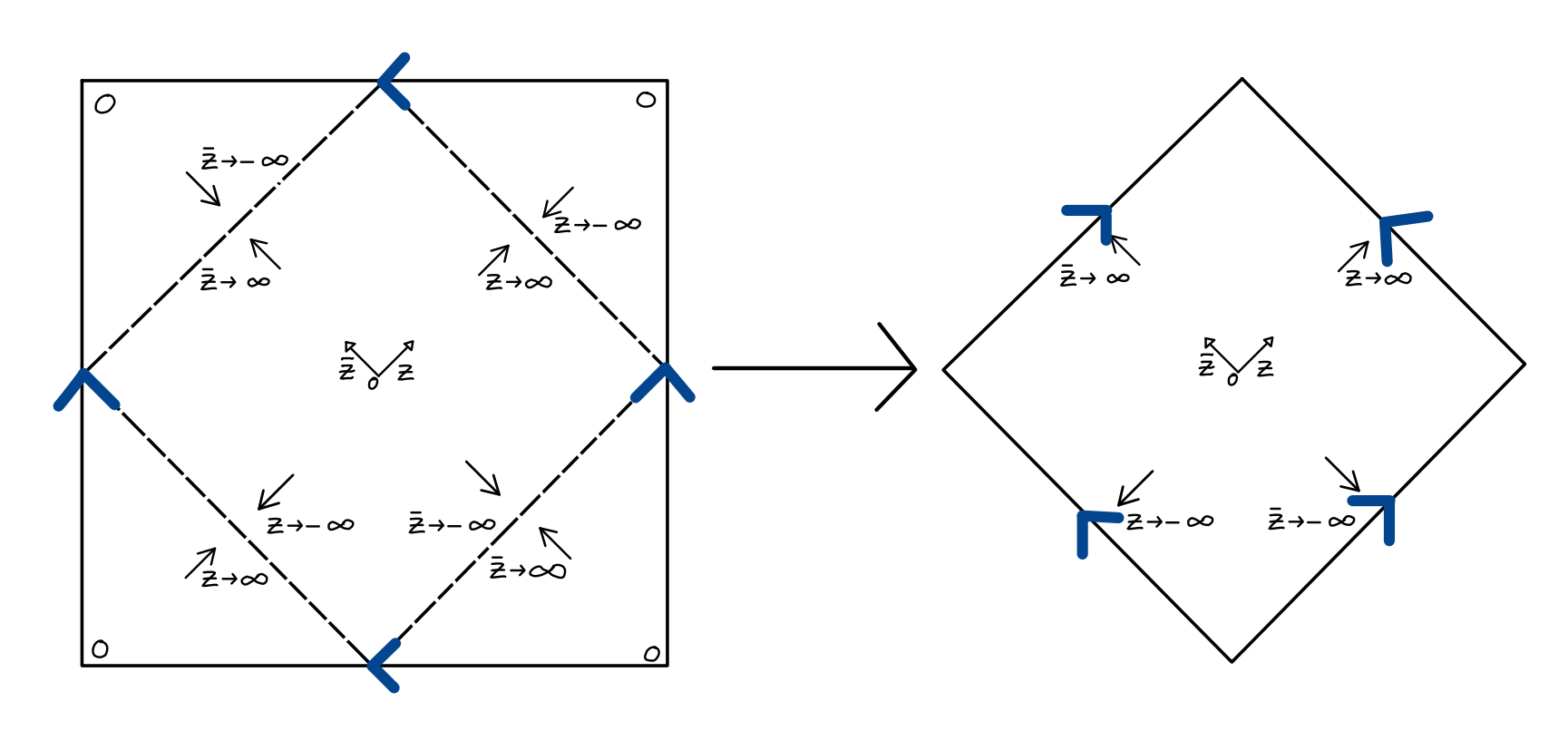}
\caption{The quotienting of the celestial torus with two patches (first diagram) by  $\lam\rightarrow-\lam$ and $\lamt \rightarrow -\lamt$ results in a smaller single patch torus (second diagram). Intuitively one obtains the second diagram by folding the edges in along the dotted lines. The blue arrows here indicate that the opposite edges are identified. }\label{fig: torusDiagram}
\end{figure}

It is worth pausing here briefly to explain the relationship between the states described above and others appearing in the celestial literature \cite{Sharma:2021gcz,Sharma:2022talk}. By including the the sign transformations $\lam\rightarrow-\lam$ and $\lamt \rightarrow -\lamt$ we are quotienting the space $\RR_*^2 \times \RR_*^2$, by $\RR_* \times \RR_*$ and not just $\RR_+ \times \RR_+ $. This means that instead of considering two patches on the celestial torus we consider only one. To see this consider just one chirality of spinor, say $\lam$. If one uses only positive rescalings in the little group, then the two affine patches:
\begin{equation}
    \begin{pmatrix}
			z \\
			1  
		\end{pmatrix}, \quad  \begin{pmatrix}
			-z \\
			-1  
		\end{pmatrix}\,,
\end{equation}
are distinct, and together cover one circle of the celestial torus. By including negative rescalings we glue these two patches on top of each other, see Figure \ref{fig: torusDiagram}. As discussed in Section \ref{sec: prelude} we do not lose information by doing this, instead we simply have more asymptotic states labelled by the representations  $\mathbb{Z}_2\times \mathbb{Z}_2$: $s_h$ and $s_{\hb}$. In the two patch approach states are labelled by the patch they belong to, but can move from one patch to the other under the action of $SL(2,\RR)\times SL(2,\RR)$. In the one patch approach however, states are labelled by their symmetry properties. These symmetries can be used to constrain the corresponding celestial amplitudes as we shall see explicitly in Section \ref{sec: celesttwistamps}.

\subsection{Chiral Mellin Transform}\label{sec: 22chiralMellin}

Based on the split signature homogeneity property \eqref{eq: 22ELGTransformation}, celestial states are homogeneous functions of a pair of real spinors with weights $(h,s_h, \hb, s_{\hb})$, and as usual we build homogeneous functions via Mellin transforms. A celestial state in (2,2) signature is then defined by the following chiral Mellin transform acting on a momentum eigenstate
\begin{align}\label{eq: 22chiralMellin}
    \boxed{\vert \lam,\lamt; h , s_h, \hb,        s_{\hb} \rangle
    :=\frac{1}{4} \int_{\mathbb{R}_*\times\mathbb{R}_*}\!\frac{d\ub}{\abs{\ub}} \wedge \frac{du}{\abs{u}} \, \abs{u}^{2h}\, \abs{\ub}^{2\hb}\,\sgn(u)^{s_h}\sgn(\ub)^{s_{\hb}}\vert  u\lam,\ub\lamt \rangle.}
\end{align}
Once again we note that this transform is in the general form of \eqref{eq: repDecombPrelude}, and again we denote a chiral Mellin with weights $h,\hb,s_h,s_{\hb}$ as $C_{h,\hb}$. In \eqref{eq: 22chiralMellin}  we integrate over $\RR_*$ scalings for each spinor, which includes both positive and negative `energies'. Since in $(2,2)$ signature the spinors are real and independent the integrals over $u,\ub$ are separable and each constitutes a Mellin transform on its own - this was not the case in  $(1,3)$ signature. It can be easily checked (for example by checking the homogeneity or expressing it in affine coordinates $z, \zb$) that the above state transforms as a conformal primary on the celestial torus.

Due to the altered (2,2) signature chiral Mellin transform, the four-dimensional \emph{bulk} conformal symmetry generators in (2,2) signature receive slight modifications to take into account the fact that they also carry the discrete weights $s_h, s_{\hb}$. These generators are described in Appendix \ref{app: conformalgens}.

We can see the even and odd states explicitly by writing our states in terms of a chiral Mellin transform over just positive energies
\begin{align}\label{eq: 22chiralMellinevenodd}
\begin{split}
    \vert \lam,\lamt; h , s_h, \hb, s_{\hb} \rangle
    &:= \int_{\mathbb{R}^+\times\mathbb{R}^+}\!\frac{d\ub}{\ub} \wedge \frac{du}{u} \, u^{2h}\, \ub^{2\hb}\,
    \vert u\lam,\ub\lamt; s_{h}, s_{\hb} \rangle\,,
    \end{split}
\end{align}
where we regard the above as performing two of the usual $\RR^+$-Mellin transforms over $\ld{}, \lamt$ on the even/odd projection states given below
\begin{align}\label{eq: symmetrisedstate}
\begin{split}
     \vert \lam,\lamt; s_{h}, s_{\hb} \rangle&\coloneqq \frac{1}{4}\sum_{\epsilon=\pm1}\sum_{\epsilonb=\pm1}\epsilon^{\,s_h}\epsilonb^{\,\,s_{\hb}}\vert \epsilon\,\lam,\epsilonb\,\lamt \rangle \,. 
\end{split}
\end{align}
Clearly the projection to an even/odd state loses information about the original state unless we consider all the above combinations in tandem\footnote{The method of summing over incoming/outgoing parameters for (2,2) celestial amplitudes has been used in \cite{De:2022gjn,Hu:2022syq}, and was even used in (1,3) signature in \cite{Fan:2021isc}. In these only a fully symmetrised sum was considered and so only a part of the amplitude was studied.}. The Mellin transform over $\RR^+$ preserves information about the state since it is invertible. We can recover the original state from the following inverse Mellin transform
\begin{align}\label{eq: 22chiralMellininverse}
\begin{split}
         \vert  \lam,\lamt \rangle=\int_{a-i\infty}^{a+i\infty} \frac{d(2h)}{2\pi i} \int_{\at-i\infty}^{\at+i\infty} \frac{d(2\hb)}{2\pi i}\,\,  \sum_{s_h=0,1}\sum_{s_{\hb}=0,1} \vert \lam,\lamt; h , s_h, \hb, s_{\hb} \rangle\,.
\end{split}
\end{align}
This is again simply a sum over representations and follows the general form of \eqref{eq: repSumPrelude}.
The (2,2) signature chiral Mellin transform \eqref{eq: 22chiralMellin} has inverse \eqref{eq: 22chiralMellininverse} since it is just a symmetrised sum of products of two independent Mellin transforms and the proof is analogous to that contained in Appendix \ref{sec: Mellininversion}.

As with (1,3) signature celestial states, in (2,2) signature we will often act with the chiral Mellin transform on a momentum eigenstate which already transforms with some helicity under the little group. For a chiral Mellin transform of a state in (2,2) signature with imaginary bulk helicity $l$ and discrete helicity $s_l$ we can simplify the symmetrisation sum in \eqref{eq: symmetrisedstate} using a little group transformation
\begin{align}\label{eq: symmetrisedstate2}
\begin{split}
     \vert \lam,\lamt; s_{h}, s_{\hb}, l, s_l \rangle&\coloneqq \frac{1}{4}(1+(-1)^{ s_h+s_{\hb}+s_l})\bigg(\vert  \lam,\lamt, l, s_l\rangle+(-1)^{s_h}\,\vert  {-}\lam,\lamt , l, s_l\rangle\bigg)\\
     &=\delta_{s_J,s_l}\frac{1}{2}\sum_{\epsilon=\pm1}\epsilon^{s_h}\vert  \epsilon\,\lam,\lamt, l, s_l\rangle\,. 
\end{split}
\end{align}

Furthermore, for such a state, the integral over the little group degrees of freedom in \eqref{eq: 22chiralMellininverse} can also be done directly
\begin{align}\label{eq: 22nonchiralMellin}
    \vert \lam,\lamt; h , s_h, \hb,        s_{\hb} , l, s_l\rangle
    =\pi i\, \delta(J - l) \delta_{s_J,s_l}\int_{0}^{\infty}\!d\omega \, \omega^{\Delta-1}\,\sum_{\epsilon=\pm1}\epsilon^{s_h}\vert  \epsilon\,\sqrt{\omega}\lam,\sqrt{\omega}\lamt, l, s_l\rangle\,.
\end{align}
We have recovered a Dirac delta from the non-compact integral, which follows from the important identity
\begin{align}\label{eq: Mellindelta}
    \int_0^{\infty} dx\, x^{\alpha-1} = 2\pi i \,\delta(\alpha)\,.
\end{align}
This holds for $\alpha=i\gamma$ pure imaginary for which we can write $\delta(\alpha)= -i\delta(\gamma)$. In the above case \eqref{eq: 22nonchiralMellin} this is true by default when we consider representations of the little group which have imaginary helicity\footnote{It is often necessary to relax this condition and allow $\alpha$ to be any complex number. In such cases we recover a generalised delta function \cite{Donnay:2020guq}. This would have been crucial when we (2,2) Mellin transform analytically extended amplitudes from (1,3) signature which have half-integer helicity, in fact we sidestep this by performing only the half-Mellin transforms of \cite{Sharma:2021gcz}.}. The constant function $f(x)=1$ is Mellin-dual to a delta function, which is the analogue of the fact that the Fourier-dual of $f(y)=1$ is also a delta function. Namely, we can show \eqref{eq: Mellindelta} by making the change of variable $x=e^y$ and using the representation of the delta function as
\begin{align}\label{eq: fourierdelta}
   \delta(\gamma) = \frac{1}{2\pi}\,\int_{-\infty}^{\infty} dy\, e^{i\gamma y}\,.
\end{align}
 Since $f(x)=1$ is always order $\mathcal{O}(1)$ the `strip of definition' where the Mellin transform is only marginally convergent is just the imaginary axis. We hence require a regularisation scheme given by
 \begin{align}
    \frac{1}{2\pi}\,\int_{-\infty}^{\infty} dy\, e^{i\gamma y} =  \frac{1}{2\pi}\,\int_{-\infty}^{\infty} dy\, e^{i\gamma y-\epsilon \abs{y}} =  \frac{1}{2\pi}\,\bigg(\frac{1}{i\gamma+\epsilon}-\frac{1}{i\gamma-\epsilon}\bigg) = \delta(\gamma)\,,
\end{align}
where $\epsilon$ is a small positive number.
Regularisation schemes like the one above will be ubiquitous in the rest of the paper as we will often come across marginally convergent Mellin transforms. Regularisation schemes are chosen such that the Mellin transform has an inverse. All schemes used in the remainder of this paper essentially boil down to the use of the above identity whose validity is guaranteed by the Fourier inversion theorem.

Given states or amplitudes already transforming under the little group, an economical alternative to the chiral Mellin transform is the \emph{half-Mellin transform}, first introduced in \cite{Sharma:2021gcz}, where we only integrate over the scale of one spinor
\begin{align}\label{eq: halfMellinstate}
\begin{split}
    \vert \lam,\lamt; h , s_h, l, s_l \rangle &:= \frac{1}{2}\int_{\RR_*} \frac{du}{\abs{u}} \abs{u}^{2h} \sgn(u)^{- s_h} \vert u\lam,\lamt; l, s_l \rangle\\
    &=\frac{1}{2}\int_{\RR^+} \frac{du}{u} u^{2h}   \sum_{\epsilon = \pm1} \epsilon^{s_h} \vert \epsilon u\,\lam,\lamt; l, s_l \rangle\,.
\end{split}    
\end{align}
Note that the above state is equivalent to \eqref{eq: 22nonchiralMellin} with the Dirac and Kronecker deltas stripped off. Indeed one can check that \eqref{eq: halfMellinstate} is homogeneous under rescalings of $\lamt$ with weight $\hb=h-l$ and $s_{\hb}=s_h-s_l \mod 2$. Additionally, the half-Mellin transformed state where we instead integrate over the scale of $\ltd{\alpha}$ is equal to \eqref{eq: halfMellinstate}. This can be shown by writing $\ub=u$ in the first line of \eqref{eq: halfMellinstate} and using a little group transformation.

 As explored at length in Section \ref{sec: celesttwistamps}, we can also take amplitudes from (1,3) signature with half-integer helicity $l$ and analytically continue them into (2,2) signature - we will always use the half-Mellin transform in such cases for its practical benefits. This entails using the above formula \eqref{eq: halfMellinstate} but with $l$ half-integer, in which case $s_l$ satisfies $s_l = 2 l \mod 2$.

The new features of split signature go beyond just the changes to the (extended) little group; as mentioned at in Section \ref{sec: shadowspace} the collection of different bases for conformal primaries on the celestial sphere expands when we consider the Lorentzian boundary CFT. In the next section we describe these new light transformed bases for conformal primaries.

 
\section{Light Transformed Basis}\label{sec: lightspace}
 
We now turn to the question of what other bases of conformal primaries exist on the celestial torus. Once again we can use the restricted Weyl group to indicate the existence of intertwining operators which map to new bases.

In (2,2) signature, the quadratic Casimir operators are still given by \eqref{eq: quadcasmiroperators} and since the Lorentz generators take the same form as in (1,3) signature (c.f. Appendix \ref{app: conformalgens}), then the Casimir eigenvalues are also given by \eqref{eq: quadcasmirvalues}, which we repeat here
\begin{equation}\label{eq: quadcasmirvalues2}
\begin{split}
    &c_2^+=2h(1-h)+2\hb(1-\hb)=\Delta(2-\Delta)-J^2 ,\\
    &c_2^{-}=2h(1-h)-2\hb(1-\hb)=2\,J(1-\Delta).
\end{split}
\end{equation}
  
In a two-dimensional Lorentzian CFT, the helicity $J$ is continuous and so can mix with $\Delta$. It is for this reason that there exists additional intertwining operators that map between representations equivalent to the Lorentzian principal continuous series. The restricted Weyl group thus expands to $\ZZ_2 \times \ZZ_2$ with the new elements corresponding to the reflections mixing $\Delta$ and $J$
\begin{equation}
\begin{split}
    h \rightarrow 1-h \iff \Delta \rightarrow 1-J,\,\,\, J\rightarrow 1-\Delta,\\
    \hb \rightarrow 1-\hb \iff \Delta \rightarrow 1+J,\,\,\, J\rightarrow \Delta-1,
\end{split}
\end{equation}
which are the light and dual light transforms. The shadow transform in a Lorentzian two-dimensional CFT is simply the product of these commuting elements, giving the interpretation of the light transforms as the `half-shadow transforms'.

With the extended little group as a guiding principle, we will now define explicit formulae for the light transforms. Before we begin, note that the discrete weights $s_h$ and $s_{\hb}$ do not appear in the Casimir eigenvalues, and so representations with differing values of $s_h$ and $s_{\hb}$ could in principle be equivalent. The only possible Weyl reflection would be to relate the even and odd $\ZZ_2$ representations so that $s_h\rightarrow s_h+1$, however, it is a trivial fact that the even and odd representations of $\ZZ_2$ are not equivalent. Simply put, the lack of an intertwining operator for even/odd representations corresponds to the fact that we cannot anti-symmetrise a function which is already even. A consequence of this is that the light transforms do not change the weights $s_h$ and $s_{\hb}$.

The light transform in homogeneous coordinates is given by a projective integral over the spinor $\lam$. We define the light transformed celestial state as
\begin{equation}\label{eq: lightcelstate}
    \boxed{\begin{split}
    \ket{\mu, \lamt;1-h, \hb, s_h,s_{\hb}}& = i^{-s_h} \frac{\Gamma(2-2h)}{\Gamma(\frac{3}{2}-h+\frac{s_h}{2})\Gamma(h-\frac{s_{h}}{2}-\frac{1}{2})}\\
    &\qquad\qquad\qquad\times\int_{\mathbb{RP}^1} \agl{\lam}{d\lam}\, \abs{\agl{ \lam}{\mu}}^{2h-2} \sgn(\agl{\lam}{\mu})^{s_h} \ket{\lam, \lamt;h, \hb, s_h,s_{\hb}}.
    \end{split}}
\end{equation}
We will denote this light transform operation symbolically as $L$.
It can be easily checked that the new state is an $\slR\times\slR$ conformal primary with shifted weights $1-h, \hb$, while the discrete weights $s_h, s_{\hb}$ are unchanged as expected. This formula for the light transform is adapted to the novel (2,2) chiral Mellin transform which builds states homogeneous under $\RR_*$ scalings. Note, however, that since \eqref{eq: lightcelstate} leaves the discrete weights $s_h, s_{\hb}$ invariant, its group theoretic behaviour is equivalent to the light transforms presented elsewhere in the literature for example \cite{Sharma:2022talk}\footnote{In \cite{Sharma:2022talk}, a definition of the light transform was given which acts on states living in two patches of a celestial torus labelled by an incoming and outgoing parameter. This is not the same as the above definition but contains completely equivalent information, just not separated into even and odd parts.}.

The definition \eqref{eq: lightcelstate} features a factor of $i^{-s_h}$ despite the fact that $s_h$ is only defined mod 2. Nevertheless, the entire formula is invariant when we shift $s_h \rightarrow s_h+2$ thanks to compensation from the gamma function factors $[\Gamma(\frac{3}{2}-h+\frac{s_h}{2})\Gamma(h-\frac{s_{h}}{2}-\frac{1}{2})]^{-1}$. In fact, the normalisation in \eqref{eq: lightcelstate} is chosen such that $L^2=1$, as shown in Appendix \ref{app: Ssquared}. 

We can recast the light transform in a compact form by combining it with a half-Mellin transform which, as explained above equation \eqref{eq: halfMellinstate}, acts on a momentum eigenstate with (2,2) helicity $l$ and $s_l$
\begin{equation}\label{eq: compactlightcelstate}
\begin{split}
    \ket{\mu, \lamt;1-h, s_h, l, s_l}& = i^{-s_h} \frac{\Gamma(2-2h)}{\Gamma(\frac{3}{2}-h+\frac{s_h}{2})\Gamma(h-\frac{s_{h}}{2}-\frac{1}{2})}\int_{\mathbb{RP}^1} \agl{\lam}{d\lam}\, \abs{\agl{ \lam}{\mu}}^{2h-2} \sgn(\agl{\lam}{\mu})^{s_h} \\
    &\qquad\times \frac{1}{2} \int_{\mathbb{R}_*}\!\frac{du}{\abs{u}} \, \abs{u}^{2h}\, \sgn(u)^{s_h}\vert  u\lam,\lamt; l, s_l \rangle\\
    =&\frac{i^{-s_h}}{2} \frac{\Gamma(2-2h)}{\Gamma(\frac{3}{2}-h+\frac{s_h}{2})\Gamma(h-\frac{s_{h}}{2}-\frac{1}{2})}\int_{\mathbb{R}^2} d^2\lam\, \abs{\agl{ \lam}{\mu}}^{2h-2} \sgn(\agl{\lam}{\mu})^{s_h} \vert  \lam,\lamt; l,s_l \rangle,
\end{split}
\end{equation}
where we have used the fact that the measure $d^2\! \lam$ breaks up into $\abs{u}\, du\, \agl{\lam}{d\lam}$. This measure is in fact the only thing responsible for the weight shift behaviour $h\rightarrow 1-h$ of the light transform, and its independence of $\sgn(u)$ is the reason why $s_h$ is unchanged.

The light transformed state is labelled with a pair spinors $\mu, \lamt$ much like a twistor. Indeed under rescalings of both spinors $\mu, \lamt$ we have the weight $-2(1-h+\hb) = -2(1-l)$ which is the usual scaling of a twistor. Of course, unlike a twistor the celestial state \eqref{eq: lightcelstate} is homogeneous under rescalings of each spinor independently. As one might expect, the bulk conformal symmetry generators in the light transformed basis in Appendix \ref{app: conformalgens} also have many similarities with those in twistor variables, for example all generators are first order differential operators.

The connection between light and twistor transforms was made concrete in \cite{Sharma:2021gcz} and we will spend some time refining this relation in the following Section \ref{sec: twistortolight}.

For completeness we record the formula for the dual light transformed state
\begin{equation}\label{eq: duallightcelstate}
\boxed{\begin{split}
    \ket{\lam, \mut;h, 1-\hb, s_h,s_{\hb}}& = i^{-s_{\hb}} \frac{\Gamma(2-2\hb)}{\Gamma(\frac{3}{2}-\hb+\frac{s_{\hb}}{2})\Gamma(\hb-\frac{s_{\hb}}{2}-\frac{1}{2})}\\
    &\qquad\qquad\qquad\times\int_{\mathbb{RP}^1} \sqr{\lamt}{d\lamt}\, \abs{\sqr{\mut}{\lamt}}^{2\hb-2} \sgn(\sqr{\mut}{\lamt})^{s_{\hb}} \ket{\lam, \lamt;h, \hb, s_h,s_{\hb}}.
\end{split}}
\end{equation}
We will also refer to this dual light transform operation as $\bar{L}$.
%


\section{From Twistor to Light}\label{sec: twistortolight}

The half-Fourier or twistor transform was  introduced in the seminal paper \cite{Witten:2003nn} as a means to directly map amplitudes in (2,2) signature (analytically continued from (1,3) signature) directly to either twistor or dual twistor space in four dimensions. The transform to twistor space takes the form
\begin{equation}\label{eq: twistortrans}
    \ket{\mu, \lamt}= \frac{1}{2\pi}\int_{\RR^2}d^2\lam\, e^{i\agl{\lam}{\mu}} \ket{\lam, \lamt}\,,
\end{equation}
while the dual twistor transform takes the form
\begin{equation}\label{eq: dualtwistortrans}
    \ket{\lam, \mut}= \frac{1}{2\pi}\int_{\RR^2}d^2\lamt\, e^{i\sqr{\mut}{\lamt}} \ket{\lam, \lamt}\,.
\end{equation}
We will again define the symbolic shorthand $T$ to denote the half-Fourier transform \eqref{eq: twistortrans} and $\bar{T}$ to denote its dual \eqref{eq: dualtwistortrans}. The above half-Fourier transforms are both self inverse since, like the Fourier transform \eqref{eq: Fourier}, they are themselves Fourier transforms with a skew-symmetric kernel.

The relationship between the twistor and the Fourier transform defined in \eqref{eq: Fourier} is that in the former we only transform half of the phase space. This is reminiscent of the relationship between light and shadow transforms. Furthermore we can see this relationship already at the level of the dilatation and helicity operators written in twistor variables
\begin{equation}
\begin{split}
    \cJ \sim \frac{1}{2}\muu{\alpha}\frac{\partial}{\partial \muu{\alpha}} + \frac{1}{2}\ltu{\alpha}\frac{\partial}{\partial \ltu{\alpha}}+1 =: d_T,\\
    d \sim -\frac{1}{2}\muu{\alpha}\frac{\partial}{\partial \muu{\alpha}} + \frac{1}{2}\ltu{\alpha}\frac{\partial}{\partial \ltu{\alpha}}=:\cJ_T,
\end{split}
\end{equation}
which, upon Mellin transforming with weights $k,\kb$, give the light transform relations $k=1-h, \kb=\hb$. We have also defined the distinct operators $\cJ_T, d_T$ generating helicity and dilatation operations of the variables $\mu_{\alpha}, \lamt_{\dot\alpha}$ based purely on their transformation properties in either the left or right hand copy of $\slR\times\slR$. That is $\cJ_T, d_T$ are not just the operators $\cJ, d$ in the twistor basis but are distinct and are related to $\cJ, d$ by the naive replacement $\lam \rightarrow \mu$.

Similar considerations hold for the dual twistor basis 
\begin{equation}
\begin{split}
    \cJ \sim -\frac{1}{2}\lu{\alpha}\frac{\partial}{\partial \lu{\alpha}} - \frac{1}{2}\mtu{\alpha}\frac{\partial}{\partial \mtu{\alpha}}-1 =: -d_{\bar T},\\
    d \sim \frac{1}{2}\lu{\alpha}\frac{\partial}{\partial \lu{\alpha}} + \frac{1}{2}\mtu{\alpha}\frac{\partial}{\partial \mtu{\alpha}}=:-\cJ_{\bar T},
\end{split}
\end{equation}
which upon Mellin transforming give the dual light transform relations $k=h, \kb=1-\hb$. In Appendix \ref{app: conformalgens} we give all the conformal symmetry generators $\cJ, d, p, k, m$ in the twistor and dual twistor basis.
Just as for the original generators and the `Fourier' ones $d_F, \cJ_F, \ldots$ etc. we find that the relation between $T$ and $\bar T$ conformal generators is described by the same automorphism of the conformal algebra, but now generated by the full Fourier transform which maps from twistor space to dual twistor space.

\begin{figure}
\[
\begin{tikzcd}
\text{Momentum Basis} \arrow[rrr, "\scalebox{1.5}{$T$}"] \arrow[dd, "\scalebox{1.5}{$C_{h,\hb}$}"'] &  &  & \text{Twistor Basis} \arrow[dd, "\scalebox{1.5}{$C_{1-h,\hb}$}"] \\                &  &  & \\                                   
\text{Mellin Basis} \arrow[rrr, "\scalebox{1.5}{$L$}"]  &  &  &  \text{Light Basis}           
\end{tikzcd}
\]
\caption{We denote the chiral Mellin transform with weights $h,\hb,s_h,s_{\hb}$ by $C_{h,\hb}$, half-Fourier by $T$ and light transform by $L$.}\label{fig: twistlight}
\end{figure}
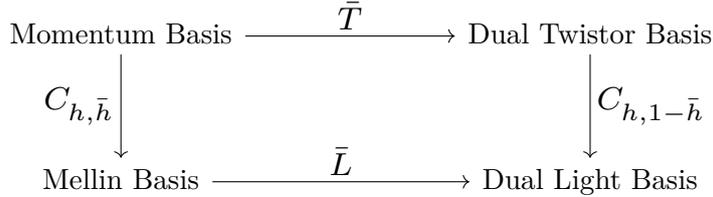
\begin{figure}
\[
\begin{tikzcd}
\text{Momentum Basis} \arrow[rrr, "\scalebox{1.5}{$\bar T$}"] \arrow[dd, "\scalebox{1.5}{$C_{h,\hb}$}"'] &  &  & \text{Dual Twistor Basis} \arrow[dd, "\scalebox{1.5}{$C_{h,1-\hb}$}"] \\                &  &  & \\                                   
\text{Mellin Basis} \arrow[rrr, "\scalebox{1.5}{$\bar L$}"]  &  &  &  \text{Dual Light Basis}           
\end{tikzcd}
\]
\caption{We denote the chiral Mellin transform with weights $h,\hb,s_h,s_{\hb}$ by $C_{h,\hb}$, dual half-Fourier by $\bar{T}$ and dual light transform by $\bar{L}$.}
\label{fig: dualtwistlight}
\end{figure}

The precise relationship between twistor and light transforms is via the commuting diagrams in Figures \ref{fig: twistlight} and \ref{fig: dualtwistlight} which state that the (dual) light transformed conformal primary is the same as the chiral Mellin transformed (dual) twistor eigenstate. These diagrams were first established in \cite{Sharma:2021gcz}, but there the diagrams commuted with the proviso that twistor transformed states corresponded to a linear combination of light transforms with different `incoming/outgoing parameters'. One of the goals of this paper has been to make it clear that we do not need to introduce any incoming or outgoing parameters in split signature since there is no sense of the past or future. In Section \ref{sec: lightspace} we made some crucial modifications to the light transforms to ensure they act properly on states which are homogeneous under $\RR_*$ scalings. These modifications lead to an exact relation between twistor and light transforms according to Figure \ref{fig: twistlight}, which we now prove. The proof for the dual light and dual twistor case is completely analogous.

Since the (2,2) chiral Mellin transform is given by two independent half-Mellin transforms, the transform on the unchanged spinor $\lamt$ passes through as usual and so we just focus on half-Mellin transforming the $\mu$ spinor
\begin{equation}
    \begin{split}
        \ket{\mu, \lamt; k, s_k}& = \frac{1}{2}\int_{\RR_*} \frac{dt}{\abs{t}} \abs{t}^{2k} \sgn(t)^{- s_k} \ket{t\mu,\lamt}\\
        & =\frac{1}{2}\int_{\RR_*} \frac{dt}{\abs{t}} \abs{t}^{2k} \sgn(t)^{- s_k}\frac{1}{2\pi}\int_{\RR^2}d^2\lam\, e^{it\agl{\lam}{\mu}} \ket{\lam, \lamt}\\
        & =\frac{1}{4\pi} \int_{\RR^2}d^2\lam\, \cI \ket{\lam, \lamt}.
    \end{split}
\end{equation}
We now perform the half-Mellin integral
\begin{equation}\label{eq: Iintegraltwistorlight}
    \cI := \int_{\RR_*} \frac{dt}{\abs{t}} \abs{t}^{2k} \sgn(t)^{- s_k} e^{it\agl{\lam}{\mu}} = \int_0^{\infty} dt\, t^{2k-1} \left(e^{it\agl{\lam}{\mu}}+(-1)^{-s_k}e^{-it\agl{\lam}{\mu}}\right),
\end{equation}
and changing variables $t \rightarrow t \abs{\agl{\lam}{\mu}}^{-1}$ we have
\begin{equation}
\begin{split}
    \cI =& \abs{\agl{\lam}{\mu}}^{-2k} \sgn(\agl{\lam}{\mu})^{-s_k} \int_0^{\infty} dt\, t^{2k-1} \left(e^{it}+(-1)^{-s_k}e^{-it}\right)\\
    =&\begin{cases} \abs{\agl{\lam}{\mu}}^{-2k} 2 \int_0^{\infty} dt\, t^{2k-1} \!\cos(t) &\text{for $s_k = 0$},\\
    \abs{\agl{\lam}{\mu}}^{-2k}\sgn(\agl{\lam}{\mu})\, 2i \int_0^{\infty} dt\, t^{2k-1}\! \sin(t) & \text{for $s_k = 1$}.
    \end{cases}
\end{split}
\end{equation}
Now as usual we need to regularise the weights, so we shift $k \in \frac{1}{2}+i\RR$\footnote{Since $\Delta \in 1+i\RR$ and $J$ is pure imaginary the weights $h,\hb \in \frac{1}{2}+i\RR$ and similarly $k,\kb \in \frac{1}{2}+i\RR$.} by a small negative parameter $-\delta$ such that $-1< \text{Re}(2(k-\delta)-1)<0$. Then sending $\delta \rightarrow 0_+$ we recover
\begin{equation}
\begin{split}
   \cI =& \begin{cases} 2\, \Gamma(2k)\,\sin(k\pi+\frac{\pi}{2}) \abs{\agl{\lam}{\mu}}^{-2k}  &\text{for $s_k = 0$},\\
    2i\,\Gamma(2k)\,\sin(k\pi)\, \abs{\agl{\lam}{\mu}}^{-2k}\sgn(\agl{\lam}{\mu}) & \text{for $s_k = 1$},
    \end{cases}
    \\
    =&2\,i^{s_k} \Gamma(2k)\sin\bigg(\bigg(k-\frac{s_k}{2}+\frac{1}{2}\bigg)\pi\bigg) \abs{\agl{\lam}{\mu}}^{-2k}\sgn(\agl{\lam}{\mu})^{-s_k}.
\end{split}
\end{equation}
Now we use the Euler reflection formula for gamma functions to yield
\begin{equation}
\begin{split}
   \cI   =&2\pi\, i^{s_k} \frac{\Gamma(2k)}{\Gamma(k-\frac{s_k}{2}+\frac{1}{2})\Gamma(\frac{1}{2}-k+\frac{s_{k}}{2})}\abs{\agl{\lam}{\mu}}^{-2k}\sgn(\agl{\lam}{\mu})^{-s_k}.
\end{split}
\end{equation}
Finally we set $k=1-h$ and $s_k=-s_h$ (equivalently $s_k=s_h$) and conclude that from a chiral Mellin transform of a Fourier conjugate state we recover the light transformed state \eqref{eq: compactlightcelstate}
\begin{equation}
    \ket{\mu, \lamt; 1-h, s_h} = \frac{i^{-s_h}}{2} \frac{\Gamma(2-2h)}{\Gamma(\frac{3}{2}-h+\frac{s_h}{2})\Gamma(h-\frac{s_{h}}{2}-\frac{1}{2})}\int_{\RR^2}d^2\lam\, \abs{\agl{\lam}{\mu}}^{2h-2}\sgn(\agl{\lam}{\mu})^{s_h} \ket{\lam, \lamt}.
\end{equation}

Alternatively, we can check the commuting diagram in Figure \ref{fig: twistlight} by inverse Mellin transforming \eqref{eq: compactlightcelstate} to recover a half-Fourier transform. That is we compute,
\begin{equation}
\begin{split}
    \int_{a-i\infty}^{a+i\infty} \frac{dh}{2\pi i} \sum_{s_h=0,1}\frac{i^{-s_h}}{2} \frac{\Gamma(2-2h)}{\Gamma(\frac{3}{2}-h+\frac{s_h}{2})\Gamma(h-\frac{s_{h}}{2}-\frac{1}{2})}\int_{\RR^2}d^2\lam\, \abs{\agl{\lam}{\mu}}^{2h-2}\sgn(\agl{\lam}{\mu})^{s_h} \ket{\lam, \lamt},
\end{split}
\end{equation}
which gives the two terms
\begin{equation}
\begin{split}
    =\int_{\RR^2}d^2\lam\int_{a-i\infty}^{a+i\infty} \frac{dh}{2\pi i} \frac{1}{2\pi} \Gamma(2-2h)\bigg(\sin(h\pi-\frac{\pi}{2})+i\sin(h\pi)\sgn(\agl{\lam}{\mu})\bigg)\abs{\agl{\lam}{\mu}}^{2h-2}\ket{\lam, \lamt},
\end{split}
\end{equation}
which we evaluate the same way by chosing $a<1$ and closing the contour to pick up the poles of the gamma function
\begin{equation}
\begin{split}
    =&\frac{1}{2\pi}\int_{\RR^2}d^2\lam\sum_{n=0}^{\infty} \frac{(-1)^n}{n!} \bigg(\sin(\pi\frac{n+1}{2})+i\sin(\pi\frac{n+2}{2})\sgn(\agl{\lam}{\mu})\bigg)\abs{\agl{\lam}{\mu}}^{n}\ket{\lam, \lamt}\\
   & =\frac{1}{2\pi}\int_{\RR^2}d^2\lam\bigg(\sum_{m=0}^{\infty} \frac{(-1)^{m}}{(2m)!} \abs{\agl{\lam}{\mu}}^{2m}+i\,\sgn(\agl{\lam}{\mu})\sum_{m=0}^{\infty}\frac{(-1)^m}{(2m+1)!}\abs{\agl{\lam}{\mu}}^{2m+1}\bigg)\ket{\lam, \lamt}\\
   & =\frac{1}{2\pi}\int_{\RR^2}d^2\lam\bigg(\cos(\abs{\agl{\lam}{\mu}})+i\,\sgn(\agl{\lam}{\mu})\sin(\abs{\agl{\lam}{\mu}})\bigg)\ket{\lam, \lamt}\\
   & =\frac{1}{2\pi}\int_{\RR^2}d^2\lam e^{i\agl{\lam}{\mu}}\ket{\lam, \lamt}.
\end{split}
\end{equation}
Hence we have recovered a half-Fourier transform. 


\section{Celestial Twistor Amplitudes}\label{sec: celesttwistamps}

In this section we will show how to compute celestial twistor amplitudes in Yang-Mills and gravity, building on  earlier work and the results and formalism developed above. In what follows, we espouse the viewpoint that the most direct way to produce light transformed conformal correlators is to perform a Mellin transform of a twistor amplitude and so traverse the commuting diagrams in Figures \ref{fig: twistlight} and \ref{fig: dualtwistlight} clockwise. Light transformed correlators have the nice property that they no longer have singular support coming from the momentum conserving delta function in the bulk
- see \cite{Sharma:2021gcz, De:2022gjn, Hu:2022syq, Guevara:2021tvr,  Banerjee:2022hgc} for discussions. These recent works have used the methodology of traversing the commuting diagrams counterclockwise, that is Mellin transforming first in momentum space and then light transforming the resultant celestial amplitude. The recent paper \cite{Hu:2022syq} in particular derived the correlators of two holomorphic and two anti-holomorphic light-ray operators, from light transforms of the four gluon celestial amplitude in $(2,2)$ signature spacetime. The results were written in terms of integrals of products of Gaussian hypergeometric functions, which when evaluated led to certain special functions which arise in Mellin-Barnes integrals. In our derivation of four-point celestial twistor amplitudes  below we find somewhat analogous, albeit simpler, interim integrals.

So far we have only studied asymptotic scattering states and classified the different bases one can take for these. Since an amplitude is a multi-particle object one can ask the important question: what is the most useful prescription for assigning bases to represent each particle leg? In principle, we can use any of the momentum space, twistor, dual twistor or Fourier bases for each leg and we can choose this for each leg independently. In the work of \cite{Mason:2009sa} the choice made was to uniformly associate to each leg a dual twistor. This led to many compact expressions for twistor amplitudes, including $n$-particle MHV, $\overline{\text{MHV}}$ and up to eight particle $\text{N}^2\text{MHV}$ amplitudes in $N=4$ SYM,  as well as MHV, $\overline{\text{MHV}}$ expressions in $N=8$ supergravity. The authors also developed a BCFW recursion relation in dual twistor space which resulted in a simple BCFW shift of the dual twistor. We leave the task of Mellin transforming amplitudes from dual twistor space to future work and instead we focus on twistor amplitudes in the ambidextrous bases of \cite{Arkani-Hamed:2009hub}.

As demonstrated in \cite{Arkani-Hamed:2009hub}, performing ambidextrous transforms to twistor space leads to remarkably simple objects - for example, for pure Yang-Mills the three and four-point ambidextrous twistor amplitudes take the constant value equal to one or minus one depending on the particular external kinematics probed. Note that `ambidextrously' here means that we can transform either $\lambda$ or $\lambdat$ for each leg independently and in general the basis chosen will not be uniform. At low particle multiplicity we will use a twistor for a plus helicity leg and a dual twistor for a minus helicity leg.
This choice leads to particularly simple integrals which decouple into products of $\sgn$ functions.

In the end we are interested in physical $(1,3)$ signature scattering amplitudes. The motivation to consider $(2,2)$ signature comes from the fact that we can analytically continue expressions for scattering amplitudes in $(1,3)$ signature into $(2,2)$ signature and consider new representations of those amplitudes in terms of twistors. This is the technique used in \cite{Mason:2009sa} and \cite{Arkani-Hamed:2009hub} to build seed amplitudes to be used in twistor BCFW to build higher point amplitudes and we also follow this methodology.  Such analytically continued $(2,2)$ signature amplitudes \textit{do not} transform in the expected representations of the $(2,2)$ little group. Rather, since they originated in $(1,3)$ signature, they are homogeneous under a little group transformation with \textit{half-integer helicity}.

Assuming we start with analytically continued (2,2) signature amplitudes with half-integer helicity (denoted simply as $l$ since $s_l = 2 l \mod 2$), we can map to Mellin space in the most practical manner by performing just a half-Mellin transform over the scale of either of the spinors. This method was introduced used in \cite{Sharma:2021gcz} and is particularly useful when applied to twistor amplitudes. In contrast to \cite{Sharma:2021gcz}, we ambidextrously perform half-Mellin transforms over the scales of the $\mu, \mut$ spinor instead of performing them on the original spinors $\lam,\lamt$. Just as was the case in momentum space, the two choices are equivalent, since under a change of variable given by an inversion $t \mapsto \frac{1}{\ub}$ the half-Mellin transformed twistor eigenstate becomes
\begin{align}
\begin{split}
     \vert \mu,\lamt; k , s_k, l \rangle :=& \frac{1}{2}\int_{\RR_*} \frac{dt}{\abs{t}} \abs{t}^{2k} \sgn(t)^{s_k} \vert t\mu,\lamt; l \rangle\\
     =& \frac{1}{2}\int_{\RR_*} \frac{d\ub}{\abs{\ub}} \abs{\ub}^{-2k} \sgn(\ub)^{s_k} \vert \ub^{-1}\mu,\lamt; l \rangle\\
     =&\frac{1}{2}\int_{\RR_*} \frac{d\ub}{\abs{\ub}} \abs{\ub}^{-2k} \sgn(\ub)^{s_k} \abs{\ub}^{-2(l-1)} \sgn(\ub)^{-2(l-1)} \vert \mu,\ub\lamt; l \rangle\\
     =&\frac{1}{2}\int_{\RR_*} \frac{d\ub}{\abs{\ub}} \abs{\ub}^{2\hb} \sgn(\ub)^{s_{\hb}} \vert \mu,\ub\lamt; l \rangle \,,\\
\end{split}
\end{align}
which is exactly the half-Mellin transform over the $\lambdat$ spinor, where we have used that $k=1-h$ and $h-\hb=l$. A similar equivalence is true for the half-Mellin transformed dual twistor eigenstate. 

\subsection{Pure Yang-Mills}

We start off by deriving the three-point and four-point celestial twistor amplitudes of pure Yang-Mills Theory.


\subsubsection{Three Points: Mellin Transform of a Sgn Function}\label{sec: threepoints}

\textbf{Three-Point} $MHV$

When we analytically continue to $(2,2)$ signature we can write down a three-point MHV amplitude with real external kinematics
\begin{align}\label{eq: 3ptmhv}
    A^{--+}_3=\frac{\agl{1}{2}^3}{\agl{2}{3}\agl{3}{1}}\,\delta^4(\lu{\alpha}_1\ltu{\alpha}_1+\lu{\alpha}_2\ltu{\alpha}_2+\lu{\alpha}_3\ltu{\alpha}_3)\,.
\end{align}
As in \cite{Arkani-Hamed:2009hub} we perform the twistor or half-Fourier transforms ambidextrously. In our conventions this corresponds to
\begin{align}\label{eq: 3pttwistortransform}
\tilde{A}^{--+}_3:= \frac{1}{(2\pi)^3} \int d^2\lambdat_1\,  d^2\lambdat_2\,  d^2\lambda_3\, e^{i\sqr{\mut_1}{ \lambdat_1}}\,e^{i[\mut_2 \lambdat_2]}\,e^{i\agl{\lambda_3}{ \mu_3}} A^{--+}_3\,.
\end{align}
To perform the half-Fourier transforms we must regulate the poles appearing from $\agl{2}{3}\agl{3}{1}$ in \eqref{eq: 3ptmhv}. Following \cite{Arkani-Hamed:2009hub}, we regulate these and subsequent poles with a principle value prescription, which is chosen exactly to preserve the little group properties of the amplitude. With this choice the three-point MHV twistor amplitude is given by
\begin{align}\label{eq: 3ptMHVtwsitor}
    \tilde{A}^{--+}_3:= \frac{(\pi i)^3}{(2\pi)^3} \sgn(\langle \lambda_1 \lambda_2 \rangle )\, \sgn(\langle \lambda_1 \mu_3 \rangle+[\lambdat_3 \mut_1]) \,\sgn(\langle \lambda_2 \mu_3 \rangle+[\lambdat_3 \mut_2])\,.
\end{align}
In order to map to Mellin space we use ambidextrous half-Mellin transforms over  the scales of the spinors $\mut_1, \mut_2, \mu_3$.

\textbf{Symmetrised Amplitudes}

As explained in Section \ref{sec: 22chiralMellin} we can separately perform a Mellin transform over positive re-scalings of the spinors and then build odd/even states via linear combination of sums and differences over the parameters $\epsilonb_1, \epsilonb_2, \epsilon_3 \in \{1,-1\}$ associated with flipping the signs of the spinors $\mut_1, \mut_2, \mu_3$. We now perform the sums first and then integrate to introduce some of the key points. At four points there are many such linear combinations, so we will perform the reverse method. That is, we integrate over positive rescalings first to derive a compact expression in \eqref{eq: 3ptMHVceltwistUnsym}, from which we can easily (anti-)symmetrise the amplitudes.

For the three-point amplitude there are 8 possible combinations of even/odd states under the flipping of the signs of $\mut_1, \mut_2, \mu_3$. Following our formalism, we denote these states as $\tilde{A}^{--+}_3\{s_{\kb_1},s_{\kb_2},s_{k_3}\}$ for $ s_{\kb_1},s_{\kb_2},s_{k_3} \in \{0,1\}$. Note that, from the form of the three-point MHV twistor amplitude \eqref{eq: 3ptMHVtwsitor}, when we flip the sign of all the spinors $\mut_1, \mut_2, \mu_3$ the amplitude remains invariant
\begin{align}\label{eq: equal3ptamps}
     \tilde{A}^{--+}_3(-\mut_1, -\mut_2, -\mu_3) = \tilde{A}^{--+}_3(\mut_1, \mut_2, \mu_3)\,.
\end{align}
The symmetrised amplitude
\begin{align}\label{eq: mhv3ptsummedamp}
    \tilde{A}^{--+}_3\{s_{\kb_1},s_{\kb_2},s_{k_3}\} = \frac{1}{2^3}\sum_{\epsilonb_1, \epsilonb_2, \epsilon_3} \epsilonb_1^{s_{\kb_1}}\epsilonb_2^{s_{\kb_2}}\epsilon_3^{s_{k_3}}\tilde{A}^{--+}_3(\epsilonb_1\mut_1, \epsilonb_2\mut_2, \epsilon_3\mu_3)\,,
\end{align}
contains pairs related  by \eqref{eq: equal3ptamps} which can sum together or subtract to give zero depending on whether $(-1)^{s_{\kb_1}+s_{\kb_2}+s_{k_3}}$ equals plus or minus one. This immediately implies if $s_{\kb_1}+s_{\kb_2}+s_{k_3} = 1 \mod 2$ then the amplitude vanishes, which can be checked explicitly. We hence have the following four non-vanishing amplitudes
\begin{align}\label{eq: 3ptMHVtwsitorevenoddamps}
    \begin{split}
        \tilde{A}^{--+}_3\{0,0,0\}\,, \,\,\,
    \tilde{A}^{--+}_3\{1,1,0\}\,,\,\,\,
    \tilde{A}^{--+}_3\{1,0,1\}\,,\,\,\,
    \tilde{A}^{--+}_3\{0,1,1\}\,, 
    \end{split}
\end{align}
where using \eqref{eq: mhv3ptsummedamp} we have
\begin{align}\label{eq: 3ptMHVtwsitorevenoddampsgeneral}
    \begin{split}
        \tilde{A}^{--+}_3\{s_{\kb_1},s_{\kb_2},s_{k_3}\}       = \frac{2(\pi i)^3}{2^3(2\pi)^3} \sgn(\langle \lambda_1 \lambda_2 \rangle )\, &\bigg(\sgn(\langle \lambda_1 \mu_3 \rangle+[\lambdat_3 \mut_1]) \,\sgn(\langle \lambda_2 \mu_3 \rangle+[\lambdat_3 \mut_2])\\&+(-1)^{s_{\kb_1}}\sgn(\langle \lambda_1 \mu_3 \rangle-[\lambdat_3 \mut_1]) \,\sgn(\langle \lambda_2 \mu_3 \rangle+[\lambdat_3 \mut_2])\\&+(-1)^{s_{\kb_2}}\sgn(\langle \lambda_1 \mu_3 \rangle+[\lambdat_3 \mut_1]) \,\sgn(\langle \lambda_2 \mu_3 \rangle-[\lambdat_3 \mut_2])\\&+(-1)^{s_{k_3}}\sgn(\langle \lambda_1 \mu_3 \rangle-[\lambdat_3 \mut_1]) \,\sgn(\langle \lambda_2 \mu_3 \rangle-[\lambdat_3 \mut_2])\bigg)\,,
    \end{split}
\end{align}
From the amplitudes \eqref{eq: 3ptMHVtwsitorevenoddamps} given by \eqref{eq: 3ptMHVtwsitorevenoddampsgeneral} we can recover the original amplitude by summing them. These amplitudes can be re-cast as
\begin{align}
\begin{split}
        &\tilde{A}^{--+}_3\{s_{\kb_1},s_{\kb_2},s_{k_3}\}       = \frac{(\pi i)^3}{(2\pi)^3} \sgn(\langle \lambda_1 \lambda_2 \rangle \agl{\lambda_1}{\mu_3} \agl{\lambda_2}{\mu_3} )\,\tilde\delta\bigg(\sum_i s_{\kibnbsmall}\bigg)\,\\
        &\qquad\qquad\qquad\qquad\qquad\qquad\times G_{s_{\kb_2}+s_{k_3}}(\theta_{31}^{-1})G_{s_{\kb_1}+s_{k_3}}(\theta_{32}^{-1})\,,
\end{split}
\end{align}
where we have included a Kronecker delta
\begin{equation}
    \tilde\delta\bigg(\sum_i s_{\kibnbsmall}\bigg)\coloneqq \delta_{s_{\bar{k}_1}+s_{\bar{k}_2}+s_{k_3},0}\,,
\end{equation}
(with argument evaluated mod 2) which enforces the aforementioned condition for the amplitude to be non-vanishing.
We have also defined the variables
\begin{align}\label{eq: defthetaij}
    \theta_{ij}=\frac{\agl{\lambda_j}{\mu_i}}{\sqr{\lambdat_i}{\mut_j}}\,,
\end{align}
and the (anti-)symmetrised $\sgn$ functions
\begin{align}
    \begin{split}
        G_0(x):=\frac{1}{2}\bigg(\sgn(1+x)+\sgn(1-x)\bigg)\,,\\
        G_1(x):=\frac{1}{2}\bigg(\sgn(1+x)-\sgn(1-x)\bigg)\,,
    \end{split}
\end{align}
which can be written in terms of Heaviside step functions as follows
\begin{align}\label{eq: HatStairThetaFunctions}
\begin{split}
    G_0(x)=&\Theta(1-\abs{x}) ,\\
    G_1(x)=&\sgn(x)\Theta(\abs{x}-1).
\end{split}
\end{align}
We will use the above definitions \eqref{eq: HatStairThetaFunctions} many times over when performing our integrals.

The (2,2) signature celestial twistor amplitude in Mellin space,  denoted with a calligraphic symbol $\cA$, is then given by an integral over $t_i$ the positive scale of $\mu_i$ with weight $2k_i$
\begin{align}\label{eq: 3ptMellintrasnforms}
\begin{split}
    \tilde{\cA}^{--+}_3\{s_{\kb_1},s_{\kb_2},s_{k_3}\}:=& \int_{0}^{\infty} \frac{d\tb_1}{\tb_1}\,\frac{d\tb_2}{\tb_2}\,\frac{dt_3}{t_3} \,
    \tb_1^{2\kb_1}\,\tb_2^{2\kb_2}\,t_3^{2k_3} \tilde{A}^{--+}_3\{s_{\kb_1},s_{\kb_2},s_{k_3}\}(t_1\mut_1, \tb_2\mut_2, t_3\mu_3).
\end{split}
\end{align}

As such we are lead to consider the following integrals
\begin{align}
\begin{split}
\cI(\pm,\pm):= &\int_{0}^{\infty} \frac{d\tb_1}{\tb_1}\,\frac{d\tb_2}{\tb_2}\,\frac{dt_3}{t_3} \,
    \tb_1^{2\kb_1}\,\tb_2^{2\kb_2}\,t_3^{2k_3} \\&\quad\times\frac{1}{4}\bigg(\sgn(1+\frac{\tb_1}{t_3}\theta_{31}^{-1})\pm\sgn(1-\frac{\tb_1}{t_3}\theta_{31}^{-1})\bigg)\bigg( \sgn(1+\frac{\tb_2}{t_3}\theta_{32}^{-1})\pm \sgn(1-\frac{\tb_2}{t_3}\theta_{32}^{-1})\bigg).
\end{split}
\end{align}

The general character of $\cI(\pm,\pm)$ (and corresponding integrals beyond three points) is very simple - it is a Mellin integral (over a positive quadrant) of an integrand which is a piece-wise \textit{constant function} in various regions bounded by the planes $\frac{\tb_i}{t_3}=\abs{\theta_{3i}}$.

Another important characteristic of the integral $\cI$, and its higher point cousins, is that it is a conformal integral and as such one of the degrees of freedom serves simply to parameterise an overall scale over which we integrate. In addition, the integrand itself is invariant under overall dilatations since it is a pure gluon amplitude. We can fix this scaling redundancy by integrating it out. We choose to write $\tb_1 = t_3 x_1$ and $\tb_2 = t_3 x_2$ and then integrate over the overall scale $t_3$
\begin{align}\label{eq: x1x2integrals}
\begin{split}
    \cI(\pm,\pm)=&\int_0^{\infty} \frac{dt_3}{t_3} t_3^{2\kb_1+2\kb_2+2k_3}\int_0^{\infty} \frac{dx_1\,dx_2}{x_1\, x_2}\, x_1^{2\kb_1}\, x_2^{2\kb_2}\, \\
    &\quad\times\frac{1}{4}\bigg(\sgn(1+x_1\theta_{31}^{-1})\pm\sgn(1-x_1\theta_{31}^{-1})\bigg)\bigg( \sgn(1+x_2\theta_{32}^{-1})\pm \sgn(1-x_2\theta_{32}^{-1})\bigg)\\
    =&\;2\pi i\,\delta\bigg(\sum_i 2\kibnbnormal\bigg)\,\int_0^{\infty} \frac{dx_1\,dx_2}{x_1\, x_2}\,x_1^{2\kb_1}\, x_2^{2\kb_2} \\&\quad\times\frac{1}{4}\bigg(\sgn(1+x_1\theta_{31}^{-1})\pm\sgn(1-x_1\theta_{31}^{-1})\bigg)\bigg( \sgn(1+x_2\theta_{32}^{-1})\pm \sgn(1-x_2\theta_{32}^{-1})\bigg),
\end{split}    
\end{align}
where we have recovered a dilatation invariance delta function using equation \eqref{eq: Mellindelta} and the fact that the weights $\kb_1$, $\kb_2$, $k_3$ are all pure imaginary. This is a special feature of the ambidextrous twistor gluon amplitude and relies on the fact we have analytically continued from (1,3) signature, so all the helicities are plus or minus one.

Now that we have integrated out an overall scale the remaining integrals over $x_1,x_2$ are separable and each has manifestly acquired a scale given by $\theta_{31}, \theta_{32}$ respectively. This separable behaviour persists at four points.

From the three-point MHV celestial amplitude we are led to consider the following Mellin transforms of the (anti-)symmetrised $\sgn$ functions $G_0, G_1$ in \eqref{eq: HatStairThetaFunctions}
\begin{align}\label{eq: x1integralevenodd}
\begin{split}
    \int_0^{\infty} \frac{dx_1}{x_1}\,x_1^{2\kb_1}G_{s}(x_1\theta_{31}^{-1})=&\int_0^{\infty} \frac{dx_1}{x_1}\,x_1^{2\kb_1} \,\frac{1}{2}\bigg(\sgn(1+x_1\theta_{31}^{-1})+(-1)^s\sgn(1-x_1\theta_{31}^{-1})\bigg)\\
    =&\;\sgn(\theta_{31})^{s}\int_0^{\infty} \frac{dx_1}{x_1}\,x_1^{2\kb_1} \Theta((-1)^s(1-x_1 \abs{\theta_{31}}^{-1})).
\end{split}
\end{align}
An important property of the function $G_0$ is that it is zero for large values $\abs{x_1}>\theta_{31}$ and so its Mellin transform has a semi-infinite strip of definition where the integral is defined for $2k_1$ with real part greater than zero. On the other hand, the function $G_1$ is zero for small values $\abs{x_1}<\theta_{31}$ and so its semi-infinite strip of definition is where $2k_1$ has real part less than zero. Note that the original $\sgn$ function, since it is $\mathcal{O}(1)$ everywhere, has a Mellin transform which only marginally converges when its weight lies on the imaginary axis and diverges otherwise. As such we can only re-organise the integral \eqref{eq: x1integralevenodd} by integrating each $\sgn$ function separately if we also give a regularisation prescription.  We demonstrate a natural regularisation in the next section in equation \eqref{eq: 1dsgnintegral}, which also commutes with the even/odd projection in \eqref{eq: x1integralevenodd}.

Continuing with our derivation, we compute the Mellin transform of $G_0$ and $G_1$ functions - since all the weights $k_i$ are pure imaginary we must regularise the weights to ensure they are within the corresponding strip of definition
\begin{align}\label{eq: hatstairsMellin}
\begin{split}
    &\int_0^{\infty} \frac{dx_1}{x_1}\,x_1^{2\kb_1+\epsilon} \,\frac{1}{2}\bigg(\sgn(1+x_1\theta_{31}^{-1})+\sgn(1-x_1\theta_{31}^{-1})\bigg)=\int_0^{\infty} \frac{dx_1}{x_1}\,x_1^{2\kb_1+\epsilon} \,\Theta\bigg(1-x_1\abs{\theta_{31}}^{-1}\bigg)\\ =&\; \Theta(\theta_{31})\frac{\theta_{31}^{2\kb_1+\epsilon}}{2\kb_1+\epsilon} + \Theta(-\theta_{31})\frac{(-\theta_{31})^{2\kb_1+\epsilon}}{2\kb_1+\epsilon}=\frac{\abs{\theta_{31}}^{2\kb_1}}{2\kb_1+\epsilon},\\
    &\int_0^{\infty} \frac{dx_1}{x_1}\,x_1^{2\kb_1-\epsilon} \,\frac{1}{2}\bigg(\sgn(1+x_1\theta_{31}^{-1})-\sgn(1-x_1\theta_{31}^{-1})\bigg)=\sgn(\theta_{31})\int_0^{\infty} \frac{dx_1}{x_1}\,x_1^{2\kb_1-\epsilon} \,\Theta\bigg(x_1\abs{\theta_{31}}^{-1}-1\bigg) \\
    =& -\Theta(\theta_{31})\frac{\theta_{31}^{2\kb_1-\epsilon}}{2\kb_1-\epsilon} + \Theta(-\theta_{31})\frac{(-\theta_{31})^{2\kb_1-\epsilon}}{2\kb_1-\epsilon}=\sgn(-\theta_{31})\frac{\abs{\theta_{31}}^{2\kb_1}}{2\kb_1-\epsilon}.
\end{split}
\end{align}
So the even case gives an `advanced' prescription for the pole in weight space and the odd case has the `retarded' prescription. Also the even/odd symmetry is  trivially carried by an overall $\sgn$ function and the remaining integral simply depends on a manifestly positive scale at $\abs{\theta_{31}}$. Finally, we have removed the appearance of $\epsilon$ in the exponents since the only role it should play is to regularise the pole.

The three-point MHV celestial twistor amplitudes are then given by a  compact formula
\begin{equation}\label{eq: 3ptMHVceltwist}
\boxed{\begin{split}
    \tilde{\cA}^{--+}_3\{s_{\kb_1},s_{\kb_2},s_{k_3}\}=&\frac{\pi}{4} \sgn(\agl{\lambda_1}{\lambda_2}\agl{\lambda_1}{\mu_3}\agl{\lambda_2}{\mu_3})\,\delta\bigg(\sum_i 2\kibnbnormal\bigg)\tilde\delta\bigg(\sum_i s_{\kibnbsmall}\bigg)\\
    &\quad\times\sgn(-\theta_{31})^{s_{\kb_1}}\frac{\abs{\theta_{31}}^{2\kb_1}}{2\kb_1+(-1)^{s_{\kb_1}}\epsilon}\sgn(-\theta_{32})^{s_{\kb_2}}\frac{\abs{\theta_{32}}^{2\kb_2}}{2\kb_2+(-1)^{s_{\kb_2}}\epsilon}.
\end{split}}
\end{equation}

The formula \eqref{eq: 3ptMHVceltwist} has similarities with light transformed conformal correlators that have appeared in the literature in \cite{Sharma:2021gcz, Hu:2022syq}, up to differences due to the newly defined light transform \eqref{eq: lightcelstate} with absolute values and sgn functions and the specific normalisation we have chosen.  We give a formula for the unsymmetrised celestial twistor amplitude in \eqref{eq: 3ptMHVceltwistUnsym} which is closer to the light transformed conformal correlators of \cite{Sharma:2021gcz}.

It is also a simple task to check conformal covariance of \eqref{eq: 3ptMHVceltwist} for each leg. Note first that we are  explicitly  using only the ambidextrous variables, in this case $\kb$ for negative helicity legs and $k$ for the positive helicity legs. In general, for conformal covariance we require that if we sum the exponents wherever a positive helicity leg $i$ appears in the $\theta_{ik}$ ratios we have overall weight $-2k_i$, while for a negative helicity leg $j$ we have overall weight $2\kb_j$. Then conformal covariance is guaranteed using the defining relations $k:=1-h$ and $\kb:=1-\hb$ and the relations $\kb_{1,2}=-h_{1,2}$ and $k_3=-\hb_3$ specific to the ambidextrous variables and pure gluon amplitudes.  Finally, for some legs we must also use dilatation invariance, for example for leg 3 we use the condition $\kb_1+\kb_2+k_3=0$.

In the derivation of \eqref{eq: 3ptMHVceltwist} we first performed the sum over incoming and outgoing parameters in order to define even and odd parity states and afterwards we performed the integrals over positive energies. In general, performing the integrals first and then the sum afterwards will give the same result only if the integrals are finite. That is, the Mellin transform of a sum of two functions is equal to the sum of their Mellin transforms only if there is an overlap in their strip of definition. We saw already that the symmetrised and anti-symmetrised $\sgn$ functions had a strip of definition in the right and left half planes respectively. However, the $\sgn$ function itself has a Mellin transform which only marginally converges when the weight lies on the imaginary axis and diverges otherwise. We show in the below section that we can define a natural regularisation prescription to define the Mellin transform of a $\sgn$ function and that this regularisation commutes with the (anti-)symmetrisation procedure.


\textbf{Unsymmetrised Amplitude}

We now look at the case where we first integrate \eqref{eq: 3ptMHVtwsitor} over positive re-scalings and then perform the even/odd projections afterwards. This will of course give the same result as in the previous section as long as we are careful in how we regularise marginally convergent Mellin transforms. We hence consider the integral
\begin{align}
    \begin{split}
        \cI_{3pt}= &2\pi i\,\delta\bigg(\sum_{i} 2\kibnbnormal\bigg) \int_0^{\infty}\frac{dx_1\,dx_2}{x_1\, x_2}\,x_1^{2\kb_1}\, x_2^{2\kb_2} \sgn(1+x_1\theta_{31}^{-1}) \sgn(1+x_2\theta_{32}^{-1}), 
    \end{split}
\end{align}
which is the unsymmetrised version of equation \eqref{eq: x1x2integrals}.

Again the integrals over $x_1$ and $x_2$  separate and are of identical form after the replacement $1\mapsto 2$. Focusing on the $x_1$ integral, since the $\sgn$ function is always $\mathcal{O}(1)$ its strip of definition is when $2\kb_1$ is pure imaginary and we must regulate in a similar manner to the integral \eqref{eq: fourierdelta} for the delta function. Instead of regulating with a factor of $e^{-\epsilon\abs{y}}$, we use a regularisation prescription which is anchored at the scale set by $\theta_{31}$. That is we define $\abs{\abs{y}}:=\sgn(y-\ln(\abs{\theta_{31}}))y$ which equals $y$ when $y>\ln(\abs{\theta_{31}})$ and $-y$ when $y<\ln(\abs{\theta_{31}})$, then define the regulated integral
\begin{align}\label{eq: 1dsgnintegral}
\begin{split}
    \int_0^{\infty} \frac{dx_1}{x_1}\,x_1^{2\kb_1}\,\sgn(1 + x_1 \theta_{31}^{-1})=& \int_{-\infty}^{\infty} dy\,e^{2\kb_1y-\epsilon \abs{\abs{y}}}\,\sgn(1 + e^{y} \theta_{31}^{-1})\\
    =&\;\Theta(-\theta_{31})\left(\int_{-\infty}^{\ln(-\theta_{31})}dy\, e^{2\kb_1 y +\epsilon y} - \int_{\ln(-\theta_{31})}^{\infty} dy\, e^{2\kb_1y-\epsilon y} \right)\\
    &\;\; +\; \Theta(\theta_{31})\left(\int_{-\infty}^{\ln\theta_{31}}dy\, e^{2\kb_1 y +\epsilon y} + \int_{\ln\theta_{31}}^{\infty} dy\, e^{2\kb_1y-\epsilon y} \right)\\
    =&\;\Theta(-\theta_{31})\bigg(\frac{(-\theta_{31})^{2\kb_1+\epsilon}}{2\kb_1+\epsilon}+\frac{(-\theta_{31})^{2\kb_1-\epsilon}}{2\kb_1-\epsilon}\bigg)+\Theta(\theta_{31})\bigg(\frac{\theta_{31}^{2\kb_1+\epsilon}}{2\kb_1+\epsilon}-\frac{\theta_{31}^{2\kb_1-\epsilon}}{2\kb_1-\epsilon}\bigg)\\
    =&\;\abs{\theta_{31}}^{2\kb_1}\bigg(\frac{1}{2\kb_1+\epsilon}+\sgn(-\theta_{31})\frac{1}{2\kb_1-\epsilon}\bigg),
\end{split}
\end{align}
where again we remove the $\epsilon$ in the exponents since it plays no regularising role there.

Note that in the case $\theta_{31}>0$ the $\sgn$ gives minus one and so the Mellin transform \eqref{eq: 1dsgnintegral} is proportional to a delta function as expected. While the case $\theta_{31}<0$ the $\sgn$ function gives plus one and the Mellin transform \eqref{eq: 1dsgnintegral} is proportional to the principal value pole. These are defined as
\begin{align}\label{eq: pvanddelta}
\begin{split}
   \delta(2\kb_1)\coloneqq \frac{1}{2\pi i}\bigg(\frac{1}{2\kb_1+\epsilon}-\frac{1}{2\kb_1-\epsilon}\bigg)\,,\\ 
     P. V.\, \frac{1}{2\kb_1}\coloneqq\frac{1}{2}\bigg(\frac{1}{2\kb_1+\epsilon}+\frac{1}{2\kb_1-\epsilon}\bigg)\,.
\end{split}
\end{align}
So we may write
\begin{align}\label{eq: 1dsgnintegraldeltaPV}
\begin{split}
    \int_0^{\infty} \frac{dx_1}{x_1}\,x_1^{2\kb_1}\,\sgn(1 + x_1 \theta_{31}^{-1})= \abs{\theta_{31}}^{2\kb_1}\bigg(\pi i\,\Theta(\theta_{31})\delta(\kb_1)+\Theta(-\theta_{31})\frac{1}{\kb_1}\bigg).
\end{split}
\end{align}
where it is understood that principal values are always implicit for a $1/\kibnbnormal$ pole.

Thus the above result \eqref{eq: 1dsgnintegral} is the Mellin transform of a sgn function depending on a single parameter. This corresponds to the decomposition of the sgn function into even and odd parts as
\begin{align}
    \sgn(1+ \theta_{31}^{-1}) =\Theta(1-\abs{\theta_{31}}^{-1})+\sgn(\theta_{31})\Theta(\abs{\theta_{31}}^{-1}-1).
\end{align}
While it is not in general true that the Mellin transform of a sum is the sum of their Mellin transforms (the strips of definition must also be compatible such that all the integrals are convergent) the regularisation we have used is precisely such that this is true and is motivated as such. In other words, given the Mellin transform of a $\sgn$ function \eqref{eq: 1dsgnintegral} it is simple to check that if we take even and odd combinations we recover the results \eqref{eq: hatstairsMellin}.

To completely confirm our results we perform the inverse Mellin transform
\begin{align}
    \begin{split}
        \int_{-i\infty}^{i\infty} \frac{d(2\kb_1)}{2\pi i} \abs{\theta_{31}}^{2\kb_1}\bigg(\frac{1}{2\kb_1+\epsilon}+\sgn(-\theta_{31})\frac{1}{2\kb_1-\epsilon}\bigg),
    \end{split}
\end{align}
where we have chosen a contour located at $a=0$ corresponding to a strip of definition on the imaginary axis. Now the $\epsilon$-prescription in each term gives poles in the $2\kb_1$ plane to either the left or right of the imaginary axis - in addition, we must condition this integral on the magnitude of $\theta_{31}$ to determine how to close the contour. The case $\abs{\theta_{31}}>1$ requires us to close the contour as an anti-clockwise semicircle to the left such that the real part of $2\kb_1$ is less than zero and so will only pick up a $2\pi i$ residue contribution from the first term, while for $\abs{\theta_{31}}<1$ we must close the contour clockwise to the right and we get a contribution from the second term and an extra minus sign. Hence we have
\begin{align}\label{eq: sgnMellininverse}
    \begin{split}
        \int_{-i\infty}^{i\infty} \frac{d(2\kb_1)}{2\pi i} \bigg(\frac{\abs{\theta_{31}}^{2\kb_1+\epsilon}}{2\kb_1+\epsilon}+\sgn(-\theta_{31})\frac{\abs{\theta_{31}}^{2\kb_1-\epsilon}}{2\kb_1-\epsilon}\bigg)
        =&\;\Theta(\abs{\theta_{31}}-1)-\sgn(-\theta_{31})\Theta(1-\abs{\theta_{31}})\\
        =&\; \sgn(1+\theta_{31}^{-1}).
    \end{split}
\end{align}

Thus we see that the $\epsilon$ prescription is crucial for the inverse Mellin transform and corresponds to a particular contour choice for each term done independently. With this specific contour choice in mind we may drop the $\epsilon$ prescription in \eqref{eq: 1dsgnintegral} which corresponds to dropping delta function terms in \eqref{eq: 1dsgnintegraldeltaPV} which have singular support. The unsymmetrised amplitude then takes a very simple form
\begin{equation}\label{eq: 3ptMHVceltwistUnsym}
\boxed{\begin{split}
    \tilde{\cA}^{--+}_3=&\frac{\pi}{4} \sgn(\agl{\lambda_1}{\lambda_2}\agl{\lambda_1}{\mu_3}\agl{\lambda_2}{\mu_3})\,\delta\bigg(\sum_i 2\kibnbnormal\bigg)\Theta(-\theta_{31})\Theta(-\theta_{32})\frac{\abs{\theta_{31}}^{2\kb_1}}{\kb_1}\frac{\abs{\theta_{32}}^{2\kb_2}}{\kb_2},
\end{split}}
\end{equation}
from which we may (anti-)symmetrise to give the amplitudes \eqref{eq: 3ptMHVceltwist} on the celestial torus. The above properties of Mellin transforms of sgn functions and their regularisation will be used again at four points.

\textbf{Three-Point} $\overline{MHV}$

Here we summarise the results of the similar derivation of the three-point $\overline{MHV}$ amplitude.
The spacetime amplitude
\begin{align}\label{eq: 3ptmhvbar}
    A^{++-}_3=\frac{\sqr{1}{2}^3}{\sqr{2}{3}\sqr{3}{1}}\,\delta^4(\lu{\alpha}_1\ltu{\alpha}_1+\lu{\alpha}_2\ltu{\alpha}_2+\lu{\alpha}_3\ltu{\alpha}_3)
\end{align}
is mapped to an ambidextrous twistor amplitude given by the following half-Fourier transforms
\begin{align}\label{eq: 3ptmhvbartwistortransform}
\tilde{A}^{++-}_3:= \frac{1}{(2\pi)^3} \int d^2\lambda_1\,  d^2\lambda_2\,  d^2\lambdat_3\, e^{i\agl{\lambda_1}{ \mu_1}}\,e^{i\agl{\lambda_2}{ \mu_2}}\,e^{i\sqr{\mut_3}{ \lambdat_3}} A^{++-}_3,
\end{align}
which gives the result
\begin{align}\label{eq: 3ptMHVbartwsitor}
    \tilde{A}^{++-}_3= \frac{(\pi i)^3}{(2\pi)^3} \sgn(\sqr{\lambdat_1}{\lambdat_2} )\, \sgn(\langle \mu_1 \lambda_3 \rangle+[\mut_3 \lambdat_1]) \,\sgn(\langle \mu_2 \lambda_3 \rangle+[\mut_3 \lambdat_2]).
\end{align}
The (2,2) signature three-point $\overline{MHV}$ celestial twistor amplitudes follow from an analogous derivation to that in the section above, and are given by 
\begin{equation}\label{eq: 3ptMHVbarceltwistamp}
\boxed{\begin{split}
    \tilde{\cA}^{++-}_3\{s_{k_1},s_{k_2},s_{\kb_3}\}=&\frac{\pi}{4} \sgn(\sqr{\lambdat_1}{\lambdat_2}\sqr{\mut_3}{\lambdat_1}\sqr{\mut_3}{\lambdat_2})\,\delta\bigg(\sum_i 2\kibnbnormal\bigg)\tilde\delta\bigg(\sum_i s_{\kibnbsmall}\bigg)\\
    &\quad\times\sgn(-\theta_{13})^{s_{k_1}}\frac{\abs{\theta_{13}^{-1}}^{2k_1}}{2k_1+(-1)^{s_{k_1}}\epsilon}\sgn(-\theta_{23})^{s_{k_2}}\frac{\abs{\theta_{23}^{-1}}^{2k_2}}{2k_2+(-1)^{s_{k_2}}\epsilon}\,,
\end{split}}
\end{equation}
where now the sums present in the delta functions are
\begin{equation}
    \sum_i 2\kibnbnormal= 2k_1+2k_2+2\kb_3,\quad \sum_i s_{\kibnbsmall}= s_{k_1}+s_{k_2}+s_{\kb_3}\,.
\end{equation}
The unsymmetrised amplitude from which the amplitudes are built is, dropping delta function terms as before, 
\begin{equation}\label{eq: 3ptMHVbarceltwistampUnSym}
\boxed{\begin{split}
    \tilde{\cA}^{++-}_3=\frac{\pi}{4} \sgn(\sqr{\lambdat_1}{\lambdat_2}\sqr{\mut_3}{\lambdat_1}\sqr{\mut_3}{\lambdat_2})\,\delta\bigg(\sum_i 2\kibnbnormal\bigg)\Theta(-\theta_{13})\Theta(-\theta_{23})\frac{\abs{\theta_{13}^{-1}}^{2k_1}}{k_1}\frac{\abs{\theta_{23}^{-1}}^{2k_2}}{k_2}\,.
\end{split}}
\end{equation}

\subsubsection{Four Points: Mellin Transform of a Product of Sgn Functions}\label{sec: fourpoints}

\textbf{Four-Point `Alternating' Amplitude}

We now move on to consider four points. We will discover that much of the same structure appears and identical methods can be used. At four points the $MHV$ and $\overline{MHV}$ amplitudes are one and the same, however we can consider different ways in which we assign helicity to each leg. We have $A_4^{+-+-}$ which we call the `alternating' helicity amplitude and $A_4^{++--}$ which we call the `separated' helicity amplitude. Following \cite{Arkani-Hamed:2009hub} we use the same ambidextrous twistor space for both of these amplitudes, in which case we find that both take their simplest, although qualitatively different form. We will first consider the alternating helicity amplitude and map it to celestial twistor space directly since the integrals are more straightforward. Once the dust has settled, we will be able to write the separated amplitude in terms of the alternating one. The alternating four-point amplitude is given by
\begin{align}\label{eq: 4ptalternating}
    A^{+-+-}_4=\frac{\agl{2}{4}^4}{\agl{1}{2}\agl{2}{3}\agl{3}{4}\agl{4}{1}}\,\delta^4(\lu{\alpha}_1\ltu{\alpha}_1+\lu{\alpha}_2\ltu{\alpha}_2+\lu{\alpha}_3\ltu{\alpha}_3+\lu{\alpha}_4\ltu{\alpha}_4)\,.
\end{align}
Once again, we map to twistor space in an ambidextrous manner by performing the following half-Fourier transforms
\begin{align}\label{eq: 4pttwistortransform}
\tilde{A}^{+-+-}_4:= \frac{1}{(2\pi)^4} \int d^2\lambda_1\,  d^2\lambdat_2\,  d^2\lambda_3\, d^2\lambdat_4\, e^{i\agl{\lambda_1}{\mu_1}}\,e^{i[\mut_2 \lambdat_2]}\,e^{i\agl{\lambda_3}{ \mu_3}}\,e^{i[\mut_4 \lambdat_4]} A^{+-+-}_4\,.
\end{align}
After regulating poles using a principal value prescription as before, the resulting amplitude takes the following form \cite{Arkani-Hamed:2009hub}
\begin{align}\label{eq: 4pttwistamp}
\begin{split}
    \tilde{A}^{+-+-}_4=&\;\frac{(\pi i)^4}{(2\pi)^4}\sgn(\agl{ \lambda_2}{\mu_1} +\sqr{\lambdat_1}{\mut_2})\,
    \sgn(\agl{ \lambda_4}{\mu_1} +\sqr{\lambdat_1}{\mut_4})\\
    &\qquad\quad\times\sgn(\agl{ \lambda_2}{\mu_3} +\sqr{\lambdat_3}{\mut_2})\,\sgn(\agl{ \lambda_4}{\mu_3} +\sqr{\lambdat_3}{\mut_4})\\
    =&\;\frac{1}{16}\sgn(\agl{ \lambda_2}{\mu_1}\agl{ \lambda_4}{\mu_1}\agl{ \lambda_2}{\mu_3}\agl{ \lambda_4}{\mu_3})\\
     &\qquad\quad\times \sgn(1 +\theta_{12}^{-1})\,
    \sgn(1 +\theta_{14}^{-1})\sgn(1 +\theta_{32}^{-1})\,\sgn(1 +\theta_{34}^{-1})\,.
\end{split}
\end{align}

\textbf{Symmetrised Amplitudes}

We are now tasked with mapping this amplitude to Mellin space via ambidextrous Mellin transforms. At four points we sum over the incoming/outgoing parameters and there are 16 combinations of parity under flipping of the sign of the spinors $\mu_1, \mut_2, \mu_3, \mut_4$. So we define
\begin{align}\label{eq: 4pttwistampevenodd}
    \tilde{A}^{+-+-}_4\{s_{k_1},s_{\kb_2},s_{k_3},s_{\kb_4}\} =\frac{1}{2^4}\sum_{\epsilon_1, \epsilonb_2, \epsilon_3, \epsilonb_4} \epsilon_1^{s_{k_1}}\epsilonb_2^{s_{\kb_2}}\epsilon_3^{s_{k_3}}\epsilonb_4^{s_{\kb_4}}\tilde{A}^{+-+-}_4(\epsilon_1\mu_1, \epsilonb_2\mut_2, \epsilon_3\mu_3,\epsilonb_4\mut_4)\,,
\end{align}
where the $\{s_{k_1},s_{\kb_2},s_{k_3},s_{\kb_4}\}$ label the parity for each spinor. As was the case at three points,  if we flip the sign of all the spinors $\mu_1, \mut_2, \mu_3, \mut_4 \mapsto -\mu_1, -\mut_2, -\mu_3, -\mut_4$ then the amplitude \eqref{eq: 4pttwistamp} is invariant thus for the amplitude \eqref{eq: 4pttwistampevenodd} to be non-zero we must have the even condition $s_{k_1}+s_{\kb_2}+s_{k_3}+s_{\kb_4}=0 \mod 2$. There are then eight non-vanishing amplitudes, which by direct calculation are given by
\begin{align}
\begin{split}
    &\tilde{A}^{+-+-}_4\{0,0,0,0\} = \frac{1}{16}\sgn(\agl{ \lambda_2}{\mu_1}\agl{ \lambda_4}{\mu_1}\agl{ \lambda_2}{\mu_3}\agl{ \lambda_4}{\mu_3})\\
     &\qquad\qquad\qquad\quad\times\bigg(G_0(\theta_{12})G_0(\theta_{14})G_0(\theta_{32})G_0(\theta_{34})+G_1(\theta_{12})G_1(\theta_{14})G_1(\theta_{32})G_1(\theta_{34})\bigg),\\
    &\tilde{A}^{+-+-}_4\{1,1,0,0\} = \frac{1}{16}\sgn(\agl{ \lambda_2}{\mu_1}\agl{ \lambda_4}{\mu_1}\agl{ \lambda_2}{\mu_3}\agl{ \lambda_4}{\mu_3})\\
     &\qquad\qquad\qquad\quad\times\bigg(G_1(\theta_{12})G_0(\theta_{14})G_0(\theta_{32})G_0(\theta_{34})+G_0(\theta_{12})G_1(\theta_{14})G_1(\theta_{32})G_1(\theta_{34})\bigg),\\
     &\qquad\qquad\vdots\\
     &\tilde{A}^{+-+-}_4\{1,1,1,1\} = \frac{1}{16}\sgn(\agl{ \lambda_2}{\mu_1}\agl{ \lambda_4}{\mu_1}\agl{ \lambda_2}{\mu_3}\agl{ \lambda_4}{\mu_3})\\
     &\qquad\qquad\qquad\quad\times\bigg(G_1(\theta_{12})G_0(\theta_{14})G_0(\theta_{32})G_1(\theta_{34})+G_0(\theta_{12})G_1(\theta_{14})G_1(\theta_{32})G_0(\theta_{34})\bigg),
\end{split}
\end{align}
where we use the expressions \eqref{eq: HatStairThetaFunctions} for the $G_0$ and $G_1$ functions. The general formula for the symmetrised amplitude is
\begin{align}\label{eq: evenodd4pttwistoramp}
\begin{split}
    \tilde{A}^{+-+-}_4\{s_{k_1},s_{\kb_2},s_{k_3},s_{\kb_4}\} =& \frac{1}{16}\sgn(\agl{ \lambda_2}{\mu_1}\agl{ \lambda_4}{\mu_1}\agl{ \lambda_2}{\mu_3}\agl{ \lambda_4}{\mu_3})\delta\bigg(\sum_{i} s_{\kibnbsmall}\bigg)\\
    &\quad\quad\times\sum_{r_{ij}}G_{r_{12}}(\theta_{12})G_{r_{14}}(\theta_{14})G_{r_{32}}(\theta_{32})G_{r_{34}}(\theta_{34}),
\end{split}
\end{align}
where the sum in \eqref{eq: evenodd4pttwistoramp} gives the two unique terms with $r_{ij} \in {0,1}$ such that
\begin{align}
    \begin{split}
        &r_{12}+r_{14}=s_{k_1} \mod 2\, , \\
        &r_{12}+r_{32}=s_{\kb_2} \mod 2\, , \\
        &r_{32}+r_{34}=s_{k_3} \mod 2\, , \\
        &r_{34}+r_{14}=s_{\kb_4} \mod 2\, .
    \end{split}
\end{align}
For example, if we want an amplitude that is odd in $\mu_1$ ($s_{k_1}{=}1$) then we require a product of an odd number of the anti-symmetric $G_1$ functions evaluated at the variables $\theta_{1i}$, which contain $\mu_1$.

One can derive the various symmetrised four-point celestial twistor gluon amplitudes by Mellin transforming \eqref{eq: evenodd4pttwistoramp}, exactly as was done at three points. We are led to consider the integrals
\begin{align}\label{eq: I4}
\begin{split}
    \cI(r_{12}, r_{14}, r_{32}, r_{34})&:=
    \int_{0}^{\infty} \frac{dt_1\,d\tb_2\,dt_3\,d\tb_4}{t_1\tb_2t_3\tb_4}\,
    t_1^{2k_1}\,\tb_2^{2\kb_2}\,t_3^{2k_3} \,\tb_4^{2\kb_4}\\
    &\qquad\qquad\times G_{r_{12}}\bigg(\frac{\tb_2}{t_1}\theta^{-1}_{12}\bigg)G_{r_{14}}\bigg(\frac{\tb_4}{t_1}\theta^{-1}_{14}\bigg)G_{r_{32}}\bigg(\frac{\tb_2}{t_3}\theta^{-1}_{32}\bigg)G_{r_{34}}\bigg(\frac{\tb_4}{t_3}\theta^{-1}_{34}\bigg) .
\end{split}
\end{align}
These integrals have a similar character to the integrals we met in the three-point case. They are  conformal Mellin integrals over a positive quadrant in four dimensions, whose integrands are piece-wise constant functions in regions bounded by hyperplanes through the origin, for example $t_1 =\tb_2\abs{\theta^{-1}_{12}}$.

Rather than continue with the derivation of the symmetrised amplitudes, we will instead derive the \textit{unsymmetrised} amplitude by Mellin transforming \eqref{eq: 4pttwistamp}. With this, we can find the symmetrised celestial twistor amplitudes through simple (anti-)symmetrisation.

\textbf{Unsymmetrised Amplitude}

We now calculate the unsymmetrised four-point celestial twistor amplitude in its entirety, extending the three-point results \eqref{eq: 3ptMHVceltwistUnsym} and \eqref{eq: 3ptMHVbarceltwistampUnSym}. At four points we are led to consider Mellin transforms of a product of $\sgn$ functions. Once again, we must regularise the integrals such that the Mellin inverse exists and the regularisation commutes with taking even and odd linear combinations.

Performing ambidextrous Mellin transforms on the un-symmetrised amplitude \eqref{eq: 4pttwistamp} leads to the integral
\begin{align}\label{eq: I4unsymtvars}
\begin{split}
    \cI_{4pt}&:=
    \int_{0}^{\infty} \frac{dt_1\,d\tb_2\,dt_3\,d\tb_4}{t_1\tb_2t_3\tb_4}\,
    t_1^{2k_1}\,\tb_2^{2\kb_2}\,t_3^{2k_3} \,\tb_4^{2\kb_4}\\
    &\qquad\qquad\times \sgn\bigg(1+\frac{\tb_2}{t_1}\theta^{-1}_{12}\bigg)\sgn\bigg(1+\frac{\tb_2}{t_3}\theta^{-1}_{32}\bigg)\sgn\bigg(1+\frac{\tb_4}{t_1}\theta^{-1}_{14}\bigg)\sgn\bigg(1+\frac{\tb_4}{t_3}\theta^{-1}_{34}\bigg).
\end{split}
\end{align}
As usual, we can integrate out an overall scale which we choose to be $t_1$. The resulting integral has a corresponding dilatation invariance delta function and now manifestly depends on the scales $\abs{\theta_{ij}}$:  
\begin{align}\label{eq: I4unsym}
\begin{split}
     \cI_{4pt}&=2 \pi i\,\delta\bigg(\sum_i 2\kibnbnormal\bigg)\int_{0}^{\infty}\frac{dx\,dy\,dz}{xyz}\,x^{2\kb_2}y^{2k_3}z^{2\kb_4}\,\sgn\bigg(1+x\,\theta^{-1}_{12}\bigg)
    \\
    &\qquad\qquad\times\sgn\bigg(1+\frac{x}{y}\theta^{-1}_{32}\bigg)\sgn\bigg(1+z\, \theta^{-1}_{14}\bigg)\sgn\bigg(1+\frac{z}{y}\theta^{-1}_{34}\bigg) ,\\
     &=2 \pi i\,\delta\bigg(\sum_i 2\kibnbnormal\bigg)\int_{0}^{\infty}\frac{dy}{y}\,y^{2k_3}\, \cJ_2(y) \cJ_4(y) ,
\end{split}
\end{align}
where we have the Mellin integrals
\begin{align}\label{eq: J2integral}
\begin{split}
    \cJ_2(y):=\int_{0}^{\infty}\frac{dx}{x}\,x^{2\kb_2}\,\sgn\bigg(1+x\,\theta^{-1}_{12}\bigg)\sgn\bigg(1+\frac{x}{y}\theta^{-1}_{32}\bigg) \\
    =\int_{-\infty}^{\infty}ds\,e^{2\kb_2s-\epsilon \abs{\abs{s}}}\,\sgn\bigg(1+e^{s}\,\theta^{-1}_{12}\bigg)\sgn\bigg(1+\frac{e^{s}}{y}\theta^{-1}_{32}\bigg) ,
    \end{split}
\end{align}
and similarly for $\cJ_4(y)$. As in the three-point case two of the integrals separate, and here the third integration over $y$ entwines them. In equation \eqref{eq: J2integral} above we also define a specific regularisation which takes into account the two parameters of the integral
\begin{align}
\begin{split}
    \abs{\abs{s}}:=  \frac{1}{2}\bigg(\sgn\bigg(s-\ln(\abs{\theta_{12}})\bigg) + \sgn\bigg(s-\ln(\abs{\theta_{32}})\bigg)\bigg)\,s,
    \end{split}
\end{align}
such that when $s$ is greater than both $\ln(\abs{\theta_{12}})$ and $\ln(\abs{\theta_{32}})$ we have $\abs{\abs{s}}=s$, when $s$ lies between them we have $\abs{\abs{s}}=0$, and when $s$ is less than both we have $\abs{\abs{s}}=-s$.

To calculate \eqref{eq: J2integral} we must condition on the relative magnitude of $\abs{\frac{\theta_{32}}{\theta_{12}}}$. Choosing $y=0$ for convenience since it can be reinstated at the end, we compute $\cJ_2(0)$ to be
\begin{align}\label{eq: J2unsymresult}
\begin{split}
    \cJ_2(0)&:=\Theta\bigg(1-\abs{\frac{\theta_{32}}{\theta_{12}}}\bigg)\bigg[ \frac{(\abs{\theta_{32}})^{2\kb_2}}{2\kb_2+\epsilon} +\sgn(\theta_{32})\bigg(\frac{\abs{\theta_{12}}^{2\kb_2}}{2\kb_2} -\frac{(\abs{\theta_{32}})^{2\kb_2}}{2\kb_2}\bigg) \\
    &\qquad\qquad\qquad\qquad-\sgn(\theta_{12}\theta_{32})\frac{\abs{\theta_{12}}^{2\kb_2}}{2\kb_2-\epsilon}\bigg] \\
    &+\Theta\bigg(\abs{\frac{\theta_{32}}{\theta_{12}}}-1\bigg)\bigg[ \frac{\abs{\theta_{12}}^{2\kb_2}}{2\kb_2+\epsilon} +\sgn(\theta_{12})\bigg( \frac{(\abs{\theta_{32}})^{2\kb_2}}{2\kb_2}-\frac{\abs{\theta_{12}}^{2\kb_2}}{2\kb_2}\bigg) \\
    &\qquad\qquad\qquad\qquad-\sgn(\theta_{12}\theta_{32})\frac{(\abs{\theta_{32}})^{2\kb_2}}{2\kb_2-\epsilon}\bigg],
\end{split}
\end{align}
where we have also dropped the regulariser $\epsilon$ when it appears in the exponents as usual.

Just as for the Mellin transform of a single $\sgn$ function, when $\theta_{12}, \theta_{32} >0$ we recover a contribution proportional to a delta function, while if $\theta_{12}, \theta_{32} <0$ we recover the principal value pole
while the other mixed cases give linear combinations of the two. 

To confirm our results we now perform the inverse Mellin transform on \eqref{eq: J2unsymresult}. In the result \eqref{eq: J2unsymresult} we have poles with various $\epsilon$-prescriptions - advanced, retarded and principal value - where poles written $\frac{1}{2\kb_2}$ are principal valued \eqref{eq: pvanddelta}.
These $\epsilon$-prescriptions are crucial for performing the inverse Mellin transform. When we integrate $2\kb_2$ along the imaginary axis we will close the contour to form a large semi-circle and pick up various possible contributions from each of these regulated poles. The direction in which we close the contour will vary term by term and is dictated by the condition that the integral dies off on the semicircular arc; this depends on the the magnitudes of $\abs{\theta_{12}}$ and $\abs{\theta_{32}}$. For example, for terms involving $\abs{\theta_{12}}$ and for $\abs{\theta_{12}}> 1$ we close to the left (negative real part) and we get a non-zero contribution if this term also has a $+\epsilon$ prescription. For terms involving $\abs{\theta_{32}}$ and for $\abs{\theta_{32}}< 1$ we close to the right (positive real part) and we get a non-zero contribution if this term also has a $-\epsilon$ prescription. The principal value terms require some care. We pick up contributions regardless of which way we close the contour. However, the principal valued poles are multiplied by a difference of terms and the residues from the two terms cancel if we close the contour in the same direction for both terms and their residues sum if we close the contour in opposite directions for each term. 

In summary, if we perform an inverse Mellin transform on \eqref{eq: J2unsymresult} we  have six cases for the magnitudes of $\abs{\theta_{12}}, \abs{\theta_{32}}$. Each case picks up a contribution from exactly one of the terms in \eqref{eq: J2unsymresult}:
\begin{enumerate}
        \item $\abs{\theta_{12}}>\abs{\theta_{32}}>1$ gives $\Theta\bigg(1-\abs{\frac{\theta_{32}}{\theta_{12}}}\bigg)\, \Theta(\abs{\theta_{12}}-1) \,\Theta (\abs{\theta_{32}}-1)$
        \item $\abs{\theta_{12}}>1>\abs{\theta_{32}}$ gives $\Theta\bigg(1-\abs{\frac{\theta_{32}}{\theta_{12}}}\bigg)\, \Theta(\abs{\theta_{12}}-1) \,\Theta (1-\abs{\theta_{32}})\, \sgn(\theta_{32})$
        \item $1>\abs{\theta_{12}}>\abs{\theta_{32}}$ gives $\Theta\bigg(1-\abs{\frac{\theta_{32}}{\theta_{12}}}\bigg) \,\Theta(1-\abs{\theta_{12}})\, \Theta (1-\abs{\theta_{32}})\, \sgn(\theta_{12}\theta_{32})$
        \item $\abs{\theta_{32}}>\abs{\theta_{12}}>1$ gives $\Theta\bigg(\abs{\frac{\theta_{32}}{\theta_{12}}}-1\bigg)\, \Theta(\abs{\theta_{12}}-1) \,\Theta (\abs{\theta_{32}}-1)$
        \item $\abs{\theta_{32}}>1>\abs{\theta_{12}}$ gives $\Theta\bigg(\abs{\frac{\theta_{32}}{\theta_{12}}}-1\bigg)\, \Theta(1-\abs{\theta_{12}}) \,\Theta (\abs{\theta_{32}}-1)\, \sgn(\theta_{12})$
        \item $1>\abs{\theta_{32}}>\abs{\theta_{12}}$ gives $\Theta\bigg(\abs{\frac{\theta_{32}}{\theta_{12}}}-1\bigg) \,\Theta(1-\abs{\theta_{12}})\, \Theta (1-\abs{\theta_{32}})\, \sgn(\theta_{12}\theta_{32})$
\end{enumerate}

The sum of the above contributions gives the required result - that is we recover the original function $\sgn(1+\theta_{12}^{-1})\,\sgn(1+\theta_{32}^{-1})$.

Once again we can discard the $\epsilon$ prescriptions, and hence delta function terms, with the proviso that there exists a series of contour deformations of our inverse Mellin transform such that the Mellin inversion theorem holds. Then we can find a simpler formula for the $\cJ_2(y)$ function, which is (reinstating $y$)
\begin{align}\label{eq: J2unsymsimpleresult}
\begin{split}
    \cJ_2(y)
    &=2\Theta(-\theta_{12})\Theta(\theta_{32})\frac{\abs{\theta_{12}}^{2\kb_2}}{2\kb_2} + 2\,\Theta(\theta_{12})\Theta(-\theta_{32})\frac{(y\abs{\theta_{32}})^{2\kb_2}}{2\kb_2}\\
    &+2\,\Theta(-\theta_{12})\Theta(-\theta_{32})\,\sgn\bigg(1-y\abs{\frac{\theta_{32}}{\theta_{12}}}\bigg)\bigg[ \frac{(y\abs{\theta_{32}})^{2\kb_2}}{2\kb_2} - \frac{\abs{\theta_{12}}^{2\kb_2}}{2\kb_2}\bigg].
\end{split}
\end{align}
Using this result we can then find the full un-symmetrised integral $\cI_{4pt}$ from which the unsymmetrised amplitude is simply
\begin{equation}
    \tilde \cA_4^{+-+-}=\frac{1}{16}\sgn(\agl{\mu_1}{\lambda_2}\agl{ \lambda_2}{\mu_3}\agl{ \mu_3}{\lambda_4}\agl{ \lambda_4}{\mu_1})\cI_{4pt}
\end{equation}
We compute $\cI_{4pt}$ by plugging in the function \eqref{eq: J2unsymsimpleresult} into integral \eqref{eq: I4unsym}. The resulting integral over $y$ consists of a collection of products of sgn functions and so is of the exact same form as the $\mathcal{J}$ integrals we just calculated. Hence we just iterate the method. 

With the usual $\epsilon$ regularisation, which we then drop, we find the unsymmetrised celestial twistor amplitude can be written compactly as
 \begin{equation}\label{eq: AmpFinalYM}
\boxed{\begin{split}
   \tilde \cA_4^{+-+-} 
   =&\frac{\pi i}{16}\sgn(\agl{\mu_1}{\lambda_2}\agl{ \lambda_2}{\mu_3}\agl{ \mu_3}{\lambda_4}\agl{ \lambda_4}{\mu_1}) \,\delta\bigg(\sum_i 2\kibnbnormal\bigg) \;\vert \theta_{12}\vert^{2\kb_2}\vert\theta_{14}\vert^{2k_3+2\kb_4}\vert\theta_{34}\vert^{-2k_3}
   \\ &   
   \times \bigg[ \,\Theta(-\theta_{12})\Theta(-\theta_{32})\Theta(-\theta_{14})\Theta(-\theta_{34})\,\sgn(1-\vert\theta\vert)
     \Big( V_{12}(\theta)  + V_{34}(\theta) -V_{32}(\theta) - V_{14}(\theta) \Big) 
   \\ &
  -\Theta(-\theta_{12})\Theta(-\theta_{32})\Theta(\theta_{34})\Theta(-\theta_{14})\;V_{34}(\theta)  
   -\Theta(\theta_{12})\Theta(-\theta_{32})\Theta(-\theta_{14})\Theta(-\theta_{34})\;V_{12}(\theta)  
   \\&
   -
  \Theta(-\theta_{12})\Theta(\theta_{32})\Theta(-\theta_{34})\Theta(-\theta_{14})\;V_{32}(\theta)  
   -\Theta(-\theta_{12})\Theta(-\theta_{32})\Theta(-\theta_{34})\Theta(\theta_{14})\;V_{14}(\theta) \bigg]\,,
\end{split}}
\end{equation}
where $\theta$ is the \textit{ratio} of conformally invariant cross-ratios
\begin{equation}
    \theta := \frac{\theta_{12}\theta_{34}}{\theta_{14}\theta_{32}} = \frac{\frac{\agl{\lam_2}{\mu_1}\agl{\lam_4}{\mu_3}}{\agl{\lam_4}{\mu_1}\agl{\lam_2}{\mu_3}}}{\frac{\sqr{\lamt_1}{\mut_4}\sqr{\lamt_3}{\mut_2}}{\sqr{\lamt_1}{\mut_2}\sqr{\lamt_3}{\mut_4}}}=: \frac{r}{\bar r}\,,
\end{equation}
and the polynomial functions $V_{12}$ etc. are given by
\begin{align}
\begin{split}
\label{eq: Hfunction}
    V_{12}(\theta) =& \frac{\abs{\theta}^{-2\kb_2}}{2\kb_2(2\kb_2+2k_3)(2\kb_2+2k_3+2\kb_4)}, \quad 
     V_{32}(\theta) = \frac{1}{2\kb_22k_3(2k_3+2\kb_4)},\\
      V_{14}(\theta) =&\frac{\abs{\theta}^{2k_3+2\kb_4}}{2\kb_4(2k_3+2\kb_4)(2\kb_2+2k_3+2\kb_4)}, \quad
       V_{34}(\theta) = \frac{\abs{\theta}^{2k_3}}{2k_32\kb_4(2\kb_2+2k_3)}.\\
\end{split}
\end{align}

Equation \eqref{eq: AmpFinalYM} is the main result of this section.
We have pulled out a conformally covariant factor $\abs{\theta_{12}}^{2\kb_2}\abs{\theta_{14}}^{2k_3+2\kb_4}\abs{\theta_{34}}^{-2k_3}$ and the remaining function has support in various \lq channels' dictated by the signs of the variables $\theta_{ij}$. Since we have accounted for conformal covariance, this function can only depend on the conformally invariant cross-ratios $r, \bar r$. In fact, in each of these channels, we have simple polynomial functions (with imaginary powers) depending only on the absolute value of the ratio of conformally invariant cross-ratios.
 On the momentum space celestial torus, momentum conservation at four points implied that the conformal cross-ratios $z,\zb$ are not independent but are actually equal $z=\zb$. Correspondingly, in twistor space, the cross-ratios $r, \bar r$ are now independent but the amplitude only depends on their ratio $\theta$.

Now we return to the task of finding the symmetrised celestial twistor amplitudes. We have sixteen parity combinations for the integrals $\mathcal{I}(r_{12}, r_{32}, r_{34}, r_{14})$, two for each amplitude in \eqref{eq: 4pttwistampevenodd}. From \eqref{eq: AmpFinalYM} we can find all of them via (anti-)symmetrisation.
For example, we can find the fully symmetric integral 
\begin{equation}\label{eq: I(OOO0)}
\begin{split}
    &\cI(0,0,0,0)\\
    &=2\pi i \,\delta\bigg(\sum_i 2\kibnbnormal\bigg)\abs{\theta_{12}}^{2\kb_2}\abs{\theta_{14}}    ^{2k_3+2\kb_4}
      \abs{\theta_{34}}^{-2k_3}\times\\
    &\!\bigg(\Theta(\abs{\theta}-1) \bigg[- \frac{\abs{\theta}^{2k_3}}{ (2k_3-\epsilon) (2\kb_4 + \epsilon)(2\kb_2 + 2k_3)}- \frac{\abs{\theta}^{-2\kb_2}}{(2\kb_2+\epsilon)(2\kb_2 + 2k_3) (2\kb_2+2k_3 +2\kb_4
       + \epsilon) }\bigg]\\ 
       + &\Theta(1 - \abs{\theta}) \bigg[-\frac{1
      }{(2\kb_2+\epsilon)(2k_3 - \epsilon) (2k_3+2\kb_4
       )}-\frac{\abs{\theta}^{2k_3+2\kb_4}
      }{(2\kb_4 +\epsilon)(2k_3+2\kb_4)(2\kb_2+2k_3+2\kb_4
       + \epsilon)}\bigg]\bigg).
\end{split}
\end{equation}
The integral $\cI(1,1,1,1)$ takes a very similar form,
\begin{equation}\label{eq: I(1111)}
\begin{split}
    &\cI(1,1,1,1)\\
    &=2\pi i \,\delta\bigg(\sum_i 2\kibnbnormal\bigg)\abs{\theta_{12}}^{2\kb_2}\abs{\theta_{14}}^{2k_3+2\kb_4}
      \abs{\theta_{34}}^{-2k_3}\,\sgn(\theta_{12}\theta_{32}\theta_{34}\theta_{14})\times\\
    &\!\bigg(\Theta(\abs{\theta}-1)\bigg[\frac{1
      }{(2\kb_2-\epsilon)(2k_3 +\epsilon) (2k_3+2\kb_4
       )}+\frac{\abs{\theta}^{2k_3+2\kb_4}
      }{(2\kb_4 -\epsilon)(2k_3+2\kb_4)(2\kb_2+2k_3+2\kb_4
       - \epsilon)}\bigg] \\ 
       + &\Theta(1 - \abs{\theta}) \bigg[ \frac{\abs{\theta}^{2k_3}}{(2k_3 
       + \epsilon)(2\kb_4 - \epsilon)(2\kb_2 + 2k_3)}+ \frac{\abs{\theta}^{-2\kb_2}}{(2\kb_2-\epsilon)(2\kb_2 + 2k_3) (2\kb_2+2k_3 +2\kb_4
       - \epsilon) }\bigg]\bigg).
\end{split}
\end{equation}
In the above formulae, we have reinstated the $\epsilon$ regularisation stemming from the result \eqref{eq: J2unsymresult}, which is required to prescribe the correct contour for the inverse Mellin transform.

The two integrals  $\cI(0,0,0,0)$ and $\cI(1,1,1,1)$ contribute to the fully symmetrised celestial twistor amplitude
\begin{equation}\label{eq: decompFullSymAmp}
    \tilde{\cA}^{+-+-}_4\{0,0,0,0\} = \frac{1}{16}\sgn(\agl{\mu_1}{\lambda_2}\agl{ \lambda_2}{\mu_3}\agl{ \mu_3}{\lambda_4}\agl{ \lambda_4}{\mu_1})\tilde\delta\bigg(\sum_i s_{\kibnbsmall}\bigg)(\cI(0,0,0,0)+\cI(1,1,1,1)),
\end{equation}
which in turn is a light transformed correlator in the celestial CFT. A similar fully symmetrised light transformed correlator was computed in \cite{De:2022gjn} by performing light transforms on a (2,2) celestial amplitude analytically continued from (1,3) with summation over incoming/outgoing parameters. In that paper the result was computed in terms of special functions, Fox H-functions and Generalised I-functions, defined by Mellin-Barnes integrals; it would be interesting to understand how these results are related to the discussion here.

We find that the general structure of the formulas for $\cI(0,0,0,0)$ and $\cI(1,1,1,1)$ is repeated for all the symmetrised integrals. They all contain the same basic functions $V_{12}, V_{32}$ etc. inherited from the unsymetrised amplitude \eqref{eq: AmpFinalYM}. Unlike the unsymetrised amplitude \eqref{eq: AmpFinalYM}, which has support on a handful of channels dictated by the signs of the $\theta_{ij}$, the symmetrised amplitudes have a clarified structure with only two channels which are given by the magnitude of $\abs{\theta}$ being greater than or less than one. 

The only differences between the various symmetrised amplitudes are: the overall sgn functions which control the even and odd symmetry, which polynomial contibutions are probed by the two \lq channels' $\abs{\theta}>1$ or $\abs{\theta}<1$, and the $\epsilon$ regularisation of the poles in the weights. We note that, there are two terms $\mathcal{I}(r_{12}, r_{32}, r_{34}, r_{14})$ which have the same even/odd parity and so contribute to the same amplitude - these two terms have opposite arguments $r_{ij}=0 \leftrightarrow r_{ij}=1$, opposite sgn function factors present, opposite $\epsilon$ prescription for the poles and opposite allocation of terms to each \lq channel'.

Finally we note that all of the symmetrised integrals $\mathcal{I}(r_{12}, r_{32}, r_{34}, r_{14})$ \eqref{eq: I4} can be encapsulated in the general formula
\begin{equation}\label{eq: AmpFinalYMtwo}
\boxed{\begin{split}
   &\mathcal{I}(r_{12}, r_{32}, r_{34}, r_{14}) = 2\pi i \;\sgn\big[\agl{\mu_1}{\lambda_2}\agl{ \lambda_2}{\mu_3}\agl{ \mu_3}{\lambda_4}\agl{ \lambda_4}{\mu_1}\big] \,\delta\bigg(\sum_i 2 \kibnbnormal \bigg) \;\vert \theta_{12}\vert^{2\kb_2}\vert\theta_{14}\vert^{2k_3+2\kb_4}\vert\theta_{34}\vert^{-2k_3}   
   \\ &   
  \times\prod_{(i,j): r_{ij}=1 }\!\!\!\!\!\!(\sgn(\theta_{ij}))\sum_{x,y}
  \bigg[\Big(
   \delta^1_x V_x(\theta)-\delta^0_y V_y(\theta)\Big)\Theta(\vert\theta\vert-1)
   + 
    \Big(\delta^1_y V_y(\theta)-\delta^0_x V_x(\theta) \Big)\Theta(1-\vert\theta\vert)\bigg] ,
\end{split}}
\end{equation}
where in the above the product is over the values of $(i,j)$ such that $r_{ij}=1$, we sum over the variables
$x\in (r_{32}, r_{14}),y\in (r_{12},r_{34})$, and we define $V_{r_{ij}}(\theta):= V_{ij}(\theta)$. As a further example, for the case $\cI(1,0,1,0)$ we have only a $\Theta(1-\vert\theta\vert)$ term 
\begin{equation}\label{eq: I(1,0,1,0)}
\begin{split}
        \cI(1,0,1,0)&=2\pi i\,\delta\bigg(\sum_i 2\kibnbnormal\bigg)\,\sgn(\theta_{12}\theta_{34})\,\Theta(1-\abs{\theta})\,\abs{\theta_{12}}^{2\kb_2}\abs{\theta_{14}}    ^{2k_3+2\kb_4}
      \abs{\theta_{34}}^{-2k_3}\\
        &\times \bigg(\frac{\abs{\theta}^{2k_3}}{(2k_3) (2\kb_4)(2\kb_2 + 2k_3)}+ \frac{\abs{\theta}^{-2\kb_2}}{(2\kb_2)(2\kb_2 + 2k_3) (2\kb_2+2k_3 +2\kb_4) }\\ 
        &-\frac{1
      }{(2\kb_2)(2k_3) (2k_3+2\kb_4
       )}-\frac{\abs{\theta}^{2k_3+2\kb_4}
      }{(2\kb_4)(2k_3+2\kb_4)(2\kb_2+2k_3+2\kb_4)}\bigg)\, ,
    \end{split}
\end{equation}
which can be checked directly by performing the Mellin integrals in \eqref{eq: I4}. Similarly for $\cI(0,1,0,1)$ we have only a $\Theta(\vert\theta\vert-1)$ term. The simplicity of the four-point results can be seen to be a consequence of BCFW recursion relations in celestial twistor space which we explore in detail in Section \ref{sec: bcfw}. The result for $\cI$ decomposed into all the even/odd parts $\mathcal{I}(r_{12}, r_{32}, r_{34}, r_{14})$ is given in a companion Mathematica file.


\textbf{Four-Point `Separated' Amplitude}

Finally, the separated four-point amplitude $A^{YM}(++--)$ is given by 
\begin{align}\label{eq: 4ptseparated}
    A^{++--}_4=\frac{\agl{3}{4}^4}{\agl{1}{2}\agl{2}{3}\agl{3}{4}\agl{4}{1}}\,\delta^4(\lu{\alpha}_1\ltu{\alpha}_1+\lu{\alpha}_2\ltu{\alpha}_2+\lu{\alpha}_3\ltu{\alpha}_3+\lu{\alpha}_4\ltu{\alpha}_4).
\end{align}
We again perform the same twistor transforms as in \eqref{eq: 4pttwistortransform} and so the twistor amplitude is in the same ambidextrous basis and takes the following form \cite{Arkani-Hamed:2009hub}
\begin{align}\label{eq: 4pttwistampseparated}
\begin{split}
    \tilde{A}^{++--}_4=&\frac{(\pi i)^4}{(2\pi)^4}\sgn(\agl{ \lambda_2}{\mu_1} +\sqr{\lambdat_1}{\mut_2})\,
    \sgn(\agl{ \lambda_4}{\mu_1} +\sqr{\lambdat_1}{\mut_4})\\
    &\qquad\quad\times\delta^{(3)}(\agl{ \lambda_2}{\mu_3} +\sqr{\lambdat_3}{\mut_2})\,\sgn(\agl{ \lambda_4}{\mu_3} +\sqr{\lambdat_3}{\mut_4})\\
    =&\frac{1}{16}\sgn(\agl{ \lambda_2}{\mu_1}\agl{ \lambda_4}{\mu_1}\agl{ \lambda_2}{\mu_3}\agl{ \lambda_4}{\mu_3})\abs{\agl{ \lambda_2}{\mu_3}}^{-4}\\
     &\qquad\quad\times \sgn(1 +\theta_{12}^{-1})\,
    \sgn(1 +\theta_{14}^{-1})\delta^{(3)}(1 +\theta_{32}^{-1})\,\sgn(1 +\theta_{34}^{-1}).
\end{split}
\end{align}

The separated amplitude is hence $\abs{\agl{ \lambda_2}{\mu_3}}^{-4}$ multiplied by the fourth derivative with respect to $\theta_{32}^{-1}$ of the alternating amplitude above. Under the half-Mellin transforms this operator becomes $\abs{\agl{ \lambda_2}{\mu_3}}^{-4}(\vert \theta_{32}\vert^2\partial_{\vert \theta_{32}\vert})^4\,\exp(-2\partial_{\kb_2})$,  which we then apply to \eqref{eq: AmpFinalYM}.
When acting on $\Theta$ functions containing $\theta_{32}$, either directly or via the cross-ratio $\theta$, these derivatives give (derivatives of) delta functions, which have only singular support and so again we drop these terms for generic kinematics. Furthermore, there are no powers of $\theta_{32}$ in the first line of \eqref{eq: AmpFinalYM}. Thus we only need to apply the derivatives to the powers of $\abs{\theta}$ inside the $V$ functions. This yields straightforward additional numerator factors, since 
\begin{align}
\label{eq: bDeriv}
(\vert \theta_{32}\vert^2\partial_{\vert \theta_{32}\vert})^4 \abs{\theta}^n = n(n-1)(n-2)(n-3)\vert \theta_{32}\vert^4 \vert\theta\vert^n.
\end{align}
Hence we can easily find the separated amplitude from the alternating one.

We leave the task of explicitly computing celestial twistor amplitudes beyond four points to future work. However, we note that the method of producing twistor amplitudes by directly half-Fourier transforming momentum space amplitudes becomes more difficult as the number of particles and MHV degree grows. In particular in the ambidextrous basis of \cite{Arkani-Hamed:2009hub} at five points and beyond one encounters Grassmann integrals which do not disentangle into sgn functions. We can Mellin transform these integrals and we recover a class of integrals related to hyper-geometric functions and their generalisations studied in \cite{De:2022gjn, Hu:2022syq}. Nevertheless, the simplicity of ambidextrous twistor amplitudes extends well beyond four points, see for example the six-point `alternating' $(+,-,+,-,+,-)$ amplitude in \cite{Arkani-Hamed:2009hub}. This simplicity is largely due to the existence of powerful BCFW formulae. The topic of BCFW recursion relations is something we explore in Section \ref{sec: bcfw} and is a promising road to higher multiplicity celestial twistor amplitudes.


\subsection{Pure Gravity}

For the gravity case, we may follow similar arguments to those used above for celestial twistor gluon amplitudes.

\subsubsection{Three Points: Mellin Transform of a Mod Function}\label{sec: grav3pts}

\textbf{Three-Point} $MHV$

For the three-point gravity amplitude we begin with the momentum space expression
\begin{align}\label{eq: 3ptmhvgravity}
    M^{--+}_3=\frac{\agl{1}{2}^6}{\agl{2}{3}^2\agl{3}{1}^2}\,\delta^4(\lu{\alpha}_1\ltu{\alpha}_1+\lu{\alpha}_2\ltu{\alpha}_2+\lu{\alpha}_3\ltu{\alpha}_3).
\end{align}
The ambidextrous twistor transform of this is 
\begin{align}\label{eq: 3ptgravitytwistortransform}
\tilde{M}^{--+}_3:= \frac{1}{(2\pi)^3} \int d^2\lambdat_1\,  d^2\lambdat_2\,  d^2\lambda_3\, e^{i\sqr{\mut_1}{ \lambdat_1}}\,e^{i[\mut_2 \lambdat_2]}\,e^{i\agl{\lambda_3}{ \mu_3}} M^{--+}_3,
\end{align}
which leads to \cite{Arkani-Hamed:2009hub}
\begin{align}\label{eq: 3ptMHVtwistorgravity}
    \tilde{M}^{--+}_3 = \frac{(\pi i)^3}{(2\pi)^3} \Big\vert\langle \lambda_1 \lambda_2 \rangle \Big\vert\, \Big\vert\langle \lambda_1 \mu_3 \rangle+[\lambdat_3 \mut_1]\Big\vert \,\Big\vert\langle \lambda_2 \mu_3 \rangle+[\lambdat_3 \mut_2]\Big\vert.
\end{align}
Analogous arguments to those given above for the YM case lead us to consider only the amplitudes
\begin{align}\label{eq: 3ptMHVtwistorevenoddampsAgainGravity}
\begin{split}
        \tilde{M}^{--+}_3\{0,0,0\}, \,\,\,       
        \tilde{M}^{--+}_3\{1,1,0\},\,\,\,
        \tilde{M}^{--+}_3\{1,0,1\},\,\,\,
        \tilde{M}^{--+}_3\{0,1,1\},
\end{split}
\end{align}
where the (anti-)symmetrised amplitudes take the general form
\begin{align}\label{eq: 3ptMHVtwistorevenoddampsAgainGravity2}
\begin{split}
        &\tilde{M}^{--+}_3\{s_{\kb_1},s_{\kb_2},s_{k_3}\}       = \frac{(\pi i)^3}{(2\pi)^3} \Big\vert\langle \lambda_1 \lambda_2 \rangle \agl{\lambda_1}{\mu_3} \agl{\lambda_2}{\mu_3}\Big\vert\,\delta\bigg(\sum_{i} s_{\kibnbsmall}\bigg)\, H_{s_{\kb_2}+s_{k_3}}(\theta_{31}^{-1})H_{s_{\kb_1}+s_{k_3}}(\theta_{32}^{-1}),
\end{split}
\end{align}
where we recall the definition used above
\begin{align}\label{eq: defthetaij2}
    \theta_{ij}=\frac{\agl{\lambda_j}{\mu_i}}{\sqr{\lambdat_i}{\mut_j}}
\end{align}
and define the (anti-)symmetrised mod functions
\begin{align}
    \begin{split}
        H_0(x):=\frac{1}{2}\bigg(\Big\vert 1+x\Big\vert+\Big\vert 1-x\Big\vert\bigg),\\
        H_1(x):=\frac{1}{2}\bigg(\Big\vert 1+x\Big\vert-\Big\vert 1-x\Big\vert\bigg).
    \end{split}
\end{align}

We then compute the following Mellin transforms
\begin{align}\label{eq: hatstairsMellinGravity}
\begin{split}
    &\int_0^{\infty} \frac{dx_1}{x_1}\,x_1^{2\kb_1} \,\frac{1}{2}\bigg(\Big\vert 1+x_1\theta_{31}^{-1}\Big\vert+\Big\vert 1-x_1\theta_{31}^{-1}\Big\vert\bigg) =\frac{\abs{\theta_{31}}^{2\kb_1}}{2\kb_1(2\kb_1+1)}\,,\\
    &\int_0^{\infty} \frac{dx_1}{x_1}\,x_1^{2\kb_1} \,\frac{1}{2}\bigg(\Big\vert 1+x_1\theta_{31}^{-1}\Big\vert-\Big\vert 1-x_1\theta_{31}^{-1}\Big\vert\bigg) =\sgn(-\theta_{31})\frac{\abs{\theta_{31}}^{2\kb_1}}{2\kb_1(2\kb_1+1)}\, .
\end{split}
\end{align}
We are again, by analogy with the Yang-Mills case, adopting a regularisation $\epsilon>0$ of the weights such that boundary terms at infinity do not contribute. In the case of the Mellin transform of the antisymmetrised mod function $H_1$, the function is $\mathcal{O}(x)$ for small $x$ and $\mathcal{O}(1)$ for large values. Hence the strip of definition is for $-1<\text{Re}(2\kb_1)<0$. In gravity, the helicity is now $-2$ for leg one and the weight has real part $\text{Re}(2\kb_1)=-1$. Hence we can regularise the weight with $2\kb_1\rightarrow 2\kb_1+\epsilon$ as usual. In other words, to invert the Mellin transform we can choose a straight line contour which goes to the right of the pole at minus one and to left of the pole at zero.

However, the case of the symmetrised $H_0$ function in gravity is more subtle. The $H_0$ function is $\mathcal{O}(1)$ for small $x$ but is $\mathcal{O}(x)$ for large $x$ and so the strip of definition does not exist and the integral is strongly divergent regardless of the value of the weights. This fact has been noted elsewhere in the literature e.g. \cite{Kalyanapuram:2020aya, Kalyanapuram:2021bvf} - in gravity the Mellin transform does not in general converge and this had led to the definition of a `regularised Mellin transform' which importantly includes an explicit factor of $e^{-\delta x}$ in its definition \cite{Kalyanapuram:2020aya, Bagchi:2022emh}. We implicitly use such a regularised Mellin transform and note that when we take $\delta \rightarrow 0_+$ we recover the symmetric result of \eqref{eq: hatstairsMellinGravity}. One can also consider taking an inverse Mellin transform of the result by integrating over the weight $2\kb_1$ to recover the original function $H_0$. In this case, a suitable contour does exist, although it is no longer a vertical straight line since there is no such strip of definition. Instead, a contour which snakes to the left of the pole at minus one and then back around to the right of the pole at zero allows us to pick up the correct residues and recover $H_0$.

The three-point MHV celestial gravity twistor amplitude is then given by 
\begin{equation}\label{eq: 3ptMHVceltwistGravity}
\boxed{\begin{split}
    \tilde{\mathcal{M}}^{--+}_3\{s_{\kb_1},s_{\kb_2},s_{k_3}\}=&\frac{\pi}{4} \Big\vert\agl{\lambda_1}{\lambda_2}\agl{\lambda_1}{\mu_3}\agl{\lambda_2}{\mu_3}\Big\vert\,\delta\bigg(\sum_{i}2\kibnbnormal \,+2\bigg)\delta \bigg(\sum_{i} s_{\kibnbsmall}\bigg)\\
    &\quad\times\sgn(-\theta_{31})^{s_{\kb_1}}\frac{\abs{\theta_{31}}^{2\kb_1}}{2\kb_1(2\kb_1+1)}\sgn(-\theta_{32})^{s_{\kb_2}}
    \frac{\abs{\theta_{32}}^{2\kb_2}}{2\kb_2(2\kb_2+1)}\, .
\end{split}}
\end{equation}
Formulae \eqref{eq: hatstairsMellinGravity} together imply that the unsymmetrised integral is given by
\begin{align}
\label{eq: SimpleGravity}
\int_0^{\infty} \frac{dx_1}{x_1}\,x_1^{2\kb_1} \,\Big\vert 1+x_1\theta_{31}^{-1}\Big\vert =2\,\Theta(-\theta_{31}) \frac{\vert\theta_{31}\vert^{2\kb_1}}{2\kb_1(2\kb_1+1)} \, ,
\end{align}
where as usual we drop terms proportional to a delta function in weight space. Similar considerations mentioned in the above cases also apply here regarding the regularisation of \eqref{eq: SimpleGravity}. Finally, we conclude that the unsymmetrised amplitude is
\begin{equation}\label{eq: 3ptMHVceltwistGravityUnsym}
    \boxed{\tilde{\mathcal{M}}^{--+}_3=\frac{\pi}{4} \Big\vert\agl{\lambda_1}{\lambda_2}\agl{\lambda_1}{\mu_3}\agl{\lambda_2}{\mu_3}\Big\vert\,\delta\bigg(\sum_{i}2\kibnbnormal \,+2\bigg) \,
    \Theta(-\theta_{31})\Theta(-\theta_{32})\frac{\abs{\theta_{31}}^{2\kb_1}}{2\kb_1(2\kb_1+1)}
    \frac{\abs{\theta_{32}}^{2\kb_2}}{2\kb_2(2\kb_2+1)}.}
\end{equation}

\textbf{Three-Point} $\overline{MHV}$

The $\tilde{\mathcal{M}}^{++-}_3$ amplitude is obtained by similar manipulations, and is
\begin{equation}\label{eq: 3ptMHVbarceltwistampGravity}
\boxed{\begin{split}
    \tilde{\mathcal{M}}^{++-}_3\{s_{k_1},s_{k_2},s_{\kb_3}\}=&\frac{\pi}{4} \Big\vert\sqr{\lambdat_1}{\lambdat_2}\sqr{\mut_3}{\lambdat_1}\sqr{\mut_3}{\lambdat_2}\Big\vert\,\delta\bigg(\sum_{i}2\kibnbnormal \,+2\bigg)\delta \bigg(\sum_{i} s_{\kibnbsmall}\bigg)\\
    &
    \quad\times\sgn(-\theta_{13})^{s_{k_1}}
    \frac{\abs{\theta_{13}^{-1}}^{2k_1}}{2k_1(2k_1+1)}\;
    \sgn(-\theta_{23})^{s_{k_2}}
    \frac{\abs{\theta_{23}^{-1}}^{2\kb_2}}{2\kb_2(2\kb_2+1)},
\end{split}}
\end{equation}
while the unsymmetrised amplitude is
\begin{equation}\label{eq: 3ptMHVbarceltwistampGravityUnsym}
    \boxed{\tilde{\mathcal{M}}^{++-}_3=\frac{\pi}{4} \Big\vert\sqr{\lambdat_1}{\lambdat_2}\sqr{\mut_3}{\lambdat_1}\sqr{\mut_3}{\lambdat_2}\Big\vert\,\delta\bigg(\sum_{i}2\kibnbnormal \,+2\bigg)\,\Theta(-\theta_{13})\Theta(-\theta_{23})
    \frac{\abs{\theta_{13}^{-1}}^{2k_1}}{2k_1(2k_1+1)}\;
    \frac{\abs{\theta_{23}^{-1}}^{2\kb_2}}{2\kb_2(2\kb_2+1)}.}
\end{equation}

\subsubsection{Four Points: Mellin Transform of a Product of Mod Functions}\label{sec: grav4pts}

\textbf{Four-Point `Alternating' Amplitude}

Turning to four points, the `alternating' four-point twistor amplitude  takes the following form \cite{Arkani-Hamed:2009hub}
\begin{align}\label{eq: 4pttwistampGravity}
\begin{split}
    \tilde{M}^{+-+-}_4=&\frac{(\pi i)^4}{(2\pi)^4}\Big\vert\agl{ \lambda_2}{\mu_1} +\sqr{\lambdat_1}{\mut_2}\Big\vert\,
    \Big\vert\agl{ \lambda_4}{\mu_1} +\sqr{\lambdat_1}{\mut_4}\Big\vert
  \Big \vert\agl{ \lambda_2}{\mu_3} +\sqr{\lambdat_3}{\mut_2}\Big\vert\,\Big\vert\agl{ \lambda_4}{\mu_3} +\sqr{\lambdat_3}{\mut_4}\Big\vert\\
    =&\frac{1}{16}\Big\vert\agl{ \lambda_2}{\mu_1}\agl{ \lambda_4}{\mu_1}\agl{ \lambda_2}{\mu_3}\agl{ \lambda_4}{\mu_3}\Big\vert
     \Big\vert 1 +\theta_{12}^{-1}\Big\vert\,
   \Big \vert 1 +\theta_{14}^{-1}\Big\vert \Big\vert 1 +\theta_{32}^{-1}\Big\vert\,\Big\vert 1 +\theta_{34}^{-1}\Big\vert.
\end{split}
\end{align}
We now perform the Mellin transforms first,  following  the Yang-Mills case for guidance, and only (anti-)symmetrise afterward. In evaluating the integrals below, we will drop boundary terms at zero and infinity by using suitable regularisations as per the discussion above.
We will need the integral
\begin{align}
\label{eq: KfullGravity}
    \cK^{(a,b)}(y, k) :=
    \int_0^\infty \frac{dx}{x} x^{2k} \Big\vert 1+\frac{x}{a} \Big\vert \Big\vert 1 + \frac{x}{y\, b}\Big\vert,
\end{align}
which is the gravity analogue of the $\cJ_2(y)$ and $\cJ_4(y)$ Yang-Mills integrals \eqref{eq: J2integral}. Defining the indefinite integral
\begin{align}
\label{eq: KfullGravityIndef}
    \cL^{(a,b)}(x, y, k) :=
    \int \frac{dx}{x} x^{2k} \Big(1+\frac{x}{a}\Big)\Big( 1 + \frac{x}{y\, b}\Big)
   & = \frac{x^{2k}}{2k}+\frac{x^{2k+1}}{2k+1}\bigg(\frac{1}{yb}+\frac{1}{a}\bigg)+\frac{x^{2k+2}}{2k+2}\frac{1}{yab}
\end{align}
we find 
\begin{align}\label{eq: Gravity_Kull}
\begin{split}
   \cK^{(a,b)}(y,k) =& 2\,\Theta(b)\Theta(-a) \cL^{(a,b)}(\vert a\vert, y, k)
   +2\,\Theta(-b)\Theta(a) \cL^{(a,b)}(\vert b\vert y, y, k)\\
   &+\Theta(-b)\Theta(-a)\,\sgn\bigg(1-y\abs{\frac{b}{a}}\bigg)
   \Big(2\,\cL^{(a,b)}(\vert b\vert y, y, k)-2\, \cL^{(a,b)}(\vert a\vert, y, k)\Big).
\end{split}
\end{align}
We note the similarity with the analogous Yang-Mills result 
\eqref{eq: J2unsymsimpleresult} the only difference being the form of the indefinite integral \eqref{eq: KfullGravityIndef}.

The gravity amplitude is then given by
\begin{align}\label{eq: GravAmp1}
\begin{split}
\tilde{\mathcal{M}}^{+-+-}_4 = &2\pi i\;\delta\bigg(\sum_{i}2\kibnbnormal \,+4\bigg)\;\big\vert\agl{ \lambda_2}{\mu_1}\big\vert\big\vert\agl{ \lambda_4}{\mu_1}\big\vert\big\vert\agl{ \lambda_2}{\mu_3}\big\vert\big\vert\agl{ \lambda_4}{\mu_3}\big\vert\;  
\\ & \qquad \times \int_0^\infty dy\; y^{2k_3-1}\cK^{(\theta_{12},\theta_{32})}(y,\kb_2)\cK^{(\theta_{14},\theta_{34})}(y,\kb_4),
\end{split}\end{align}
where just like at three points we have a shifted dilatation invariance delta function. Dilatation invariance for both Yang-Mills and gravity is linked to the mass dimension of a $n$-point amplitude which must be $-n$ in both cases. However, for gravity not just the momentum but also the coupling, $\kappa\coloneqq\sqrt{32\pi G_N}$, carries mass dimension  minus one. Hence, for an $n$ point amplitude with $n-2$ powers of $\kappa$, the total power of spinors appearing is shifted by $2(n-2)$, and so the dilatation invariance delta function for gravity is shifted by $2(n-2)$, matching our results at three and four points.

Before stating the integrated result of \eqref{eq: GravAmp1}, we first define the general form of the weight dependent coefficients that appear in the gravity case
\begin{align}
\label{eq: GravFactor}
C_{k,l,m}=- \frac{1}{2k(2k+1)}\;\frac{1}{2l(2l+1)}\;\frac{1}{2m(2m+1)}\, .
\end{align}
We also define the following functions of the conformal cross-ratios
\begin{align}\label{eq: GravFactor2}
\begin{split}
W^{\pm}_{12}(\theta) =&\; C_{\kb_2,\kb_2+k_3-\frac{1}{2},\kb_2+k_3+\kb_4}\; \vert\theta\vert^{-2\kb_2}\pm
C_{\kb_2+\frac{1}{2},\kb_2+k_3,\kb_2+k_3+\kb_4+\frac{1}{2}}\; \abs{\theta}^{-2\kb_2-1}  ,
\\
W^{\pm}_{32}(\theta) =&\; \pm C_{\kb_2+\frac{1}{2},k_3-1,k_3+\kb_4-\frac{1}{2}}\;\vert\theta\vert\;  +
C_{\kb_2,k_3-\frac{1}{2},k_3+\kb_4} \, ,
\\
W^{\pm}_{14}(\theta) =&\; \pm C_{\kb_4+\frac{1}{2},k_3+\kb_4,\kb_2+k_3+\kb_4+\frac{1}{2}} \; \vert\theta\vert^{2k_3+2\kb_4+1} +
C_{\kb_4,k_3+\kb_4-\frac{1}{2},\kb_2+k_3+\kb_4}\; \vert\theta\vert^{2k_3+2\kb_4}\,  ,
\\
W^{\pm}_{34}(\theta) =&\; C_{\kb_4,k_3-\frac{1}{2},\kb_2+k_3} \; \vert\theta\vert^{2k_3} \pm
C_{\kb_4+1/2,k_3-1,\kb_2+k_3-\frac{1}{2}} \;\vert\theta\vert^{2k_3-1}\, .
  \end{split}
\end{align}
Finally, computing \eqref{eq: GravAmp1} the four-point `alternating' unsymmetrised celestial twistor gravity amplitude is then
\begin{equation}\label{eq: GravAmpFinal}
\boxed{\begin{split}
\tilde{\mathcal{M}}^{+-+-}_4&=\; \pi i\;\delta\bigg(\sum_{i}2\kibnbnormal \,+4\bigg) \;\big\vert(\agl{ \lambda_2}{\mu_1}\agl{ \lambda_4}{\mu_1}\agl{ \lambda_2}{\mu_3}\agl{ \lambda_4}{\mu_3}\big\vert\;  \vert \theta_{12}\vert ^{2\kb_2}\vert \theta_{14}\vert ^{2k_3+2\kb_4}\vert \theta_{34}\vert ^{-2k_3}
   \\ &   
   \times\bigg[ \,\Theta(-\theta_{12})\Theta(-\theta_{32})\Theta(-\theta_{14})\Theta(-\theta_{34})\,\sgn(1-\vert\theta\vert)
     \Big( W^-_{12}(\theta)  + W^-_{34}(\theta) -W^-_{32}(\theta) - W^-_{14}(\theta) \Big) 
   \\ &\,\,\,\,
  -\Theta(-\theta_{12})\Theta(-\theta_{32})\Theta(\theta_{34})\Theta(-\theta_{14})\;W^+_{34}(\theta)  
   -\Theta(\theta_{12})\Theta(-\theta_{32})\Theta(-\theta_{14})\Theta(-\theta_{34})\;W^+_{12}(\theta)  
   \\
   &\,\,\,\,
   -
  \Theta(-\theta_{12})\Theta(\theta_{32})\Theta(-\theta_{34})\Theta(-\theta_{14})\;W^+_{32}(\theta)  
   -\Theta(-\theta_{12})\Theta(-\theta_{32})\Theta(-\theta_{34})\Theta(\theta_{14})\;W^+_{14}(\theta) \bigg].
  \end{split}}
\end{equation}
From this, we can derive the symmetrised amplitudes by considering (anti-)symmetrisations which satisfy $\sum s_{\kibnbsmall}=0 \mod 2$; the others are zero. Note the remarkably similar structure
to that of the Yang-Mills amplitude \eqref{eq: AmpFinalYM}. All the symmetrised amplitudes in gravity may be captured by an analogous formula to \eqref{eq: AmpFinalYMtwo} in the Yang-Mills case.
Finally, the `separated' amplitude with helicities $(++--)$ is then obtained from this by derivatives with respect to $\vert \theta_{32}\vert$, as discussed in the Yang-Mills case above.

 \section{BCFW Recursion}
\label{sec: bcfw}

The BCFW-type recursion relations of \cite{Britto:2005fq} have  been developed in twistor space in \cite{Arkani-Hamed:2009hub}, and here we will apply them in the celestial context. In this section, we will generate celestial twistor recursion relations and show how these can be used in practice to re-derive our results at four points.

For an $n$-point twistor amplitude, where the deformed legs $i$ and $j$ have helicities $(+,-)$, the recursion relation of \cite{Arkani-Hamed:2009hub} is
\begin{align}\label{eq: BCFW1}
\begin{split}
\tilde A_n =& \sum_{L,R} \sgn(\langle\mu_i\lambda_j\rangle+[\tilde\mu_j\tilde\lambda_i])\int d^2\lambda\;d^2\lamt\;
d^2\mu\;d^2\tilde\mu\;
e^{i[\tilde\lambda\tilde\lambda_i]}e^{i\langle\lambda\lambda_j\rangle}\;
\delta^{'''}\!\big(\agl{\lam}{\mu}+\sqr{\lamt}{\mut}\big)\;
\\&
\qquad\qquad\;
\Bigg(\tilde A_L\Big[(\mu_i,\tilde\lambda_i, +);(\lambda,\tilde\mu, -)\Big]
\tilde A_R\Big[(\lambda_j,\tilde\mu_j, -);(\mu,\tilde\lambda, +)\Big]\\
&
\qquad\qquad\qquad\qquad\qquad+ \tilde A_L\Big[(\mu_i,\tilde\lambda_i, +);(\mu,\tilde\lambda, +)\Big]
\tilde A_R\Big[(\lambda_j,\tilde\mu_j, -);(\lambda,\tilde\mu, -)\Big]
\Bigg) ,
\end{split}
\end{align}
where we are writing the twistors in terms of spinors as $W=(\mu,\tilde\lambda),Z=(\lambda,\tilde\mu)$ and any external leg variables that are not shifted in the amplitudes are not written explicitly. Note we work in an ambidextrous basis which follows the helicity of the particles, that is we use $W_i,Z_j$ for the $(+,-)$ deformed legs. We also sum over the helicity of the glued leg which gives the two terms in \eqref{eq: BCFW1} with different twistor variables - $(\mu,\tilde\lambda)$ for plus helicity and $(\lambda,\tilde\mu)$ for minus helicity.

We will focus on the case of Yang-Mills in the following. To obtain the recursion relation for the Mellin-transformed twistor amplitudes, we  perform Mellin transforms on all the external variables, and inverse Mellin transforms on all the variables inside the amplitudes on the right-hand side of the above equation. Using the scaling properties of the amplitudes, the Mellin and inverse Mellin transforms cancel for all the external variables on the right-hand side except those labelled by $i$ and $j$. We also have the inverse Mellin transform over $K,\bar K$ for the glued legs of each left and right amplitude, we can choose a single pair of weights $K, \bar K$ since the integral transform acts linearly on the BCFW expression. We thus obtain the following BCFW relation
\begin{align}\label{eq: BCFW2}
\begin{split}
\tilde \cA_n =& \sum_{L,R} \int d^2\lambda\;d^2\tilde\lambda\;
d^2\mu\;d^2\tilde\mu \int\;\frac{dK\;d\bar K}{(2\pi i)^2} \sum_{s_K, s_{\bar K} \in \{0,1\}}\;
e^{i[\tilde\lambda\tilde\lambda_i]}e^{i\langle\lambda\lambda_j\rangle}\;
\delta^{'''}\Big(\agl{\lam}{\mu}+\sqr{\lamt}{\mut}\Big)\; 
\\&
\;
\int\;\frac{d\kap_i\;d\kapb_j}{(2\pi i)^2} \!\!\sum_{s_{\kap_i}, s_{\kapb_j} \in \{0,1\}}\!\!\!\!\!\! I(i,j)\Bigg(
\;\tilde \cA_L\Big[(\mu_i,\tilde\lambda_i,\kap_i);
(\lambda,\tilde\mu,\bar K)\Big]
\tilde \cA_R\Big[(\lambda_j,\tilde\mu_j,\kapb_j);(\mu,\tilde\lambda, K)\Big]
\\
&
\qquad\qquad\qquad\qquad+ \;\tilde \cA_L\Big[(\mu_i,\tilde\lambda_i,\kap_i);(\mu,\tilde\lambda,  K)\Big]
\tilde \cA_R\Big[(\lambda_j,\tilde\mu_j,\kapb_j);(\lambda,\tilde\mu, \bar K)\Big]
\Bigg) ,
\end{split}
\end{align}
where for notational simplicity we have left implicit the discrete weights $s_{K}$ etc. which follow suit with the continuous ones. We have also defined the Mellin integrals $I(i,j)$ for legs $i$ and $j$ which we evaluate first using the scale covariance of the celestial amplitudes
\begin{align}\label{eq: Mellinoverij}
\begin{split}
 \cI(i,j) \coloneqq& \int_{\RR_*} \frac{dt_i}{\abs{t_i}}\; \abs{t_i}^{2(k_i-\kap_i)}\sgn(t_i)^{s_{k_i}-s_{\kap_i}}\; \int_{\RR_*} \frac{d\tb_j}{\abs{\tb_j}} \; \abs{\tb}_j^{2(\kb_j-\kapb_j)}\sgn(\tb_j)^{s_{\kb_j}-s_{\kapb_j}}\\
 &\qquad\qquad\qquad\qquad\qquad \sgn    
 \big(t_i\langle\mu_i\lambda_j\rangle+\tb_j[\tilde\lambda_i\tilde\mu_j]\big)
 \\=&\;2\pi i\, \delta(2(k_i+\kb_j-\kap_i-\kapb_j))\, \tilde \delta(s_{k_i}+s_{\kb_j}-s_{\kap_i}-s_{\kapb_j}+1)\, \sgn(\agl{\mu_i}{\lam_j})\abs{\theta_{ij}}^{2(\kb_j-\kapb_j)}\\
 &\qquad\qquad\sgn(-\theta_{ij})^{s_{\kb_j}-s_{\kapb_j}}\frac{1}{2(\kb_j-\kapb_j)+(-1)^{s_{\kb_j}-s_{\kapb_j}}\epsilon}\\
 =&:2\pi i\, \delta(2(k_i+\kb_j-\kap_i-\kapb_j))\, \tilde \delta(s_{k_i}+s_{\kb_j}-s_{\kap_i}-s_{\kapb_j}+1)\\
 &\qquad\qquad\qquad\,\sgn(\agl{\mu_i}{\lam_j})\, \cG(\kb_j-\kapb_j, s_{\kb_j}-s_{\kapb_j}, \theta_{ij}).
\end{split}
\end{align}
which we have evaluated in terms of the function $\cG(k, s, x)$ which is just defined to be the Mellin transform over $\RR_+$ of  $G_{s}(x)$ appearing in \eqref{eq: x1integralevenodd}. Note again that, for readability, we are using the notation $\tilde \delta$ for a Kronecker delta evaluated mod 2.

We continue performing all the Mellin transforms such that the BCFW relation is purely in terms of projective integrals. We first write $\delta^{'''}(x)\sim\int ds\; s^3\; e^{isx}$ and then by rescaling $\mu,\mut$ by $1/s$ we perform the integral over $s$ which generates the delta functions below in \eqref{eq: BCFWweightintegral}. In addition, we also break up the BCFW measure as $d^2\lam=du \abs{u} \agl{\lam}{d\lam}$, $d^2\mu=dt \abs{t} \agl{\mu}{d\mu}$ etc., and perform the integrals over the scales $u, \ub, t, \tb$ which we can extract from the left and right celestial subamplitudes. Now recall that we are using ambidextrous variables for the glued legs and so we always have the relations $\hb=-k$ or $h=-\kb$ which we used at three and four-point and are specific to gluon amplitudes. Therefore we pull out the following factors in the integrals over $u, \ub, t, \tb$ \footnote{Performing these integrals requires analytic continuation. That is we relax the conditions from the delta function and compute the integrals in the region of $K,\bar K$ space where they converge. We then analytically continue the final expression to the region supported by the delta functions.}
\begin{equation}\label{eq: BCFWweightintegral}
\begin{split}
    \cI (\bar K,K)\coloneqq &\;2\pi i\,\delta(2K+2\bar K)\, \tilde\delta(s_K+s_{\bar K}+1)\int_{\RR_*}\! du\,d\ub\,dt\,d\tb\, \abs{u}^{1+2\bar K} \sgn(u)^{-s_{\bar K}}\abs{\ub}^{1+ 2K} \sgn(\ub)^{-s_{K}}\\
    &\qquad\times\abs{\tb}^{1-2\bar K} \sgn(\tb)^{-s_{\bar K}}\abs{t}^{1- 2K} \sgn(t)^{-s_{K}}\,e^{i u\agl{\lam}{\lam_j}}\,e^{i\ub\sqr{\lamt}{\lamt_i}}\,e^{i\, tu\agl{\lam}{\mu}}\,e^{i\,\tb\ub\sqr{\lamt}{\mut}}.
\end{split}
\end{equation}
These integrals are of an identical form to \eqref{eq: Iintegraltwistorlight}, encountered when relating a half-Fourier transform to a light transform. From \eqref{eq: Iintegraltwistorlight} we define
\begin{equation}\label{eq: Ndef}
     \cN(a,b,x):=\int_{\RR_*}\, dt \,\abs{t}^{a-1} \sgn(t)^{b} e^{itx}=2\pi\, i^{-b} \frac{\Gamma(a)}{\Gamma(\frac{a+b+1}{2})\Gamma(\frac{1-a-b}{2})}\abs{x}^{-a}\sgn(x)^{b},
\end{equation}
then we have
\begin{equation}
\begin{split}
    \cI (\bar K,K)=&\;2\pi i\,\delta(2K+2\bar K)\, \tilde\delta(s_K+s_{\bar K}+1)\cN(2-2K,-s_K,\agl{\lam}{\mu})\cN(2-2\bar K,-s_{\bar K},\sqr{\lamt}{\mut})\\
    &\times\int_{\RR_*}\! du\,d\ub\, \abs{u}^{2K+2\bar K-1} \sgn(u)^{-s_K-s_{\bar K}}\abs{\ub}^{2\bar K+ 2K-1} \sgn(\ub)^{-s_{K}-s_{\bar K}}\,
    e^{i u\agl{\lam}{\lam_j}}\,e^{i\ub\sqr{\lamt}{\lamt_i}}.
\end{split}
\end{equation}

Now we perform the final integrals and simplify on the constraint of the delta functions,
\begin{equation}
\begin{split}
    \cI (\bar K,K)=&\;(2\pi i)^3\,\delta(2K+2\bar K)\, \tilde\delta(s_K+s_{\bar K}+1)\,\cN(2-2K,-s_K,\agl{\lam}{\mu})\\
    &\qquad\times\cN(2-2\bar K,-s_{\bar K},\sqr{\lamt}{\mut})\, \sgn(\agl{\lam}{\lam_j}\sqr{\lamt}{\lamt_i})\\
    =&:(2\pi i)^3\,\delta(2K+2\bar K)\, \tilde\delta(s_K+s_{\bar K}+1)\, \sgn(\agl{\lam}{\lam_j}\sqr{\lamt}{\lamt_i})\,\cR(2\bar K,2K, s_{\bar K},s_K).
\end{split}
\end{equation}
Thus we obtain a BCFW relation given by
\begin{align}\label{eq: BCFW3}
\begin{split}
\tilde \cA_n =& \;
(2\pi i)^4\sgn(\agl{\mu_i}{\lam_j})\sum_{L,R}\int\frac{dK\;d\bar K}{(2\pi i)^2}\!\! \sum_{s_K, s_{\bar K} \in \{0,1\}}\!\!\!\!\delta(2K+2\bar K)\, \tilde\delta(s_K+s_{\bar K}+1)\\
&\times\int\!\frac{d\kap_i\;d\kapb_j}{(2\pi i)^2} \!\!\!\!\sum_{s_{\kap_i}, s_{\kapb_j} \in \{0,1\}}\!\!\!\!\!\!\delta(2(k_i\!+\!\kb_j\!-\!\kap_i\!-\!\kapb_j))\, \tilde \delta(s_{k_i}\!+\!s_{\kb_j}\!-\!s_{\kap_i}\!-\!s_{\kapb_j}\!+1)\\
&\times\cG(\kb_j-\kapb_j, s_{\kb_j}-s_{\kapb_j}, \theta_{ij})\int\, \agl{\lam}{d\lam}\, \sqr{\lamt}{d\lamt}\, \agl{\mu}{\,d\mu}\, \sqr{\mut}{\,d\mut} \, \sgn(\agl{\lam}{\lam_j}\sqr{\lamt}{\lamt_i})\\
&\times
\Bigg(\cR(2\bar K,2K, s_{\bar K}, s_K)
\tilde \cA_{L}\Big[(\mu_i,\tilde\lambda_i,\kap_i);
(\lambda,\tilde\mu,\bar K)\Big]
\tilde \cA_{R}\Big[(\lambda_j,\tilde\mu_j,\kapb_j);(\mu,\tilde\lambda,K)\Big]
\\
& 
+ \cR( 2K, 2\bar K, s_K, s_{\bar K})\tilde \cA_{L}\Big[(\mu_i,\tilde\lambda_i, \kap_i);(\mu,\tilde\lambda,K)\Big]
\tilde \cA_{R}\Big[(\lambda_j,\tilde\mu_j,\kapb_j);(\lambda,\tilde\mu,\bar K)\Big]
\Bigg).
\end{split}
\end{align}
Having done all the scale integrals we now have a projective form of the BCFW recursion relations in celestial twistor space. We can actually simplify the above expression greatly since the integrals which glue the weights can be performed against the delta functions. The left and right Yang-Mills subamplitudes come with dilatation invariance delta functions and also Kronecker deltas enforcing overall evenness. For example, for the first product of amplitudes we have the delta functions
\begin{equation}\label{eq: subampdeltas}
\begin{split}
    \delta\bigg(2\bigg(\kap_i+\sum_{m\in L,m\not=i}\!\!\!\kbnbnormal{m}+\bar K\bigg)\bigg)\,\tilde\delta\bigg(s_{\kap_i}+\sum_{m\in L,m\not=i}\!\!s_{\kbnbsmall{m}}+s_{\bar K}\bigg)\\
   \times \delta\bigg(2\bigg(\kapb_j+\sum_{m\in R,m\not=j}\!\!\!\kbnbnormal{m}+ K\bigg)\bigg)\,\tilde\delta\bigg(s_{\kapb_j}+\sum_{m\in R,m\not=j}\!\!s_{\kbnbsmall{m}}+s_{K}\bigg),
\end{split}
\end{equation}
while for the second term we simply swap $K$ and $\bar K$. We now use the delta functions in \eqref{eq: subampdeltas} to localise the sums and integrals over $\kap_i,\kapb_j, s_{\kap_i}, s_{\kapb_j}$. In addition we have the four external delta functions in \eqref{eq: BCFW3} and plugging in the localised values of $\kap_i,\kapb_j, s_{\kap_i}, s_{\kapb_j}$ these give the overall dilatation invariance delta functions of the amplitude $\tilde \cA_n$
\begin{equation}\label{eq: externadeltas}
\begin{split}
    &\delta(2K+2\bar K)\, \tilde\delta(s_K+s_{\bar K}+1)\delta\bigg(\sum_{a=1}^{n} 2\kbnbnormal{a}+2\!K+2\bar K\bigg)\, \tilde \delta\bigg(\sum_{a=1}^{n} s_{\kbnbsmall{a}}+\!s_{K}\!+\!s_{\bar K}\!+1\bigg)\\
    =&\;\delta\bigg(\sum_{a=1}^{n} 2\kbnbnormal{a}\bigg)\, \tilde\delta\bigg(\sum_{a=1}^{n} s_{\kbnbsmall{a}}\bigg)\delta\bigg(2\!K+2\bar K\bigg)\, \tilde \delta\bigg(\!s_{K}\!+\!s_{\bar K}\!+1\bigg).
\end{split}
\end{equation}
Hence, we have the following BCFW recursion relation
\begin{align}\label{eq: BCFW4}
\begin{split}
\tilde \cA_n =& 
(2\pi i)^2\sgn(\agl{\mu_i}{\lam_j}) \delta\bigg(\sum_a 2\kbnbnormal{a}\bigg)\,\tilde\delta\bigg(\sum_a s_{\kbnbsmall{a}}\bigg)\sum_{L,R}\int \frac{dK\;d\bar K}{(2\pi i)^2} \!\!\!\sum_{s_{K}, s_{\bar K} \in \{0,1\}}\\
&\times\delta\bigg(2\!K+2\bar K\bigg)\, \tilde \delta\bigg(\!s_{K}\!+\!s_{\bar K}\!+1\bigg)\,\cG(\kb_j+k_R+\bar K, s_{\kb_j}+s_{k_R}+s_{\bar K}, \theta_{ij})\\
&\times\int\, \agl{\lam}{d\lam}\, \sqr{\lamt}{d\lamt}\, \agl{\mu}{\,d\mu}\, \sqr{\mut}{\,d\mut} \, \sgn(\agl{\lam}{\lam_j}\sqr{\lamt}{\lamt_i})\,\cR( 2K, 2\bar K, s_K, s_{\bar K})\\
&\times
\Bigg(
\tilde \cA_{L}\Big[(\mu_i,\tilde\lambda_i,-k_L-K);
(\lambda,\tilde\mu,K)\Big]
\tilde \cA_{R}\Big[(\lambda_j,\tilde\mu_j,-k_R-\bar K);(\mu,\tilde\lambda,\bar K)\Big]
\\
& 
\quad + \tilde \cA_{L}\Big[(\mu_i,\tilde\lambda_i, -k_L-K);(\mu,\tilde\lambda, K)\Big]
\tilde \cA_{R}\Big[(\lambda_j,\tilde\mu_j,-k_R-\bar K);(\lambda,\tilde\mu,\bar K)\Big]
\Bigg),
\end{split}
\end{align}
where in the above we have sub-amplitudes that are stripped of their dilatation invariance delta functions and also we have redefined summation and integration variables $(K, \bar K, s_K, s_{\bar K})$ for each term to pull out common factors. We would also like to reiterate that all the weights in the above are for the $\mu, \mut$ variables - the weights for the $\lam, \lamt$ being found using the fact the helicity is $\pm 1$ which implies the relations $h=-\kb$ and $\hb=-k$. Although not written explictly above, the localised values of the discrete weights follow that of the continuous ones.

We now perform the inverse Mellin transform over $K$ and $s_{K}$ using the delta functions and arrive at the following BCFW recursion relation for celestial twistor Yang-Mills amplitudes
\begin{equation}\label{eq: BCFW5}
\boxed{\begin{split}
\tilde \cA_n =&\; 
2\pi i\,\sgn(\agl{\mu_i}{\lam_j}) \delta\bigg(\sum_a 2\kbnbnormal{a}\bigg)\,\tilde\delta\bigg(\sum_a s_{\kbnbsmall{a}}\bigg)\sum_{L,R}\int \frac{d \bar K}{2\pi i} \!\!\sum_{s_{\bar K} \in \{0,1\}}\\
&\times\int\, \agl{\lam}{d\lam}\, \sqr{\lamt}{d\lamt}\, \agl{\mu}{\,d\mu}\, \sqr{\mut}{\,d\mut} \, \sgn(\agl{\lam}{\lam_j}\sqr{\lamt}{\lamt_i})\\
&\qquad\times\cR\bigg( -2\bar K, 2\bar K, -s_K-1, s_{K}\bigg)\, \cG(\kb_j+k_R+\bar K, s_{\kb_j}+s_{k_R}+s_{\bar K}, \theta_{ij})\\
&\times
\Bigg(
\tilde \cA_{L}\Big[(\mu_i,\tilde\lambda_i,-k_L+\bar K);
(\lambda,\tilde\mu,-\bar K, - \hat s_{\bar K})\Big]
\tilde \cA_{R}\Big[(\lambda_j,\tilde\mu_j,-k_R-\bar K);(\mu,\tilde\lambda,\bar K,  s_{\bar K})\Big]
\\
& 
+ \tilde \cA_{L}\Big[(\mu_i,\tilde\lambda_i, -k_L+\bar K);(\mu,\tilde\lambda, -\bar K, -\hat s_{\bar K})\Big]
\tilde \cA_{R}\Big[(\lambda_j,\tilde\mu_j,-k_R-\bar K);(\lambda,\tilde\mu,\bar K, s_{\bar K})\Big]
\Bigg),
\end{split}}
\end{equation}
where we have again left the discrete weights implicit in the above expression but they follow the same pattern as the continuous weights with one exception - in the left sub-amplitudes the weight $s_{K}$ is localised to the shifted value $-\hat s_{\bar K} := -s_{\bar K}-1$.

We are left with a contour integral over $\bar K$ and a sum over the weights $s_{\bar K}$, together with four projective integrals over $\lam, \lamt, \mu, \mut$. Carrying these out will generate the full amplitude - we demonstrate this in an example below by computing the four-point amplitude from a product of two three-point amplitudes, after which we will make some general comments valid at $n$-points. 

In fact we can carry out the integrals over $\mu, \mut$ above immediately, since the pre-factor $\cR( -2\bar K, 2 \bar K, -s_{\bar K}-1, s_{\bar K})$ is given by
\begin{equation}
\begin{split}
    \cR& = 4\pi^2i\frac{\Gamma(2+2 \bar K)\Gamma(2-2\bar K)\abs{\agl{\lam}{\mu}}^{2\bar K-2}\sgn(\agl{\lam}{\mu})^{s_{\bar K}}\abs{\sqr{\lamt}{\mut}}^{-2\bar K-2}\sgn(\sqr{\lamt}{\mut})^{s_{\bar K}+1}}{\Gamma(\frac{3}{2}-\bar K-\frac{s_{\bar K}}{2})\Gamma(\bar K+\frac{s_{\bar K}}{2}-\frac{1}{2})\Gamma(2+\bar K+\frac{s_{\bar K}}{2})\Gamma(-\bar K-\frac{s_{\bar K}}{2}-1)},
\end{split}
\end{equation}
and contains exactly the correct factors of $\abs{\agl{\lam}{\mu}}$ etc. and the correct normalisation such that the $\mu$ (respectively $\mut$) integral is a light transform \eqref{eq: lightcelstate} (respectively dual light transform \eqref{eq: duallightcelstate}) of the glued legs. The light and dual light transforms are self inverse so we conclude that after the $\mu,\mut$ integrals, the glued legs are expressed with variables $\lam,\lamt$ and are only half-Mellin transformed without any light transforms acting at all. We can hence write down an alternative form of celestial BCFW recursion which takes as input amplitudes with \emph{all but one leg light transformed}, and glues the un-transformed legs resulting in an amplitude with all legs light transformed. From \eqref{eq: BCFW5} the alternative BCFW recursion relation is
\begin{equation}\label{eq: BCFW6alt}
\boxed{\begin{split}
\tilde \cA_n =& \;
2\pi i\,\sgn(\agl{\mu_i}{\lam_j}) \delta\bigg(\sum_a 2\kbnbnormal{a}\bigg)\,\tilde\delta\bigg(\sum_a s_{\kbnbsmall{a}}\bigg)\sum_{L,R}\int \frac{d \bar K}{2\pi i} \!\!\sum_{s_{\bar K} \in \{0,1\}}\\
&\times\cG(\kb_j+k_R+\bar K, s_{\kb_j}+s_{k_R}+s_{\bar K}, \theta_{ij})\int\, \agl{\lam}{d\lam}\, \sqr{\lamt}{d\lamt} \, \sgn(\agl{\lam}{\lam_j}\sqr{\lamt}{\lamt_i})\\
&\times
\Bigg(
\tilde \cA_{L}\Big[(\mu_i,\tilde\lambda_i,-k_L+\bar K, +);
(\lambda,\lamt,\hb_L,s_{\hb_L}, -)\Big]
\tilde \cA_{R}\Big[(\lambda_j,\tilde\mu_j,-k_R-\bar K, -);(\lam,\tilde\lambda,h_R,s_{h_R}, +)\Big]
\\
+& \tilde \cA_{L}\Big[(\mu_i,\tilde\lambda_i, -k_L+ \bar K, +);(\lam,\tilde\lambda, h_L,s_{h_L}, +)\Big]
\tilde \cA_{R}\Big[(\lambda_j,\tilde\mu_j,-k_R- \bar K, -);(\lambda,\lamt,\hb_R,s_{\hb_R}, -)\Big]
\Bigg),
\end{split}}
\end{equation}
where the weights of the glued legs are
\begin{equation}
\begin{split}
    \hb_L=1+ \bar K=h_L \quad \hb_R=1- \bar K=h_R\,,\\
    s_{\hb_L}=-s_{\bar K}-1=s_{h_L} \quad s_{\hb_R}=s_{\bar K}=s_{h_R}.
\end{split}
\end{equation}
We conclude that the most economical BCFW recursion relation for celestial twistor amplitudes happens to involve gluing sub-amplitudes with one leg in the Mellin basis and the rest in the light transformed basis. This form of the recursion may not necessarily be the most useful at high multiplicity since it requires recursion on an additional set of amplitudes beyond those we are immediately interested in. Nevertheless, the recursion relation \eqref{eq: BCFW6alt} proves very useful for demonstrating the four-point BCFW recursion relation as we will see below, since the three-point sub-amplitudes can be easily computed directly.

The BCFW recursion relations for gravity amplitudes take a very similar form. The only difference to the BCFW expression comes from the changed dilatation invariance delta function which, as we saw at three and four points, is shifted by $2(n-2)$ due to the mass dimension carried by the powers of the coupling $\kappa$ in an $n$ point gravity amplitude. Hence, in the gravity case the delta functions of the sub-amplitudes \eqref{eq: subampdeltas} are shifted and so the localised values of the continuous weights of legs $i$ and $j$ on the left and right subamplitudes are shifted by $2(n_{L,R}-2)$ also, while the localised values for the discrete weights are the same as for the Yang-Mills case.

\textbf{Four-Point BCFW}

To see how the recursion relation works in an explicit example, consider the four-point amplitude $A(1^+, 2^-, 3^+, 4^-)$ with a BCFW shift of legs $(1^+, 4^-)$. In this case the amplitude pairs that appear in the BCFW recursion are
\begin{equation}
    A^{\overline{MHV}}_L(1^+, 2^-, P^+)A^{MHV}_R(P^-, 3^+, 4^-) + A^{MHV}_L(1^+, 2^-, P^-)A^{\overline{MHV}}_R(P^+, 3^+, 4^-).
\end{equation}
As usual the second pair of amplitudes do not contribute since they vanish for generic kinematics and so we consider just the first pair.
The relevant three-point $MHV$ and $\overline{MHV}$ celestial amplitudes with two legs light transformed are derived in Appendix \ref{app: almostceltwistamps} and for the case at hand are given by
\begin{equation}
\begin{split}
    &\tilde \cA_L^{\overline{MHV}}\Big[(\mu_1, \lamt_1, -\kb_2+\bar K, +) , (\lam_2, \mut_2,\kb_2,-), (\lam, \lamt,1+ \bar K,-s_{\bar K}-1, +)\Big]\\
    =&\;\delta(\agl{\lam}{\lam_2})\,\sgn(\sqr{\lamt_1}{\lamt}\sqr{\lamt_1}{\mut_2})\,\sgn(-\theta_{12})^{-s_{\kb_2}+s_{\bar K}+1} \frac{\abs{\theta_{12}^{-1}}^{-2\kb_2+2\bar K}}{-2\kb_2+2\bar K+(-1)^{-s_{\kb_2}+s_{\bar K}+1}\epsilon}\\
    &\qquad\qquad \times\cN\big(2\bar K,-s_{\bar K},\sqr{\lamt}{\mut_2}\big),
\end{split}
\end{equation}
and
\begin{equation}
\begin{split}
    &\tilde \cA_R^{MHV}\Big[(\mu_3, \lamt_3, k_3, +) , (\lam_4, \mut_4,-k_3-\bar K,-), (\lam, \lamt,1- \bar K,s_{\bar K}, -)\Big]\\
    =&\;\delta(\sqr{\lamt}{\lamt_3})\,\sgn(\agl{\lam_4}{\lam}\agl{\mu_3}{\lam_4})\,\sgn(-\theta_{34})^{-s_{k_3}-s_{\bar K}} \frac{\abs{\theta_{34}}^{-2k_3-2\bar K}}{-2k_3- 2\bar K+(-1)^{-s_{k_3}-s_{\bar K}}\epsilon}\\
    &\qquad \qquad\times\cN\big(-2\bar K,s_{\bar K}+1,\agl{\mu_3}{\lam}\big).
\end{split}
\end{equation}
where we have stripped off the dilatation invariance delta functions since they have already been used to localise the weights.  

We now consider the product of these two amplitudes and perform the integrals over $\lam, \lamt$. Using the delta functions which come from three-point momentum conservation, we identify collinear pairs $\lam=\lam_2$ and $\lamt=\lamt_3$\footnote{The other BCFW pair gives vanishing contribution for four particle kinematics. We can see this directly since it has delta functions which give collinear pairs $\lam=\lam_4$ and $\lamt=\lamt_1$ which forces the vanishing of the  $\sgn(\agl{\lam}{\lam_4})\,\sgn(\sqr{\lamt}{\lamt_1})$ factors in the BCFW relation.}. The four-point amplitude is then given by
\begin{equation}\label{eq: BCFW4pt}
\begin{split}
\tilde \cA_4^{+-+-} =& \;
2\pi i\,\sgn(\agl{\mu_1}{\lam_4}) \delta\bigg(\sum_a 2\kbnbnormal{a}\bigg)\,\tilde\delta\bigg(\sum_a s_{\kbnbsmall{a}}\bigg)\int \frac{d \bar K}{2\pi i} \!\!\sum_{s_{\bar K} \in \{0,1\}}\\
&\times \cG(k_3+\kb_4+\bar K, s_{k_3}+s_{\kb_4}+s_{\bar K}, \theta_{14}) \, \sgn(\agl{\lam_2}{\lam_4})\,\sgn(\sqr{\lamt_3}{\lamt_1}) \tilde \cA_L^{\overline{MHV}}\, \tilde \cA_R^{MHV}\\
=&\;2\pi i\,\sgn(\sqr{\lamt_1}{\mut_2}\agl{\mu_3}{\lam_4}\agl{\mu_1}{\lam_4}) \delta\bigg(\sum_a 2\kbnbnormal{a}\bigg)\,\tilde\delta\bigg(\sum_a s_{\kbnbsmall{a}}\bigg)\int \frac{d \bar K}{2\pi i} \!\!\sum_{s_{\bar K} \in \{0,1\}}\\
&\times\cG(k_3+\kb_4+\bar K, s_{k_3}+s_{\kb_4}+s_{\bar K}, \theta_{14})\,\sgn(-\theta_{12})^{-s_{\kb_2}+s_{\bar K}+1}\sgn(-\theta_{34})^{-s_{k_3}-s_{\bar K}}\\
&\quad\times \, \frac{\abs{\theta_{12}^{-1}}^{-2\kb_2+2\bar K}}{-2\kb_2+2\bar K+(-1)^{-s_{\kb_2}+s_{\bar K}+1}\epsilon}\, \frac{\abs{\theta_{34}}^{-2k_3-2\bar K}}{-2k_3-2\bar K+(-1)^{-s_{k_3}-s_{\bar K}}\epsilon}\\
&\quad\quad \times \, \cN\big(2\bar K,-s_{\bar K},\sqr{\lamt_3}{\mut_2}\big)\,\cN\big(-2\bar K,s_{\bar K}+1,\agl{\mu_3}{\lam_2}\big).
\end{split}
\end{equation}
Now, using the Legendre Duplication Formula for gamma functions \eqref{eq: legendreduplication} we find that 
\begin{equation}
\begin{split}
    &\cN\big(2\bar K,-s_{\bar K},\sqr{\lamt_3}{\mut_2}\big)\,\cN\big(-2\bar K,s_{\bar K}+1,\agl{\mu_3}{\lam_2}\big)\\
    =& \;2\pi i\frac{1}{2\bar K+(-1)^{s_{\bar K}}\epsilon} \abs{\sqr{\lamt_3}{\mut_2}}^{-2\bar K} \sgn(\sqr{\lamt_3}{\mut_2})^{-s_{\bar K}} \abs{\agl{\mu_3}{\lam_2}}^{2\bar K} \sgn(\agl{\mu_3}{\lam_2})^{s_{\bar K}+1}\\
    =&\;2\pi i\frac{1}{2\bar K+(-1)^{s_{\bar K}}\epsilon} \sgn(\agl{\mu_3}{\lam_2})\, \sgn(-\theta_{32})^{s_{\bar K}}\abs{\theta_{32}}^{2\bar K},
\end{split}
\end{equation}
where we have also included a regularisation of the pole in $2\bar K$. As usual, when we integrate over $\bar K$, this regularisation corresponds to a contour along the imaginary axis which avoids the pole either to the left or right. Using the above and the definition of the $\cG$ function in \eqref{eq: Mellinoverij}, the four-point amplitude is given by
\begin{equation}\label{eq: BCFW4pt2}
\boxed{\begin{split}
\tilde \cA_4^{+-+-}
=&\;4\pi^2\,\sgn(\agl{\mu_1}{\lam_2}\agl{\mu_3}{\lam_2}\agl{\mu_3}{\lam_4}\agl{\mu_1}{\lam_4}) \delta\bigg(\sum_a 2\kbnbnormal{a}\bigg)\,\tilde\delta\bigg(\sum_a s_{\kbnbsmall{a}}\bigg)\int \frac{d \bar K}{2\pi i} \!\!\sum_{s_{\bar K} \in \{0,1\}}\\
&\times \sgn(\theta_{12})^{s_{\kb_2}+s_{\bar K}}\sgn(\theta_{32})^{s_{\bar K}}\sgn(\theta_{34})^{s_{k_3}+s_{\bar K}}\sgn(\theta_{14})^{s_{k_3}+s_{\kb_4}+s_{\bar K}}\\
&\quad\times \, \frac{\abs{\theta_{12}}^{2\kb_2}\abs{\theta_{14}}^{2k_3+2\kb_4}
      \abs{\theta_{34}}^{-2k_3}}{2k_3+2\kb_4+2\bar K+(-1)^{s_{k_3}+s_{\kb_4}+s_{\bar K}}\epsilon}\, \frac{\abs{\theta}^{-2\bar K}}{2\bar K+(-1)^{s_{\bar K}}\epsilon}\\
&\quad\quad \times \frac{1}{2k_3+2\bar K+(-1)^{s_{k_3}+s_{\bar K}+1}\epsilon}\, \frac{1}{-2\kb_2+2\bar K+(-1)^{s_{\kb_2}+s_{\bar K}+1}\epsilon } .
\end{split}}
\end{equation}
The sum over $s_{\bar K}$ gives two terms which exactly correspond to the decomposition \eqref{eq: decompFullSymAmp} which we previously found. The two terms are the unique $\cI(r_{12}, r_{32}, r_{34}, r_{14})$ with the correct even/odd symmetry and, as already noticed, these terms have the opposite $\epsilon$ prescriptions and sgn function factors.

The contour integral over $\bar K$ has contour along the pure imaginary axis and the expression has poles at $\bar K = (0, \kb_2, -k_3-\kb_4, -k_3)$ with varying $\epsilon$ prescription away from the imaginary axis. It is straightforward to check that the residues on these poles generate four terms involving the functions $V_{32}, V_{12}, V_{14}, V_{34}$ given in \eqref{eq: Hfunction} which make up all four-point symmetrised amplitudes. With regard to the choice of contours, the $\bar K$ dependence in the above is $\vert\theta\vert^{-\bar 2K}$, where $\theta$ is the ratio of cross-ratios. In order to avoid contributions at infinity the contour must be chosen to close to the right with large positive $\bar K$ for $\vert\theta\vert>1$, while for $\vert\theta\vert<1$ it must close to the left with large negative $\bar K$. So for each case we pick up a subset of the poles, namely those with $\epsilon$ prescription such that the poles lie in either the left or right hand half plane. In this way the various functions $V_{32}, V_{12}, V_{14}, V_{34}$ are assigned to the channels $\vert\theta\vert>1$ or $\vert\theta\vert<1$.

For example, the fully symmetrised amplitude has all the $s_{\kbnbsmall{a}}$ equal to zero and so we have two terms
\begin{equation}\label{eq: BCFW4ptSym}
\begin{split}
\tilde \cA_4^{+-+-}\{0,0,0,0\}&=4\pi^2\,\sgn(\agl{\mu_1}{\lam_2}\agl{\mu_3}{\lam_2}\agl{\mu_3}{\lam_4}\agl{\mu_1}{\lam_4}) \delta\bigg(\sum_a 2\kbnbnormal{a}\bigg) \\
&\times \abs{\theta_{12}}^{2\kb_2}\abs{\theta_{14}}^{2k_3+2\kb_4}
      \abs{\theta_{34}}^{-2k_3} [I(0,0,0,0)+\sgn(\theta_{12}\theta_{32}\theta_{34}\theta_{14})\,I(1,1,1,1)]
\end{split}
\end{equation}
where, focusing on the contour integral $I(0,0,0,0)$ we have
\begin{equation}
\begin{split}
    I(0,0,0,0)&:=\int \frac{d \bar K}{2\pi i}\, \frac{\abs{\theta}^{-2\bar K}}{2k_3+2\kb_4+2\bar K+\epsilon}\, \frac{1}{2\bar K+\epsilon} \frac{1}{2k_3+2\bar K-\epsilon}\, \frac{1}{-2\kb_2+2\bar K-\epsilon }
\end{split}
\end{equation}
which gives four terms, one for each residue, with two terms appearing in each channel as dictated by the $\pm \epsilon$ prescription. This exactly matches the four terms \eqref{eq: I(OOO0)} appearing in the function $\cI(0,0,0,0)$ found in the four-point amplitude. The $I(1,1,1,1)$ contour integral has opposite $\pm\epsilon$ prescription and takes a similar form which matches the terms in $\cI(1,1,1,1)$ in \eqref{eq: I(1111)}.

Beyond four points we will encounter similar steps and the BCFW recursion relations will allow us to iteratively build amplitudes using previous results as the inputs. Given the amplitudes with up to $n$ particles, we can build the $n+1$-point amplitude by a sum of terms involving the gluing of two sub-amplitudes, with integrals over the homogeneous coordinates  of the glued legs on the celestial torus. Given the similarity of these integrals to the integrals in the twistor BCFW recursion relations of \cite{Arkani-Hamed:2009hub}, we might expect a diagrammatic formalism à la Hodges \cite{Hodges:2005bf} to exist. Once the gluing is complete the contour integral would simply build terms in a combinatorial fashion by evaluating residues on the poles of the BCFW expression. We leave the exploration of such higher multiplicity questions to future work.


\section{Conclusions}
\label{sec: conclusions}

In this paper we have computed ambidextrous light transformed correlators by deriving celestial twistor amplitudes. This involved the formulation and understanding of the Fourier, twistor, shadow, light and Mellin transforms in their appropriate signature spacetimes. We proved relations between these transforms, summarised by the commuting diagrams in Figures \ref{fig: fouriershadow}, \ref{fig: twistlight}, \ref{fig: dualtwistlight},  before their application to derive three and four-point YM and gravity amplitudes, and a general BCFW recursion relation in sections \ref{sec: celesttwistamps} and \ref{sec: bcfw}. In concluding we would like to highlight some areas for further research which are prompted by our results.

We found a concise expression for the four-point YM amplitude in \eqref{eq: AmpFinalYMtwo} and it would be interesting to relate this to the four-point results of \cite{De:2022gjn} for four light-transformed fields. For gravity, the four-point alternating amplitude, given in \eqref{eq: GravAmpFinal}, shows a striking similarity in structure to its YM analogue - clarifying the physics behind this may lead to insights applicable to more general cases. The application of the double copy for celestial twistor amplitudes (c.f. \cite{Casali:2020vuy, Sharma:2021gcz}) should also be explored in this context.

Extending these results to more general amplitudes is clearly a next step. Higher point results could naturally be derived from the  twistor celestial recursion relation, given in \eqref{eq: BCFW5} or \eqref{eq: BCFW6alt}. The structure of contour integrals and cross-ratios that was apparent in our BCFW derivation of the four-point amplitude would be expected to persist at higher points and it would be interesting to pursue this via direct calculations. In particular the $n$-point MHV amplitudes may be amenable to this approach.
It would also be of interest to understand if the integrands in these recursion relations  have some geometric interpretation, perhaps analogous to the
  diagrammatic rules and identities described in \cite{Arkani-Hamed:2009hub} for the twistor amplitudes.

Developing the OPEs of light transformed correlators via celestial twistor amplitudes should provide insights into the construction of higher point amplitudes as well as elucidate the structure of celestial conformal field theory, along with factorisation, conformal blocks, etc. These OPEs should follow from the consequences of collinear limits for the amplitudes presented here. The extension of this work to supersymmetry can be done using the results in \cite{Brandhuber:2021nez, Jiang:2021xzy} - since the anticommuting superspace coordinate scales under the little group, the chiral Mellin transform is modified for each of the superspace component fields and hence amplitudes. The twistor space super-amplitudes described in \cite{Arkani-Hamed:2009hub} neatly package together their component sub-amplitudes and one might expect similar simplifications for their celestial analogues. There would then exist super-shadow and super-light transforms that are related to super-Fourier and super-twistor transforms. Thus the celestial twistor superamplitudes could be found by Mellin transforming twistor space superamplitudes.

The relative simplicity and interesting structure of the twistor celestial amplitudes derived in this paper would seem to provide encouraging signs for developing this approach in these directions. 

{\bf Acknowledgements:} We would like to thank Andreas Brandhuber and Gabriele Travaglini for initial collaboration on this project and many interesting discussions throughout. GRB and JG would like to thank the Kavli Institute for Theoretical Physics at the University of California, Santa Barbara, where their research was supported in part by the National Science Foundation under Grant No. PHY- 1748958. This work was supported by the Science and Technology Facilities Council (STFC) Consolidated Grants ST/P000754/1 “String theory, gauge theory \& duality” and ST/T000686/1 “Amplitudes, strings and duality”, and by the European Union’s Horizon 2020 research and innovation programme under the Marie Sk\l{}odowska-Curie grant agreement No. 764850 “SAGEX”. The work of GRB and JG is supported by STFC quota studentships.


\appendix

\section{All But One Leg Celestial Twistor Amplitudes}\label{app: almostceltwistamps}

In this appendix, we present three-point amplitudes with all but one leg half-Fourier transformed and also their celestial versions found via ambidextrous half-Mellin transforms. These amplitudes feature in a form of the BCFW recursion relations \eqref{eq: BCFW6alt} for celestial twistor amplitudes and we have used them above to derive the four-point ambidextrous celestial twistor amplitude via BCFW recursion.

Ambidextrously half-Fourier transforming legs one and two of an $\overline{MHV}$ three-point amplitude we have
\begin{equation}
\begin{split}
    A^{\overline{MHV}}\big[(\mu_1, \lamt_1,+) , (\lam_2, \mut_2,-), (\lam, \lamt, +)\big]=&\;\delta(\agl{\lam}{\lam_2})\,\sgn(\sqr{\lamt_1}{\lamt})\,\sgn(\agl{\mu_1}{\lam_2}+\sqr{\mut_2}{\lamt_1})\\
   \qquad &\times \frac{\agl{\lam_2}{\xi}}{\agl{\xi}{\lam}} \,\exp(i\frac{\agl{\lam}{\xi}}{\agl{\xi}{\lam_2}}\sqr{\mut_2}{\lamt})
\end{split}
\end{equation}
which can be derived by an identical method to that laid out in \cite{Arkani-Hamed:2009hub}. We have written the result in terms of a reference spinor $\xi$ on which the amplitude does not depend. This property is a result of the presence of the delta function  $\delta(\agl{\lam}{\lam_2})$ which is the remnant of three-point momentum conservation. We now half-Mellin transform over the scales of the spinors $\mu_1, \mut_2, \lam$ with weights $k_1, \kb_2, h$ and their corresponding discrete weights. Since we have already performed many such Mellin transforms of exponentials and sgn functions we can refer to previous calculations, for example \eqref{eq: hatstairsMellin} and \eqref{eq: Iintegraltwistorlight}, and we find
\begin{equation}
\begin{split}
    &A^{\overline{MHV}}\big[(\mu_1, \lamt_1, k_1, +) , (\lam_2, \mut_2,\kb_2,-), (\lam, \lamt,h, +)\big]\\
    \qquad=&\;\delta(\agl{\lam}{\lam_2})\,\sgn(\sqr{\lamt_1}{\lamt}\sqr{\mut_2}{\lamt_1})\,\frac{\agl{\lam_2}{\xi}}{\agl{\xi}{\lam}}\sgn(-\theta_{12})^{s_{k_1}} \frac{\abs{\theta_{12}^{-1}}^{2k_1}}{2k_1+(-1)^{s_{k_1}}\epsilon}\\
    &\qquad \times\delta\bigg(\sum_{i=1}^{2} 2\kibnbnormal+2-2h\bigg)\,\tilde\delta\bigg(\sum_{i=1}^{2} s_{\kibnbsmall} + s_h\bigg)\cN\Big(2h-2,s_{h}+1, \frac{\agl{\lam}{\xi}}{\agl{\xi}{\lam_2}}\sqr{\mut_2}{\lamt}\Big),
\end{split}
\end{equation}
where the function $\cN(a,b,x)$ is defined in the main text \eqref{eq: Ndef}.

Since in a celestial amplitude the spinors act as homogeneous coordinates, the delta function $\delta(\agl{\lam}{\lam_2})$ actually implies $\lam,\lam_2$ are equivalent as homogeneous coordinates and so we may write
\begin{equation}\label{eq: almostceltwist3ptMHVbar}
\begin{split}
    &A^{\overline{MHV}}\big[(\mu_1, \lamt_1, k_1, +) , (\lam_2, \mut_2,\kb_2,-), (\lam, \lamt,h, +)\big]\\
    =\;&\delta(\agl{\lam}{\lam_2})\,\sgn(\sqr{\lamt_1}{\lamt}\sqr{\lamt_1}{\mut_2})\,\sgn(-\theta_{12})^{s_{k_1}} \frac{\abs{\theta_{12}^{-1}}^{2k_1}}{2k_1+(-1)^{s_{k_1}}\epsilon}\\
    &\qquad \times\delta\bigg(\sum_{i=1}^{2} 2\kibnbnormal+2-2h\bigg)\,\tilde\delta\bigg(\sum_{i=1}^{2} s_{\kibnbsmall} + s_h\bigg)\cN\Big(2h-2,s_{h}+1,\sqr{\lamt}{\mut_2}\Big),
\end{split}
\end{equation}
and the apparent dependence on the reference spinor drops out. The $MHV$ amplitude is completely analogous. We half-Fourier transform legs 3 and 4 and then ambidextrously half-Mellin transform and find
\begin{equation}\label{eq: almostceltwist3ptMHV}
\begin{split}
    &A^{MHV}\big[(\mu_3, \lamt_3, k_3, +) , (\lam_4, \mut_4,\kb_4,-), (\lam, \lamt,\hb, -)\big]\\
    =\;&\delta(\sqr{\lamt}{\lamt_3})\,\sgn(\agl{\lam_4}{\lam}\agl{\mu_3}{\lam_4})\,\sgn(-\theta_{34})^{s_{\kb_4}} \frac{\abs{\theta_{34}}^{2\kb_4}}{2\kb_4+(-1)^{s_{\kb_4}}\epsilon}\\
    &\qquad \times\delta\bigg(\sum_{i=1}^{2} 2 \kibnbnormal+2-2\hb\bigg)\,\tilde\delta\bigg(\sum_{i=1}^{2} s_{\kibnbsmall}+s_{\hb}\bigg)\cN\Big(2\hb-2,s_{\hb}+1,\agl{\mu_3}{\lam}\Big).
\end{split}
\end{equation}



\section{From Shadow to Fourier}\label{app: fourierequalsshadow}

In this Appendix we complete the proof that a chiral Mellin transform of a Fourier transform gives the shadow transform. This requires us to compute the integral $\cI$ in \eqref{eq: Iintegralfouriershad} given by
\begin{equation}
    \cI :=\int_{\mathbb{C}^*}  \frac{d\bar{\sigma}}{\bar{\sigma}} \wedge \frac{d\sigma}{\sigma} \sigma^{2-2h}\bar{\sigma}^{2-2\hb}e^{i(\sigma+\bar{\sigma})}\,.
\end{equation}
Computing $\cI$ completes the proof of the commuting diagram Figure \eqref{fig: fouriershadow}. However, as an additional check, we also prove the diagram in the reverse direction by inverse chiral Mellin transforming the shadowed conformal primary to recover a Fourier transformed momentum eigenstate.

\newpage 

\textbf{Computing} $\cI$

To compute $\cI$ we first change variables $\sigma= Re^{i\phi}$ 
\begin{equation}
    \begin{split}
        \cI =& \;2i\int_0^\infty dR\, R^{3-2\Delta}\int_0^{2\pi} d\phi\, \left(e^{i\phi}\right)^{-2J} e^{2iR \cos{\phi}}\\
            =& \;2i\int_0^\infty dR\, R^{3-2\Delta} \cI_{\phi}\, .
    \end{split}
\end{equation}
We can compute the angular integral as a contour integral using $z = e^{i\phi}$, giving 
\begin{equation}
    \cI_{\phi} = -i\oint \frac{dz}{z}\, z^{-2J} e^{iR(z+z^{-1})} = -i \sum_{n=0}^\infty \frac{(iR)^n}{n!}\oint \frac{dz}{z}\,z^{-2J}\left(z+z^{-1}\right)^n\,.
\end{equation}
This is  also a modified Bessel function of the first kind. Now we use the binomial theorem and find that the contour integal has a residue contribution of  $2\pi i$ at $z=0$ only when $n+2J$ is even and $n-\vert 2J\vert \geq0$ and for the single term in the binomial sum with $k=J+\frac{n}{2}\geq 0$. The integral is then given by the sum
\begin{equation}\label{eq: angularSum}
    \cI_\phi = 2\pi\sum_{\substack{n\geq \lvert 2J\lvert \\ n+2J\, \text{even}}}^\infty \!\! \frac{ (iR)^n}{(\frac{n}{2}+J)!(\frac{n}{2}-J)!}  =2\pi i^{2\abs{J}} \sum_{l=0}^\infty(-1)^l \frac{R^{2l+2\abs{J}}}{l!(l+2\abs{J})!}\,,
\end{equation}
which we have rewritten with $l:=\frac{n}{2}-\abs{J}$. This sum corresponds to a generalised hypergeometric function, since the ratio of any two consecutive terms is a rational function in $l$. Labelling the terms in the above sum by $a_l$ we have
\begin{equation}
    \frac{a_{l+1}}{a_l} = \frac{-R^2}{(l+1)(l+2\abs{J}+1)}, \quad a_{0}= 2\pi i^{2\abs{J}}R^{2\abs{J}} ,
\end{equation}
hence the angular integral is given by
\begin{equation}
    \cI_{\phi}= 2\pi i^{2\abs{J}}R^{2\abs{J}}\, {}_0F_1(\_;2\abs{J}+1;-R^2)\,.
\end{equation}
The sum \eqref{eq: angularSum} converges for all finite values of $R^2$ by the ratio test. 

To evaluate the $R$ integral we will find it easier to work with the explicit sum form of $\cI_{\phi}$
\begin{equation}
    \begin{split}
        \cI= 4\pi i\, i^{2\abs{J}} \int_0^\infty dR R^{3-2\Delta} \sum_{l=0}^\infty (-1)^l\frac{R^{2l+2\abs{J}}}{l!(l+2\abs{J})!}\,.
    \end{split}
\end{equation}
We now regulate the $R$ integral using $e^{-\delta R}$ and then take $\delta\rightarrow 0^{+}$\footnote{Ramanujan's master theorem also gives the same answer \cite{RamaujanMT}.}. We also commute the infinite sum with the integral since both now converge
\begin{equation}
    \begin{split}
        \cI =& \;4\pi i\, i^{2\abs{J}} \sum_{l=0}^\infty \frac{(-1)^l}{l!(l+2\abs{J})!}\int_{0}^\infty R^{3-2\Delta +2l + 2\abs{J}}e^{-\delta R}\\
            =& \;4\pi i\, i^{2\abs{J}}\sum_{l=0}^\infty \frac{(-1)^l }{l!(l+2\abs{J})!} \frac{\Gamma(4-2\Delta +2l +2\abs{J})}{\delta^{4-2\Delta+2l+2\abs{J}}}\,.
    \end{split}
\end{equation}
Once again this sum corresponds to a hypergeometric function: labelling the terms in the sum by $b_l$ we have
\begin{equation}
    \frac{b_{l+1}}{b_{l}}= \frac{(5-2\Delta + 2l +2\abs{J})(4-2\Delta + 2l + 2\abs{J})}{(l+1)(l+2\abs{J}+1)} \frac{-1}{\delta^2}, \quad b_{0}=\frac{4\pi i\, i^{2\abs{J}} \Gamma(4-2\Delta +2\abs{J})}{(2\abs{J})!\delta^{4-2\Delta+2\abs{J}}}\,.
\end{equation}
Thus the integral $\cI$ is given by a type ${}_2F_1$ hypergeometric function. The sum converges for $\abs{\frac{-4}{\delta^2}}<1/2$ and is on the principal branch
\begin{equation}\label{eq: Iint3}
    \cI = \frac{4\pi i \, i^{2\abs{J}} \Gamma(4-2\Delta +2\abs{J})}{(2\abs{J})!\delta^{4-2\Delta+2\abs{J}}} {}\; _2F_1\left(\ontop{\frac{5}{2} -\Delta +\abs{J},\,\, 2-\Delta+ \abs{J}}{2\abs{J}+1} ; \frac{-4}{\delta^2}\right) .
\end{equation}
In order to tame this expression we must use a series of transformations and identities. The first of these is a Pfaff transformation,
\begin{equation}\label{eq: Pfaff}
    {}_2F_{1} \left(\ontop{a,b}{c}; z\right) = (1-z)^{-b} {}_2F_{1} \left(\ontop{b,c-a}{c}; \frac{z}{z-1}\right)\,,
\end{equation}
and applying this to \eqref{eq: Iint3} gives 
\begin{equation}
    \begin{split}
        \cI = \frac{4\pi i \, i^{2\abs{J}} \Gamma(4-2\Delta +2\abs{J})}{(2\abs{J})!(\delta +2)^{4-2\Delta+2\abs{J}}} {}_2F_1\left(\ontop{2-\Delta+ \abs{J},\,\, -\frac{3}{2}+ \Delta +\abs{J}}{2\abs{J}+1} ; \frac{1}{1+\frac{\delta^2}{4}}\right)\,.
    \end{split}
\end{equation}
At this point we are safe to take the limit $\delta\rightarrow 0$  and we obtain
\begin{equation}
    \cI = 4\pi i\,  \frac{i^{\abs{J}}}{(2\abs{J})!} 2^{2\Delta -2\abs{J}-4} \Gamma(4-2\Delta +2\abs{J}) {}_2F_1\left(\ontop{2-\Delta+ \abs{J},\,\, -\frac{3}{2}+ \Delta +\abs{J}}{2\abs{J}+1} ; 1\right)\,.
\end{equation}
We can now evaluate the ${}_2F_1$ at the point $1$ in terms of objects we are very familiar with: gamma functions
\begin{equation}
    {}_2F_{1} \left(\ontop{a,b}{c}; 1\right) = \frac{\Gamma(c)\Gamma(c-a-b)}{\Gamma(c-a)\Gamma(c-b)}\,,
\end{equation}
which holds if $\Re(c-a-b)>0$. This condition is satisfied for us since $c-a-b =1/2$, hence we can write 
\begin{equation}
    \begin{split}
        \cI =& \;4\pi i \, \frac{i^{2\abs{J}}}{(2\abs{J})!} 2^{2\Delta -2\abs{J}-4} \Gamma(4-2\Delta +2\abs{J}) \frac{\Gamma(2\abs{J}+1)\Gamma(\half)}{\Gamma(-1 +\Delta +\abs{J})\Gamma(\frac{5}{2}+ \abs{J}-\Delta) }\\
        =& \;4\pi i \, i^{2\abs{J}} 2^{2\Delta -2\abs{J}-4}  \frac{\Gamma(4-2\Delta +2\abs{J}) \sqrt{\pi}}{\Gamma(-1 +\Delta +\abs{J})\Gamma(\frac{5}{2}+ \abs{J}-\Delta) }\,.
    \end{split}    
\end{equation}
Further, we can simplify this expression using the Legendre duplication formula \cite[Eq.~5.5(iii)]{NIST:DLMF}
\begin{equation}\label{eq: legendreduplication}
    \sqrt{\pi}\,2^{1-2z}\Gamma(2z)= \Gamma(z)\Gamma(z+\half),
\end{equation}
and the Euler reflection formula
\begin{equation}\label{eq: eulerreflection}
    \Gamma(z)\Gamma(1-z)=\frac{\pi}{\sin(\pi z)}.
\end{equation}
Using these we arrive at
\begin{equation}\label{eq: Iintegralthreeforms}
    \cI= 2\pi i\, i^{2\abs{J}}\frac{\Gamma(2-\Delta+\abs{J})}{\Gamma(-1 + \Delta + \abs{J})} = 2\pi i\, i^{2J}\frac{\Gamma(2-2\hb)}{\Gamma(2h-1)} = 2\pi i\, i^{-2J}\frac{\Gamma(2-2h)}{\Gamma(2\hb-1)} .
\end{equation}
Each of the three expressions above is equivalent and we are free to use whichever we prefer. We then use this result in the main text to recover the shadow transform.

\textbf{From Shadow to Fourier}

We begin with our expression for the shadow transformed state \eqref{eq: 13compactshadcelstate} which we will rewrite in terms of $k=1-h$ and $\hb=1-\hb$
\begin{equation}
\begin{split}
   \vert\mu,\mut;k,\kb\rangle = \frac{i^{-2k+2\kb}\Gamma(2\kb)}{4\pi ^2\Gamma(1-2k)}\int_{\mathbb{C}_*^{2}} d^2\lambda\wedge d^2\!\lamt \, \agl{\lambda}{\mu}^{-2k} \sqr{\mut}{\lamt}^{-2\kb} \vert\lam,\lamt \rangle\,,
\end{split}
\end{equation}
and perform an inverse chiral Mellin transform \eqref{eq: InverseChiralMellin} to recover a Fourier transformed state. We can choose to integrate with respect to $\Delta'= k+\kb$ and $J' = k-\kb$ since this is just a choice of integration variables. We then have 
\begin{equation}
\begin{split}
     &\frac{1}{2\pi i}  \sum_{2J'\in \mathbb{Z}}  \int_{a-i\infty}^{a+i\infty} d\Delta' \,  \vert\mu,\mut;k,\kb \rangle \\
     =&
     \frac{1}{4\pi^2} \int_{\mathbb{C}_*^{2}} \!d^2\lambda\wedge d^2\!\lamt \,\sum_{2J'\in \mathbb{Z}} i^{-2J'}(-1)^{\vert J' \vert+J'}  \left(\frac{\sqr{\mut}{\lamt}}{\agl{\lam}{\mu}}\right)^{J'}  \int_{a-i\infty}^{a+i\infty} d\Delta' \, \frac{\Gamma(\Delta'+\vert J' \vert)(\agl{\lam}{\mu}\sqr{\mut}{\lamt})^{-\Delta}}{2\pi i\,\Gamma(1 - \Delta' + \vert J' \vert)}  \vert\lam,\lamt \rangle,
\end{split}
\end{equation}
where we have used the form of \eqref{eq: Iintegralthreeforms} written in terms of $\Delta$ and $J$. The integrand has poles at $\Delta'= -\vert J' \vert -n$ for $n\in \mathbb{Z}_{\geq 0}$ due to the gamma function $\Gamma(\Delta'+\vert J'\vert)$. Thus the integral over $\Delta$ can be performed by choosing the contour with $a>- \vert J\vert$ closed anticlockwise. Hence we have 
\begin{equation}
    \begin{split}
        =&\frac{1}{4\pi^2} \int_{\CC_*^2} \!d^2\lambda\wedge d^2\!\lamt \,\sum_{2J'\in \mathbb{Z}} i^{-2J'}(-1)^{\vert J' \vert+J'}  \left(\frac{\sqr{\mut}{\lamt}}{\agl{\lam}{\mu}}\right)^{J'} \sum_{n=0}^\infty  \, \frac{(-1)^{n}(\agl{\lam}{\mu}\sqr{\mut}{\lamt})^{\vert J' \vert +n}}{n!\Gamma(1 +2 \vert J' \vert+ n)} \, \vert\lam,\lamt \rangle\\
        =&\frac{1}{4\pi^2} \int_{\CC_*^2} \!d^2\lambda\wedge d^2\!\lamt \,\sum_{2J'\in \mathbb{Z}}  (i\sqr{\mut}{\lamt})^{\vert J'\vert+J'} (i\agl{\lam}{\mu})^{\vert J' \vert-J'} \sum_{n=0}^\infty  \, \frac{((i\agl{\lam}{\mu})(i\sqr{\mut}{\lamt}))^{n}}{n!(2 \vert J' \vert+ n)!} \, \vert\lam,\lamt \rangle\, .
    \end{split}
\end{equation}
These two sums look slightly unwieldy but we can massage them in to something more familiar as follows 
\begin{equation}
    \begin{split}
        =&\frac{1}{4\pi^2}\int_{\CC_*^2} \!d^2\lambda\wedge d^2\!\lamt \sum_{n=0}^\infty \frac{((i\agl{\lam}{\mu})(i\sqr{\mut}{\lamt}))^{n}}{n!}\left(\frac{1}{n!} + \sum_{m=1}^\infty\frac{(i\sqr{\mut}{\lamt})^m}{(m+n)!}   + \sum_{m=1}^\infty\frac{(i\agl{\lam}{\mu})^m}{(m+n)!} \right)\vert\lam,\lamt \rangle\\
        =& \frac{1}{4\pi^2}\int_{\CC_*^2} \!d^2\lambda\wedge d^2\!\lamt\bigg( \sum_{n=0}^\infty \frac{(i\agl{\lam}{\mu})^n}{n!}\frac{(i\sqr{\mut}{\lamt})^{n}}{n!} \\
        & \qquad\qquad\qquad\qquad + \sum_{n=0}^\infty\sum_{m=1}^\infty\left(\frac{(i\sqr{\mut}{\lamt})^{m+n}}{(m+n)!}\frac{(i\agl{\lam}{\mu})^n}{n!}   +  \frac{(i\sqr{\mut}{\lamt})^{n}}{n!}\frac{(i\agl{\lam}{\mu})^{m+n}}{(m+n)!}\right) \bigg)\vert\lam,\lamt \rangle\\
        =&\frac{1}{4\pi^2}\int_{\CC_*^2} \!d^2\lambda\wedge d^2\!\lamt\left( \sum_{n=0}^\infty \sum_{k=0}^n \frac{(i\agl{\lam}{\mu})^{n-k}}{(n-k)!}\frac{(i\sqr{\mut}{\lamt})^{k}}{k!} \right)\vert\lam,\lamt \rangle\\
        =& \frac{1}{4\pi^2}\int_{\CC_*^2} \!d^2\lambda\wedge d^2\!\lamt e^{i\agl{\lam}{\mu}} e^{i\sqr{\mut}{\lamt}}\vert\lam,\lamt \rangle
    \end{split}
\end{equation}
which is exactly the Fourier transformed state \eqref{eq: Fourier}.

\section{Bulk Conformal Generators}\label{app: conformalgens}

In this Appendix we provide formulae for the conformal generators of the four-dimensional bulk spacetime. We present these generators in a variety of bases: momentum space, Mellin space, Fourier and twistor space and finally shadow and light space. These different bases for the generators reflect the various integral transforms that we have at our disposal. The generators themselves are always defined in the same way - we simply commute them with the integral transform which defines the basis and we always use $\sim$ to denote this relation. We also in general favor presenting the celestial generators in homogeneous coordinates since they are much more compact, nevertheless we also give some examples in terms of affine coordinates.

To check the generators satisfy the conformal algebra it is easiest to check that the spinor and spinor derivative from which we build generators satisfy the canonical commutation relations. The only non-vanishing commutator is
\begin{equation}
    [\partial_{\alpha}, \lambda_{\beta}]=
\epsilon_{\beta\alpha}
\end{equation}
and similarly for the conjugate spinor and its derivative.

The generators of the four-dimensional conformal algebra written in terms of spinors, and the commutator relations they satisfy are given in \cite{Henn:2014sjd}. The form of these generators is the same in either Lorentzian or split signature with the only difference being the form of the metric $\eta^{\mu\nu}$ in the structure constants of the algebra. This then informs the reality conditions on the spinors.

\begin{landscape}
\begin{table}
    \centering
    {\renewcommand{\arraystretch}{2}%
    \begin{tabular}{ |P{3cm}||P{6cm}|P{6cm}|P{6cm}| }
    \hline
     \multicolumn{4}{|c|}{Table of Operators in (1,3)} \\
    \hline
    Momentum & Fourier & Mellin &  Shadow\\
    \hline
    $p_{\alpha\dot\alpha}$   &  $-\frac{\partial}{\partial \muu{\alpha}}\frac{\partial}{\partial \mtu{\alpha}} =: -k^{F}_{\alpha\dot\alpha}$ & $\ld{\alpha}\ltd{\alpha} e^{\frac{\partial_h}{2}+ \frac{\partial_{\hb}}{2}}$   &$-\frac{\partial}{\partial \muu{\alpha}}\frac{\partial}{\partial \mtu{\alpha}}e^{-\frac{\partial_k}{2}- \frac{\partial_{\kb}}{2}}$\\[11pt]
    \hline
    $k_{\alpha\dot\alpha}$   &  $- \mud{\alpha}\mtd{\alpha} =: -p^{F}_{\alpha\dot\alpha}$ & $\pld{\alpha}\pltd{\alpha} e^{-\frac{\partial_h}{2}- \frac{\partial_{\hb}}{2}}$   &$- \mud{\alpha}\mtd{\alpha}e^{\frac{\partial_k}{2}+ \frac{\partial_{\kb}}{2}}$\\[11pt]
    \hline
    $d$   &  $-\frac{1}{2}\muu{\alpha}\frac{\partial}{\partial \muu{\alpha}}-\frac{1}{2}\mtu{\alpha}\frac{\partial}{\partial \mtu{\alpha}}-1 =: -d_{F}$ & $-h-\hb+1$&$k+\kb-1$\\[11pt]
    \hline
    $\cJ$   &  $\frac{1}{2}\muu{\alpha}\frac{\partial}{\partial \muu{\alpha}}-\frac{1}{2}\mtu{\alpha}\frac{\partial}{\partial \mtu{\alpha}} =: -\cJ_{F}$ & $h-\hb$   &$-k+\kb$\\[11pt]
    \hline
    $m_{\alpha\beta}$   &  $-\mu_{(\alpha}\frac{\partial}{\partial\mu^{\beta)}} =: -m_{\alpha\beta}^{F}$ & $\lam_{(\alpha}\partial_{\beta)}$   &$-\mu_{(\alpha}\frac{\partial}{\partial\mu^{\beta)}}$\\[11pt]
    \hline
    $\bar m_{\dot\alpha\dot\beta}$   &  $-\tilde\mu_{(\dot\alpha}\frac{\partial}{\partial\tilde\mu^{\dot\beta)}} =: -m_{\dot\alpha\dot\beta}^{F}$ & $\lamt_{(\dot\alpha}\tilde\partial_{\dot\beta)}$   &$-\tilde\mu_{(\dot\alpha}\frac{\partial}{\partial\tilde\mu^{\dot\beta)}}$\\[11pt]
    \hline
    \end{tabular}}
    \caption{Table of conformal generators in each basis in (1,3). In the above $k=1-h, \kb=1-\hb$ such that $e^{\partial_k} = e^{-\partial_h}$ etc.}
    \label{tab:tableofgens}
\end{table}
\end{landscape}

\textbf{Mellin Space}

The generators of the bulk conformal algebra in the Mellin basis in affine coordinates were presented in \cite{Stieberger:2018onx} and re-derived from the (1,3) signature chiral Mellin transform in \cite{Brandhuber:2021nez} along with the $N=4$ superconformal generators. In Table \ref{tab:tableofgens} we give the (1,3) celestial conformal generators in homogeneous coordinates; an example derivation of $\partial_{\alpha} e^{-\partial_h}$ is given in the main text, equation \eqref{eq: spinorderivhomo}.
        
\textbf{Fourier Space}

We define conformal generators in the Fourier basis by commuting them with the Fourier transform \eqref{eq: Fourier} using integration by parts. The basic building block operators have the following Fourier conjugates
\begin{equation}\label{fourierconjugatespinors}
    \begin{split}
    &\mud{\alpha} \sim i\partial_{\alpha}, \,\muu{\alpha} \sim -i\partial^{\alpha},\, \frac{\partial}{\partial \muu{\alpha}} \sim - i \ld{\alpha},\, \frac{\partial}{\partial \mud{\alpha}} \sim  i \lu{\alpha},\\
        &\mtd{\alpha} \sim i\tilde\partial_{\dot\alpha},\, \,\mtu{\alpha} \sim -i\tilde\partial^{\dot\alpha},\, \frac{\partial}{\partial \mtu{\alpha}} \sim  -i\ltd{\alpha},\, \frac{\partial}{\partial \mtd{\alpha}} \sim  i\ltu{\alpha},
        \end{split}
\end{equation}
such that
\begin{align}
    \ld{\alpha}\pld{\beta} \sim -\epsilon_{\alpha\beta} - \mud{\beta}\frac{\partial}{\partial \muu{\alpha}}, \quad \ltd{\alpha}\pltd{\beta} \sim -\epsilon_{\dot\alpha\dot\beta} - \mtd{\beta}\frac{\partial}{\partial \mtu{\alpha}},
\end{align}
implying
\begin{align}
    \lu{\alpha}\pld{\alpha} \sim -2 - \muu{\alpha}\frac{\partial}{\partial \muu{\alpha}}, \quad \ltu{\alpha}\pltd{\alpha} \sim -2 - \mtu{\alpha}\frac{\partial}{\partial \mtu{\alpha}},\\
    \lam_{(\alpha}\partial_{\beta)} \sim  - \mu_{(\alpha}\frac{\partial}{\partial \mu^{\beta)}}, \quad \lamt_{(\dot\alpha}\partial_{\dot\beta)} \sim - \mut_{(\dot\alpha}\frac{\partial}{\partial \mut^{\dot\beta)}},
\end{align}
from which the helicity and dilatation operator featuring in the main text, as well as all the other conformal generators in Table \ref{tab:tableofgens}, are derived.

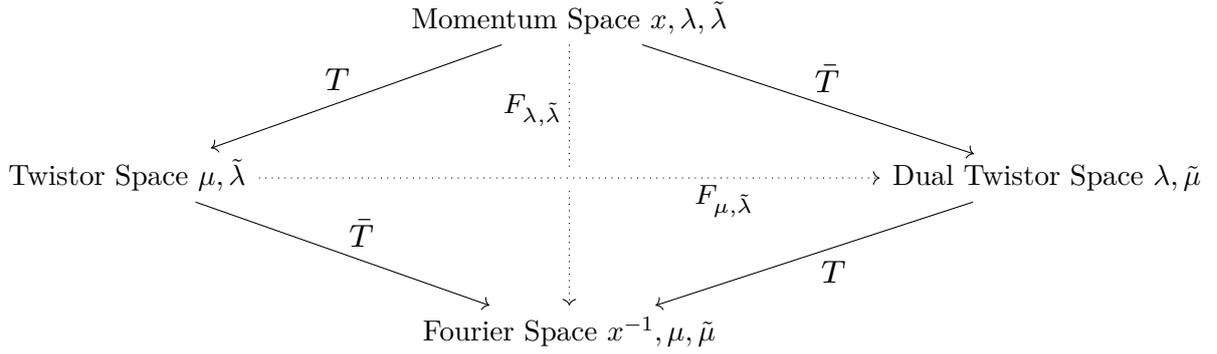
\begin{figure}
    \centering
    \[
\begin{tikzcd}
&&\text{Momentum Space  $x, \lam,\lamt$} \arrow[drdr, "\scalebox{1.5}{$\bar T$}"] \arrow[dldl, "\scalebox{1.5}{$T$}"'] \arrow[dddd,dotted,near start, "\scalebox{1.3}{$F_{\lam, \lamt}$}"'] &  & \\
&&&&\\
\text{Twistor Space  $\mu,\lamt$} \arrow[drdr, "\scalebox{1.5}{$\bar T$}"] \arrow[rrrr, crossing over, dotted, near end,"\scalebox{1.3}{$F_{\mu, \lamt}$}"'] & & & &\text{Dual Twistor Space  $\lam,\mut$} \arrow[dldl, "\scalebox{1.5}{$T$}"] \\               
&&  &  & \\   
& &  \text{Fourier Space  $x^{-1}, \mu,\mut$}    &&      
\end{tikzcd}
\]
    \caption{A `bulk diamond' of relationships between twistor and Fourier transforms, including the action of conformal inversions on position space. A chiral Mellin transform maps this to a celestial Diamond a lá \cite{Pasterski:2021fjn, Pasterski:2021dqe}.}
    \label{fig: fouriersandinversion}
\end{figure}

In Table \ref{tab:tableofgens} we also define another set of operators $\cJ_F, d_F, p_F, k_F, m_F$ which generate the conformal symmetries of $\mu, \mut$ space directly and without reference to their origin as conjugate spinors to $\lam, \lamt$. We find that the original set are related to these by an automorphism of the conformal algebra given by
\begin{equation}\label{eq: conformalinversionautomorphism}
    \begin{split}
        & d \mapsto -d, \quad \cJ \mapsto -\cJ, \quad p \mapsto -k, \\
        & k \mapsto  -p,\quad m \mapsto  -m, \quad \bar{m} \mapsto -\bar m.
    \end{split}
\end{equation}
One can check that this preserves the commutator relations of the conformal algebra. This automorphism of the conformal algebra is generated by conjugation with the discrete operator which generates a conformal inversion $I$ in position space. For example, we have the well known description of special conformal transformations as $k_{\mu}=-I\,p_{\mu}I$. We can make this more concrete by considering the conformal symmetry generators in position space defined by the Fourier transform 
\begin{equation}
    \int d^4x \,e^{-i p\cdot x}
\end{equation}
giving the conformal generators
\begin{equation}\label{eq: positionconformalgens}
\begin{split}
    p_{\mu}\sim-i\frac{\partial}{\partial x^{\mu}}, \quad k_{\mu}\sim i\bigg(x^2\frac{\partial}{\partial x^{\mu}} - x_{\mu}x^{\nu}\frac{\partial}{\partial x^{\nu}}\bigg),\\
    d\sim -ix^{\mu}\frac{\partial}{\partial x^{\mu}}, \quad m_{\mu\nu}\sim i\bigg(x_{\mu}\frac{\partial}{\partial x^{\nu}}-x_{\nu}\frac{\partial}{\partial x^{\mu}}\bigg).
\end{split}    
\end{equation}
We can then take these expression and consider their representation in a new copy of Minkowski space with coordinate $y^{\mu}$ (with units of inverse length) and related to $x^{\mu}$ by a conformal inversion $y^{\mu}=\frac{x^{\mu}}{x^2}$. Then, by changing variables in \eqref{eq: positionconformalgens} we find exactly the relations \eqref{eq: conformalinversionautomorphism}. This shows that the special conformal dual $k_{\alpha\dot\alpha}=\mud{\alpha}\mtd{\alpha}$ can be regarded as the `momentum' associated to a conformally inverted position space. This situation is summarised in the diagram contained in Figure \ref{fig: fouriersandinversion}.

\textbf{Shadow Space}

Since the shadow basis is simply given by a chiral Mellin transform of the Fourier basis, the shadowed bulk conformal generators in Table \ref{tab:tableofgens} are just celestial versions of the Fourier transformed generators. This can be checked using the formula for the shadowed celestial state \eqref{eq: 13compactshadcelstate} and the relations $k=1-h$ and $\kb=1-\hb$. For the spinor derivative $\pld{\alpha}$, using integration by parts we find
\begin{equation}
\begin{split}
    &\frac{i^{2h-2\hb}\Gamma(2-2\hb)}{(2\pi)^2\Gamma(2h-1)}\int_{\mathbb{C}_*^2} d^2\lamt\wedge\, d^2\lam \, \agl{ \lam}{\mu}^{2h-2} \sqr{ \mut}{\lamt}^{2\hb-2} \pld{\alpha} \ket{\lam, \lamt}\\
    =& \frac{i^{2h-2\hb}\Gamma(2-2\hb)}{(2\pi)^2\Gamma(2h-1)}\int_{\mathbb{C}_*^2} d^2\lamt\wedge\, d^2\lam \, (-\mud{\alpha} (2h-2)) \agl{ \lam}{\mu}^{2h-3} \sqr{ \mut}{\lamt}^{2\hb-2}  \ket{\lam, \lamt}\\
    =&\frac{i^{2h-2\hb}\Gamma(2-2\hb)}{(2\pi)^2\Gamma(2h-1)}(-\mud{\alpha}(2h-2) e^{-\partial_h/2})\int_{\mathbb{C}_*^2} d^2\lamt\wedge\, d^2\lam \, \agl{ \lam}{\mu}^{2h-2} \sqr{ \mut}{\lamt}^{2\hb-2}  \ket{\lam, \lamt}\\
    =&-i\mud{\alpha}e^{-\partial_h/2}\frac{i^{2h-2\hb}\Gamma(2-2\hb)}{(2\pi)^2\Gamma(2h-1)}\int_{\mathbb{C}_*^2} d^2\lamt\wedge\, d^2\lam \, \agl{ \lam}{\mu}^{2h-2} \sqr{ \mut}{\lamt}^{2\hb-2} \ket{\lam, \lamt},
\end{split}
\end{equation}
so the shadowed spinor derivative is given by $\pld{\alpha} \sim -i\mud{\alpha}e^{-\partial_h/2}=-i\mud{\alpha}e^{\partial_k/2}$.

\begin{landscape}
\begin{table}
    \centering
    {\renewcommand{\arraystretch}{2}%
    \begin{tabular}{ |M{1.9cm}||M{3cm}|M{4.3cm}|M{4cm}|M{4.4cm}|M{4.4cm}|}
    \hline
     \multicolumn{6}{|c|}{Table of Operators in (2,2)} \\
    \hline
    Momentum & Twistor & Dual Twistor & Mellin & Light & Dual Light \\
    \hline
    $\quad p_{\alpha\dot\alpha}$   &     $i\ltd{\alpha}\frac{\partial}{\partial \mu^{\alpha}}=:p_{\alpha\dot\alpha}^T$&$i\ld{\alpha}\frac{\partial}{\partial \mut^{\dot\alpha}}=-k_{\alpha\dot\alpha}^{\bar T}$&$\ld{\alpha}\ltd{\alpha} e^{\frac{\partial_h}{2}+\frac{\partial_{\hb}}{2}}e^{\partial_{s_h}+\partial_{s_{\hb}}}$&$i\ltd{\alpha}\frac{\partial}{\partial \mu^{\alpha}}e^{-\frac{\partial_k}{2}+\frac{\partial_{\hb}}{2}}e^{-\partial_{s_k}+\partial_{s_{\hb}}}$&$i\ld{\alpha}\frac{\partial}{\partial \mut^{\dot\alpha}}e^{\frac{\partial_h}{2}-\frac{\partial_{\kb}}{2}}e^{\partial_{s_h}-\partial_{s_{\kb}}}$\\[25pt]
    \hline
    $\quad k_{\alpha\dot\alpha}$   &     $-i\mud{\alpha}\pltd{\alpha}=:k_{\alpha\dot\alpha}^T$&$-i\mtd{\alpha}\pld{\alpha}=-p_{\alpha\dot\alpha}^{\bar T}$&$\pld{\alpha}\pltd{\alpha} e^{-\frac{\partial_h}{2}-\frac{\partial_{\hb}}{2}}e^{-\partial_{s_h}-\partial_{s_{\hb}}}$&$-i\mud{\alpha}\pltd{\alpha}e^{\frac{\partial_k}{2}-\frac{\partial_{\hb}}{2}}e^{\partial_{s_k}-\partial_{s_{\hb}}}$&$-i\mtd{\alpha}\pld{\alpha}e^{-\frac{\partial_h}{2}+\frac{\partial_{\kb}}{2}}e^{-\partial_{s_h}+\partial_{s_{\kb}}}$\\[25pt]
    \hline
    $d$ &  $-\frac{1}{2}\muu{\alpha}\frac{\partial}{\partial \muu{\alpha}} + \frac{1}{2}\ltu{\alpha}\pltd{\alpha}=:\cJ_T$& $\frac{1}{2}\lu{\alpha}\pld{\alpha} + \frac{1}{2}\mtu{\alpha}\frac{\partial}{\partial \mtu{\alpha}}=:-\cJ_{\bar T}$ & $-h-\hb+1$ & $k-\hb$ &$-h+\kb$ \\[25pt]
    \hline
    $\cJ$& $\frac{1}{2}\muu{\alpha}\frac{\partial}{\partial \muu{\alpha}} + \frac{1}{2}\ltu{\alpha} \pltd{\alpha}+1 =: d_T$ & $-\frac{1}{2}\lu{\alpha} \pld{\alpha} - \frac{1}{2}\mtu{\alpha}\frac{\partial}{\partial \mtu{\alpha}}-1 =: -d_{\bar T}$& $h-\hb$ & $1-k-\hb$ & $h+\kb-1$\\[25pt]
    \hline
    $m_{\alpha\beta}$&$-\mu_{(\alpha}\frac{\partial}{\partial\mu^{\beta)}}=:-m_{\alpha\beta}^T$&$\lam_{(\alpha}\partial_{\beta)}=: m^{\bar T}_{\alpha\beta}$&$\lam_{(\alpha}\partial_{\beta)}$&$-\mu_{(\alpha}\frac{\partial}{\partial\mu^{\beta)}}$&$\lam_{(\alpha}\partial_{\beta)}$\\[25pt]
    \hline
    $\bar m_{\dot\alpha\dot\beta}$&$\lamt_{(\dot\alpha}\tilde\partial_{\dot\beta)}=:\bar m^{T}_{\dot\alpha\dot\beta}$&$-\mut_{(\dot\alpha}\frac{\partial}{\partial\mut^{\dot\beta)}}=:-\bar m_{\alpha\beta}^{\bar T}$&$\lamt_{(\dot\alpha}\tilde\partial_{\dot\beta)}$&$\lamt_{(\dot\alpha}\tilde\partial_{\dot\beta)}$&$-\mut_{(\dot\alpha}\frac{\partial}{\partial\mut^{\dot\beta)}}$\\[25pt]
    \hline
    \end{tabular}}
    \caption{Table of conformal generators in each basis in (2,2). In the above $k=1-h$ etc. such that $e^{\partial_k} = e^{-\partial_h}$ etc.}
    \label{tab:tableofgens2}
\end{table}
\end{landscape}

\textbf{(2,2) Mellin}

Conformal generators derived from the (2,2) signature chiral Mellin transform are only slightly different from those in (1,3) signature, since we have discrete weights $s_h, s_{\hb}$ which are also carried by spinors and spinor derivatives. As such we must also have discrete weight shifting operators which shift $s_h, s_{\hb}$; by an abuse of notation we also denote these by exponentiated differential operators although, since $s_h, s_{\hb}$ are discrete, they clearly they do not take such a form. Consequently, split signature celestial generators take the same form as their Lorentzian signature cousins but with the universal replacement $e^{\frac{1}{2}\partial_h}\rightarrow e^{\frac{1}{2}\partial_h}e^{\partial_{s_h}}$. The inclusion of the $e^{\partial_{s_h}}$ operators has no effect on the closing of the conformal algebra but is needed to properly shift the (2,2) weights.

For example, consider the spinor derivative $\partial_{\alpha}$ and act with the (2,2) chiral Mellin transform \eqref{eq: 22chiralMellin}
\begin{equation}
  \int_{\RR_*\times \RR_*} \frac{d\ut}{\ut}\wedge \frac{du}{u} \, \abs{u}^{2h}\, \abs{\ut}^{2\hb}\,\sgn(u)^{s_h}\sgn(\ut)^{s_{\hb}}  \frac{1}{u} \pld{\alpha} \ket{u \lam,\ut \lamt}
  = \pld{\alpha}e^{-\frac{1}{2}\partial_h}e^{-\partial_{s_h}}  \ket{\lam , \lamt;h, s_h, \hb, s_{\hb}}\,.
\end{equation}
Hence the spinor derivative acting in Mellin space is simply $\pld{\alpha}e^{-\frac{1}{2}\partial_h-\partial_{s_h}} $.

Compare this with the affine coordinate version
\begin{equation}
    \begin{split}
        &\pld{\alpha}e^{-\frac{1}{2}\partial_h}e^{-\partial_{s_h}}  \ket{\ld{\alpha} , \ltd{\alpha};h, s_h, \hb, s_{\hb}}\\
        =& \;\pld{\alpha}e^{-\frac{1}{2}\partial_h}e^{-\partial_{s_h}}\abs{\ld{2}}^{-2h}\abs{\ltd{2}}^{-2\hb}  \sgn(\ld{2})^{-s_h}\sgn(\ltd{2})^{-s_{\hb}} \ket{\begin{pmatrix}
			z \\
			1  
		\end{pmatrix},\begin{pmatrix}
			\zb \\
			1  
		\end{pmatrix};h, s_h, \hb, s_{\hb}}\\
        =&\frac{1}{\ld{2}}\begin{pmatrix}
            \ld{2}\partdiv{}{\ld{2}}-z\partial_{z}\\
            -\partial_{z}
        \end{pmatrix} e^{-\frac{1}{2}\partial_h}e^{-\partial_{s_h}}\abs{\ld{2}}^{-2h}\abs{\ltd{2}}^{-2\hb}  \sgn(\ld{2})^{-s_h}\sgn(\ltd{2})^{-s_{\hb}}\ket{\begin{pmatrix}
			z \\
			1  
		\end{pmatrix},\begin{pmatrix}
			\zb \\
			1  
		\end{pmatrix}; h, s_h, \hb, s_{\hb}}\\
        =& \;\abs{\ld{2}}^{-2h}\abs{\ltd{2}}^{-2\hb}  \sgn(\ld{2})^{-s_h}\sgn(\ltd{2})^{-s_{\hb}} \begin{pmatrix}
            -2h+1-z\partial_{z}\\
            -\partial_{z}
        \end{pmatrix}e^{-\frac{1}{2}\partial_h-\partial_{s_h}}\ket{\begin{pmatrix}
			z \\
			1  
		\end{pmatrix},\begin{pmatrix}
			\zb \\
			1  
		\end{pmatrix}; h, s_h, \hb, s_{\hb}}\,,
    \end{split}
\end{equation}
where we changed variables $\{\ld{1},\ld{2}\}\rightarrow\{z,\ld{2}\}$ with $z=\ld{1}/\ld{2}=-\lu{2}/\lu{1}$ and recovered an affine expression for the spinor derivative $\begin{pmatrix}
            -2h+1-z\partial_{z}\\
            -\partial_{z}
        \end{pmatrix}e^{-\frac{1}{2}\partial_h}e^{-\partial_{s_h}}$. Note that the derivative $\ld{2}\partdiv{}{\ld{2}}$ gave a factor $-2h+1$. In principle it also acted on the factors $\sgn(\ld{2})^{-s_h}$ but these give terms proportional to $\ld{2}\,\delta(\ld{2})$ which vanish. In other words, unlike the continuous weights $h, \hb$, the discrete weights cannot be measured with a differential operator $\ld{2}\partdiv{}{\ld{2}}$.

\textbf{Twistor Space}

The conformal generators in the twistor and dual twistor bases are listed in Table \ref{tab:tableofgens2} and are found by using the replacements \eqref{fourierconjugatespinors}, but only for either the $\lam$ or $\lamt$ spinor and its derivative. In Table \ref{tab:tableofgens2} we also define two additional sets of conformal generators: $\cJ_T, d_T, p_T, k_T, m_T$ which gives the conformal symmetry generators which treat $\mu_{\alpha}, \lamt_{\dot\alpha}$ in a manner based purely on whether they transform under the left or right hand factor of $\slR\times \slR$. That is the $\cJ_T, d_T, p_T, k_T, m_T$ are simply the original momentum space conformal generators but with the na\"ive replacement $\lam  \rightarrow \mu$ and similarly for the $\cJ_{\bar T}, d_{\bar T}, p_{\bar T}, k_{\bar T}, m_{\bar T}$ for dual twistor space via $\lamt \rightarrow \mut$. These are distinct operators from the original set and are used as a comparison to study the action of the twistor and dual twistor transforms. We find that these new operators are related to each other by the automorphism \eqref{eq: conformalinversionautomorphism} which can be understood as being generated from the full Fourier transformation from twistor to dual twistor space
\begin{equation}\label{eq: fulltwistortrans}
\int d^2\mu \wedge d^2 \lamt e^{i(\agl{\lam}{\mu}+\sqr{\mut}{\lamt})},
\end{equation}
which has the same kernel as the Fourier transformation \eqref{eq: Fourier} but is integrated over an orthogonal region of the full phase space $\lam, \lamt, \mu, \mut$. This correspondence is summarised in the diagram in Figure \ref{fig: fouriersandinversion}. Note that unlike the Fourier transform from $\lam, \lamt$ to $\mu, \mut$ space the measure of the transform \eqref{eq: fulltwistortrans} from twistor to dual twistor space does not carry any dilatation weight and so does not change the dilatation/mass dimension of the object it acts on. As discussed in Section \ref{sec: fourieramps}, this means the full twistor transform \eqref{eq: fulltwistortrans} does not correspond to an active conformal inversion. As explained in \cite{Korchemsky:2009jv, Korchemsky:2010ut, Drummond:2008vq}, an active conformal inversion of the moduli coordinate associated to an amplitude supported on a set of twistor lines, or equivalently an active inversion of the dual coordinates, generates an inversion \emph{within} twistor space which exchanges $\ld{\alpha} \leftrightarrow \mtd{\alpha}$ in our conventions.

\textbf{Light Space}

The light basis can be thought of as a celestial twistor basis and so the bulk conformal generators in the light basis in Table \ref{tab:tableofgens2} are simply Mellin transformed versions of the twistor conformal generators. We can also derive them by directly commuting with the light transform given in say \eqref{eq: compactlightcelstate}; for example for the spinor derivative we find
\begin{equation}
    \begin{split}
        &\frac{i^{-s_h}}{2} \frac{\Gamma(2-2h)}{\Gamma(\frac{3}{2}-h+\frac{s_h}{2})\Gamma(h-\frac{s_{h}}{2}-\frac{1}{2})}\int_{\mathbb{R}^2} d^2\lam\, \abs{\agl{ \lam}{\mu}}^{2h-2} \sgn(\agl{\lam}{\mu})^{s_h} \pld{\alpha} \vert  \ld{\alpha},\ltd{\alpha} \rangle\\
        =&\frac{i^{-s_h}}{2} \frac{\Gamma(2-2h)}{\Gamma(\frac{3}{2}-h+\frac{s_h}{2})\Gamma(h-\frac{s_{h}}{2}-\frac{1}{2})}\int_{\mathbb{R}^2} d^2\lam\, (-\mud{\alpha}\sgn(\agl{\lam}{\mu})(2h-2))\abs{\agl{ \lam}{\mu}}^{2h-3} \sgn(\agl{\lam}{\mu})^{s_h} \vert  \ld{\alpha},\ltd{\alpha} \rangle\\
        =&\frac{i^{-s_h}}{2} \frac{\Gamma(2-2h)}{\Gamma(\frac{3}{2}-h+\frac{s_h}{2})\Gamma(h-\frac{s_{h}}{2}-\frac{1}{2})}(-\mud{\alpha}(2h-2)e^{-\frac{\partial_h}{2}-\partial_{s_h}})\int_{\mathbb{R}^2} d^2\lam\,\abs{\agl{ \lam}{\mu}}^{2h-2} \sgn(\agl{\lam}{\mu})^{s_h} \vert  \ld{\alpha},\ltd{\alpha} \rangle\\
        =&(-i\mud{\alpha}e^{-\frac{\partial_h}{2}-\partial_{s_h}})\frac{i^{-s_h}}{2} \frac{\Gamma(2-2h)}{\Gamma(\frac{3}{2}-h+\frac{s_h}{2})\Gamma(h-\frac{s_{h}}{2}-\frac{1}{2})}\int_{\mathbb{R}^2} d^2\lam\,\abs{\agl{ \lam}{\mu}}^{2h-2} \sgn(\agl{\lam}{\mu})^{s_h} \vert  \ld{\alpha},\ltd{\alpha} \rangle,
    \end{split}
\end{equation}
where in the second line we used integration by parts and dropped the term where the derivative acts on the sgn function $\pld{\alpha}\sgn(\agl{\lam}{\mu}) \sim \delta(\agl{\lam}{\mu})$, with vanishing contribution due to the delta function.

\section{Proofs of \texorpdfstring{$S^2=Id$}{} and \texorpdfstring{$L^2=Id$}{}}\label{app: Ssquared}

Since $S^2$ (and $L^2$) are entwining operators between the same irreducible representations then they must be proportional to the identity. In this appendix we prove directly that the shadow transform \eqref{eq: affineShadow} and light transform \eqref{eq: lightcelstate} are actually \emph{self-inverse}. This is already implied by their relationship with the self-inverse Fourier transform \eqref{eq: Fourier} and half-Fourier \eqref{eq: twistortrans}. Indeed the normalisation chosen such that $S$ and $L$ are self-inverse was exactly that which appeared when linking to the Fourier and half-Fourier transforms. Note, however, that the self-inverse property is not just due to our choice of normalisation. Take the shadow transform for example, since by the square of the shadow transform we mean $S^2:= S_{1-h, 1-\hb} \circ S_{h,\hb}$, then normalising $S$ by any factor at all can only change the normalisation of $S^2$ by a function invariant under $h\rightarrow 1-h, \,\hb\rightarrow 1- \hb$. We find that such a function is enough to normalise $S^2= Id$ and that this is a consequence of the structure of the shadow transform in two dimensions, in particular the use of the two point structure $(w-z)^{2h-2}(\wb-\zb)^{2\hb-2}$ (inherited from the anti-symmetric spinor brackets) which picks up a factor of $(-1)^{-2J}$ when we swap $w \leftrightarrow z$, as we shall now explain.

\textbf{Self Inverse Shadow}

We consider the square of the shadow transform in affine coordinates \eqref{eq: affineShadow} acting on a function with weights $h, \hb$
\begin{equation}\label{eq: affineShadowSquared}
\begin{split}
    S^2 \{f(z, \zb; h, \hb)\} := &\frac{i^{-2h+2\hb}\Gamma(2\hb)}{2\pi i\,\Gamma(1-2h)} \int_{\mathbb{C} \mathbb{P}^1} d\wb\wedge dw\, (z'-w)^{-2h} (\zb'-\wb)^{-2\hb}\\
    &\times\frac{i^{2h-2\hb}\Gamma(2-2\hb)}{2\pi i\,\Gamma(2h-1)} \int_{\mathbb{C} \mathbb{P}^1} d\zb\wedge dz\, (w-z)^{2h-2} (\wb-\zb)^{2\hb-2} f(z, \zb;h, \hb)\,.
\end{split}
\end{equation}
So we are lead to compute the integral,
\begin{align}
\begin{split}
\cI :=\frac{1}{2\pi i} \int d\wb\wedge dw\, (y-w)^{-2h}(\bar{y}-\wb)^{-2\hb}(w-z)^{2h-2}(\wb-\zb)^{2\hb-2}.
\end{split}
\end{align}
A similar integral was studied in \cite{Dolan:2011dv} denoted $I_2$ and for $\sum \alpha_i = \sum \bar\alpha_i =2$ gave the result\footnote{We have written the result of \cite{Dolan:2011dv} using our complex measure $d\wb \wedge dw = 2i\, d(\text{Re}(w)) d(\text{Im}(w))$ which explains the additional normalisation by a factor $\frac{1}{2i}$. Similarly we write the result using a delta function of the complex variable $z'-z$ with an extra factor of $2i$.}  \begin{equation}\label{DolanA.3}
I_2 := \frac{1}{2\pi i} \int d\wb\wedge dw \prod_{i=1,2}(w-w_i)^{-\alpha_i}(\wb-\wb_i)^{-\bar\alpha_i} =2\pi i\, (-1)^{\alpha_1-\bar\alpha_1} K \delta^2(z'-z),
\end{equation}
where we have defined
\begin{equation}
    K := \frac{\Gamma(1-\alpha_1)\Gamma(1-\alpha_2)}{\Gamma(\bar\alpha_1)\Gamma(\bar\alpha_2)}.
\end{equation}
Solving the constraints on the $\alpha_i$ with $\alpha_1= 2h, \, \alpha_2 = 2-2h$ and similarly for the barred variables, we then have an integral very close to $\cI$, and $K$ is such that it exactly cancels the normalisation factors in \eqref{eq: affineShadowSquared}. It remains to relate $I_2$ to our integral $\cI$. The crucial difference between $\cI$ and $I_2$ is that the ordering of the variables in the brackets in $\cI$ is flipped since they reflect the order in which the shadow transforms were applied. We write $\cI$ in a manner such that the branch cut structure is transparent
\begin{align}
\begin{split}
&\cI=\frac{1}{2\pi i} \int d\wb\wedge dw\, (y-w)^{-2\hb-2J}(\bar{y}-\wb)^{-2\hb}(w-z)^{2\hb+2J-2}(\wb-\zb)^{2\hb-2}\\
=&\frac{1}{2\pi i} \int d\wb\wedge dw\, (y-w)^{-2J}(w-z)^{2J}(\vert y-w\vert^2)^{-2\hb}(\vert w-z\vert^2)^{2\hb-2}\\
=&(-1)^{2J}\frac{1}{2\pi i} \int d\wb\wedge dw\, (w-y)^{-2J}(w-z)^{2J}(\vert w-y\vert^2)^{-2\hb}(\vert w-z\vert^2)^{2\hb-2}= (-1)^{2J} I_{2} ,
\end{split}
\end{align}
where we have extracted a factor of $(-1)^{2J}$ and recovered an integral with an ordering of variables that matches $I_2$. Now, using the result \eqref{DolanA.3} we conclude that $S^2=\text{Id}$. We would like to emphasise again that this result is specific to the shadow transform in two dimensions with the two point structure $(w-z)^{2h-2}(\wb-\zb)^{2\hb-2}$ \footnote{Other two point structures are available in $d$-dimensions, for example in \cite{Kravchuk:2018htv}, whose corresponding shadow transform may not be self-inverse.}.

We can also show that $S^2=\text{Id}$ in an independent and revealing way without the computation of the integral $\cI$. We first note that the shadow transform in affine coordinates \eqref{eq: affineShadow} is a convolution of a conformal primary with a kernel given by the two point structure $(w-z)^{2h-2}(\wb-\zb)^{2\hb-2}$. In particular, the kernel is translation invariant. This means that in Fourier space conjugate to $z$ the shadow transform is not an integral transform at all, but a purely multiplicative operation. This insight is thanks to the authors of \cite{Karateev:2018oml}. Here the Fourier transform we will use is a \textit{two-dimensional} one and again we will employ the Fourier kernel which is better adapted to the two-dimensional complex nature of the CFT. For a function $f(z,\zb; h, \hb)$ with weights $h, \hb$ on the celestial sphere we define the 2d Fourier transform
\begin{equation}\label{2Dfouriertransform}
    f( z,\zb;h,\hb ) := \frac{1}{2\pi} \int_{\CC} d\sbar \wedge ds\,\, e^{-i(sz+\sbar\zb)} \tilde f(s,\sbar; h, \hb).
\end{equation}
Note that this is distinct from the Fourier transform we considered in Section \ref{sec: fourier2shad}, since the above is only defined in the affine patch $\lam_{2}\neq0$.

Now let us consider a shadow transform of the conformal primary $f( z,\zb;h,\hb )$ as written in \eqref{2Dfouriertransform},
\begin{align}
\begin{split}
    &S\{ f( z,\zb;h,\hb )\}\\
    =& \frac{1}{2\pi} \frac{i^{2J}}{2\pi i}\frac{\Gamma(2-2\hb)}{\Gamma(2h-1)}\int_{\CC} d\sbar \wedge ds \int_{\CC} d\zb \wedge dz\,\, (\bar{w}-\zb)^{2\hb-2}(w-z)^{2h-2}e^{-i(sz+\sbar\zb)} \tilde f(s,\sbar; h, \hb)\\
    =& \frac{1}{2\pi} \frac{i^{2J}}{2\pi i}\frac{\Gamma(2-2\hb)}{\Gamma(2h-1)}\int_{\CC} d\sbar \wedge ds \,\,e^{-i(sw+\sbar\wb)}s^{1-2h} \sbar^{1-2\hb}\int_{\CC} d\zb \wedge dz\,\, \zb^{2\hb-2}\,z^{2h-2}e^{i(z+\zb)} \tilde f(s,\sbar; h, \hb),
\end{split}
\end{align}
where in the second line we have performed a simple change of variables, shifting $z$ by $w$ and re-scaling by $s$. We can now recycle our calculation of \eqref{eq: Iintegralfouriershad} to perform the integral over $z,\zb$, noting the different exponents appearing in the case at hand
\begin{align}\label{eq: 2dfouriershadowfunc}
\begin{split}
    S\{ f( z,\zb;h,\hb )\}&
    = \frac{1}{2\pi} \frac{i^{2J}\Gamma(2-2\hb)i^{-2J}\Gamma(2\hb-1)}{\Gamma(2h-1)\Gamma(2-2h)}\int_{\CC} d\sbar \wedge ds \,\,e^{-i(sw+\sbar\wb)}s^{1-2h} \sbar^{1-2\hb} \tilde f(s,\sbar; h, \hb)\\
    & = \frac{(-1)^{2J}}{2\pi} \int_{\CC} d\sbar \wedge ds \,\,e^{-i(sw+\sbar\wb)}s^{1-2h} \sbar^{1-2\hb} \tilde f(s,\sbar; h, \hb),
\end{split}
\end{align}
so we see that in Fourier space the shadow transform acts in the simplest manner possible as multiplicative operation,
\begin{equation}\label{shadowinfourierspace}
    \tilde f(s,\sbar; h, \hb) \overset{S}{\longmapsto} (-1)^{2J}s^{1-2h}\sbar^{1-2\hb} \tilde f(s,\sbar; h, \hb).
\end{equation}
The factors of $s,\sbar$ appearing in \eqref{shadowinfourierspace} match the results of \cite{Karateev:2018oml} up differences in the definition of the two point structure used in the shadow transform.

From \eqref{shadowinfourierspace} it is immediate that the shadow transform is self inverse. That is, by performing a shadow transform again with the weights given now by $1-h,1-\hb$ we will cancel the above multiplicative factors and so return the original function. We can check this explicitly by performing a second shadow transform on the function given by the final line of \eqref{eq: 2dfouriershadowfunc}
\begin{align}
\begin{split}
    &\cS^2\{ f( z,\zb;h,\hb )\}= (-1)^{2J}\frac{1}{2\pi} \frac{1}{2\pi i}\frac{i^{-2J}\Gamma(2\hb)}{\Gamma(1-2h)}\int_{\CC} d\sbar \wedge ds \int_{\CC} d\wb \wedge dw\\ &\times(\bar{y}-\wb)^{-2\hb}(y-w)^{-2h}e^{-i(sw+\sbar\wb)}s^{1-2h}\sbar^{1-2\hb} \tilde f(s,\sbar; h, \hb)\\
    =& (-1)^{2J}\frac{1}{2\pi} \frac{1}{2\pi i}\frac{i^{-2J}\Gamma(2\hb)}{\Gamma(1-2h)}\int_{\CC} d\sbar \wedge ds \,\,e^{-i(sy+\sbar\bar{y})}\int_{\CC} d\wb \wedge dw\,\, \wb^{-2\hb}\,w^{-2h}e^{i(w+\wb)} \tilde f(s,\sbar; h, \hb),
\end{split}
\end{align}
where in the second line we have performed the same simple change of variables as before. Again we can now recycle our calculation of \eqref{eq: Iintegralfouriershad} to do the integrals over $w,\wb$, noting the different exponents appearing,
\begin{align}
\begin{split}
    S^2\{ f( z,\zb;h,\hb )\}&
    = \frac{1}{2\pi}(-1)^{2J} \frac{i^{-2J}\Gamma(2\hb)i^{2J}\Gamma(1-2\hb)}{\Gamma(1-2h)\Gamma(2h)}\int_{\CC} d\sbar \wedge ds \,\,e^{-i(sy+\sbar\bar{y})} \tilde f(s,\sbar; h, \hb)\\
    & = \frac{1}{2\pi} \int_{\CC} d\sbar \wedge ds \,\,e^{-i(sy+\sbar\bar{y})} \tilde f(s,\sbar; h, \hb) = f( y,\bar{y};h,\hb ).
\end{split}
\end{align}

\textbf{Self Inverse Light}

We now show that the light transform is also self inverse (the dual light transform case is completely analogous). The light transform acting in affine coordinates on a function with weights $h, s_h, \hb, s_{\hb}$ on the celestial torus is given by
\begin{equation}\label{eq: affinelight}
\begin{split}
    L\{ f(z, \zb; h, s_h, \hb, s_{\hb})\} =&\; i^{-s_h} \frac{\Gamma(2-2h)}{\Gamma(\frac{3}{2}-h+\frac{s_h}{2})\Gamma(h-\frac{s_{h}}{2}-\frac{1}{2})}\\
    &\qquad\qquad\qquad \times \int_{\mathbb{RP}^1} dz\, \abs{w-z}^{2h-2} \sgn(w-z)^{s_h} f(z, \zb; h, s_h, \hb, s_{\hb}),
\end{split}
\end{equation}
and so is also a convolution with translation invariant kernel $(w-z)^{2h-2} \sgn(w-z)^{s_h}$. In Fourier space with coordinate $s$ conjugate to $z$ the function becomes
\begin{equation}
     f(z, \zb; h, s_h, \hb, s_{\hb}) := \frac{1}{\sqrt{2\pi}}\int_{\RR} ds\, e^{-isz} f(s, \zb; h, s_h, \hb, s_{\hb})
\end{equation}
and the light transform acting on this gives after the usual change of variable 
\begin{equation}
\begin{split}
     L\{ f(z, \zb; h, s_h, \hb, s_{\hb})\}=&\; \kappa_{h, s_h, \hb, s_{\hb}} \frac{1}{\sqrt{2\pi}}\int_{\RR} ds\, \abs{s}^{1-2h} \sgn(s)^{-s_h} e^{-isw}\\
     &\qquad\qquad\qquad  \times\int_{\RR} dz\, \abs{z}^{2h-2} \sgn(z)^{s_h} e^{iz} f(s, \zb; h, s_h, \hb, s_{\hb}) ,
\end{split}
\end{equation}
where $\kappa_{h, s_h, \hb, s_{\hb}}$ is the normalisation pre factor in \eqref{eq: affinelight}. The $z$ integral we calculated already in \eqref{eq: Iintegraltwistorlight}, note however we have different exponents in the case at hand and so we find
\begin{equation}
\begin{split}
     L\{ f(z, \zb; h, s_h, \hb, s_{\hb})\}=&\; i^{-s_h} \frac{\Gamma(2-2h)}{\Gamma(\frac{3}{2}-h+\frac{s_h}{2})\Gamma(h-\frac{s_{h}}{2}-\frac{1}{2})} 2\pi i^{s_h} \frac{\Gamma(2h-1)}{\Gamma(h-\frac{s_h}{2})\Gamma(1-h+\frac{s_{h}}{2})}\\
     &\qquad \qquad \qquad \times \frac{1}{\sqrt{2\pi}}\int_{\RR} ds\, \abs{s}^{1-2h} \sgn(s)^{-s_h} e^{-isw} f(s, \zb; h, s_h, \hb, s_{\hb}) .
\end{split}
\end{equation}
Now, using the Euler reflection formula and various trigonometric identities, the prefactor is given by
\begin{equation}
\begin{split}
    &2\pi \frac{\Gamma(2-2h)}{\Gamma(\frac{3}{2}-h+\frac{s_h}{2})\Gamma(h-\frac{s_{h}}{2}-\frac{1}{2})} \frac{\Gamma(2h-1)}{\Gamma(h-\frac{s_h}{2})\Gamma(1-h+\frac{s_{h}}{2})}\\
    =&\; 2 \frac{\sin((h-\frac{s_h}{2}-\frac{1}{2})\pi)\sin((h-\frac{s_h}{2})\pi)}{\sin((2h-1)\pi)} = -  \frac{2\,\cos((h-\frac{s_h}{2})\pi)\sin((h-\frac{s_h}{2})\pi)}{\sin((2h-1)\pi)}\\
    =& - \frac{\sin(2(h-\frac{s_h}{2})\pi)}{\sin((2h-1)\pi)}= (-1)^{s_h}
\end{split}
\end{equation}
and so in Fourier space with coordinate $s$ the light transform acts multiplicatively via
\begin{equation}\label{lightinfourierspace}
    \tilde f( s,\zb;h,s_h,\hb , s_{\hb}) \overset{L}{\longmapsto} (-1)^{s_h}\abs{s}^{1-2h} \sgn(s)^{s_h} \tilde f( s,\zb;h,s_h,\hb , s_{\hb}),
\end{equation}
where we would like to highlight the similarity of the form of this action with that of the shadow in \eqref{shadowinfourierspace}. Now if we act with a second light transform, with shifted weights $h \rightarrow1-h$ in its definition, we cancel the above factors and recover the original state.
\section{All Leg Fourier Gluon Amplitudes in Split Signature}\label{app: all legfouriergluon}

In this appendix, we perform two half-Fourier transforms on every leg of a  tree-level momentum space gluon amplitude. This is the split signature analogue of the all-leg Fourier transform considered in $\phi^4$ theory in Section \ref{sec: fourieramps}. We will perform these transformations in two steps by first performing ambidextrous half-Fourier transforms on each leg of the amplitude according to its helicity and then performing the remaining set of half-Fourier transforms. We will find that the amplitude in the Fourier space $\mu,\mut$ takes an identical form. We begin with the $n$-point $N^mMHV$ tree-level amplitude in split signature expressed using the CHY formula \cite{Cachazo:2013gna, Cachazo:2013hca, Cachazo:2013iea}
\begin{equation}
    A_n=\int \prod_{a=1}^n \frac{d^2\sigma_a}{(a\,a+1)} \frac{1}{GL_2} \prod_{I=1}^{m} \delta^2 \bigg( \lamt_I - \sum_{i'=m+1}^n \frac{\lamt_{i'}}{(I\,i')}\bigg)\,\prod_{i=m+1}^{n} \delta^2 \bigg( \lam_i - \sum_{I'=1}^m \frac{\lam_{I'}}{(i\,I')}\bigg)
\end{equation}
where legs $I=1, ..., m$ have negative helicity and legs $i=m+1, ..., n$ have positive helicity and $(a b):= \sigma_a-\sigma_b$. We can write the CHY formula as
\begin{equation}
\begin{split}
    A_n=&\int \prod_{a=1}^n \frac{d^2\sigma_a}{(a\,a+1)} \frac{1}{GL_2} \prod_{I=1}^{m} \frac{1}{(2\pi)^2}\int d^2\tilde\chi_I \exp\bigg[i\tilde\chi_I\bigg( \lamt_I - \sum_{i'=m+1}^n \frac{\lamt_{i'}}{(I\,i')}\bigg)\bigg]\\
    &\qquad\times\prod_{i=m+1}^{n} \frac{1}{(2\pi)^2}\int d^2\chi_i \exp\bigg\langle i\bigg( \lam_i - \sum_{I'=1}^m \frac{\lam_{I'}}{(i\,I')}\bigg)\chi_i\bigg \rangle.
\end{split}
\end{equation}
and then perform ambidextrous half-Fourier transforms over the coordinates $\lamt_I$ for $I=1, ..., m$ and $\lam_i$ for $i=m+1, ..., n$. The half-Fourier integrals can be done trivially and we find
\begin{equation}
\begin{split}
    T\{A_n\}=&\int \prod_{a=1}^n \frac{d^2\sigma_a}{(a\,a+1)} \frac{1}{GL_2} \prod_{I=1}^{m} \frac{1}{2\pi} \exp\bigg(i\,\bigg[\mut_I\!\sum_{i'=m+1}^n  \frac{\lamt_{i'}}{(I\,i')}\bigg]\bigg)\\
    &\qquad\times\prod_{i=m+1}^{n} \frac{1}{2\pi} \exp\bigg(i\bigg\langle\sum_{I'=1}^m \frac{\lam_{I'}}{(i\,I')}\mu_i\bigg\rangle\bigg).
\end{split}
\end{equation}
and so the CHY formula in an ambidextrous twistor basis consists of just exponentials with no delta function support. We can now ambidextrous half-Mellin transform this expression and we recover the $n$-point formula for ambidextrous light transformed celestial amplitudes found in \cite{Sharma:2021gcz}. This is further demonstration of the commuting diagram Figure \ref{fig: twistlight}.

To find the all-leg Fourier amplitude in $\mu, \mut$ space we need to perform the complementary half-Fourier transforms over the variables $\lam_I'$ for $I=1', ..., m$ and $\lamt_i'$ for $i'=m+1, ..., n$. We first re-write the above formula by noting that we can reorganise the exponentials using the trivial relation
\begin{equation}
    \sum_{I=1}^m \bigg[\mut_I \sum_{i'=m+1}^n\frac{\lamt_{i'}}{(I\,i')}\bigg] = \sum_{i'=m+1}^n \bigg[\sum_{I=1}^m \frac{\mut_I}{(I\,i')}\lamt_{i'}\bigg] \
\end{equation}
such that
\begin{equation}
\begin{split}
   T\{A_n\}=&\int \prod_{a=1}^n \frac{d^2\sigma_a}{(a\,a+1)} \frac{1}{GL_2} \prod_{i'=m+1}^n \frac{1}{2\pi} \exp\bigg(i\bigg[\sum_{I=1}^{m} \frac{\mut_I}{(I\,i')}\lamt_{i'}\bigg]\bigg)\\
    &\qquad\times\prod_{I'=1}^m \frac{1}{2\pi} \exp\bigg(i\bigg\langle\lam_{I'}\sum_{i=m+1}^{n} \frac{\mu_i}{(i\,I')}\bigg\rangle\bigg),
\end{split}
\end{equation}
from which we can half-Fourier transform the remaining $\lam, \lamt$ variables to give
\begin{equation}
\begin{split}
    F\{A_n\}=&\int \prod_{a=1}^n \frac{d^2\sigma_a}{(a\,a+1)} \frac{1}{GL_2} \prod_{i'=m+1}^n \delta^2\bigg(\mut_{i'}-i\sum_{I=1}^{m} \frac{\mut_I}{(i'\,I)}\bigg)\\
    &\qquad\times\prod_{I'=1}^m  \delta^2\bigg(\mu_{I'}-i\sum_{i=m+1}^{n} \frac{\mu_i}{(I'\,i)}\bigg),
\end{split}
\end{equation}
where we have used the antisymmetry of the bracket $(\cdot, \cdot)$. We conclude that for all tree-level gluon amplitudes in (2,2) signature the Fourier transform acting on all legs leaves the structure of the amplitude invariant, and simply replaces $\lam \rightarrow \mu, \lamt \rightarrow \mut$. Using the commuting diagram Figure \ref{fig: fouriershadow}, the  same conclusion applies to the all leg shadow transformed amplitudes in (2,2) signature - they take the  same form as the usual Mellin space celestial amplitudes. Whether or not this result can be extended to gluon amplitudes in Lorentzian signature is not clear. From our own initial investigations, performing Fourier transforms directly on gluon amplitudes in (1,3) signature results in different integrals, due to the different integration measure for complex spinors, and these require careful regularisation.  
In addition, it is not obvious how one would ``analytically continue'' the above result to (1,3) signature. We leave the further study of these questions to later work. 
\section{Complexified Spacetime}\label{app: complexify}

In this appendix we provide some more motivation for the form of the Lorentzian and split signature transforms we have encountered in the main text, highlighting their differences and similarities. We do this by exploring a conceptually important technology that is often used in the study of scattering amplitudes, namely that of complexifying spacetime. This provides a unified view of spacetime of any signature - we can then restrict to a slice of complexified spacetime in order to hone in on (1,3) or (2,2) signature spacetimes. In what follows we present a condensed version of the general methodology we have already applied to Lorentzian and split signature.

In complexified spacetime we have independent complex spinors $\lambda_\alpha, \lambdat_{\dot\alpha}$ such that the momentum $p_{\alpha\dot{\alpha}}=\ld{\alpha}\ltd{\alpha}$ is complex. The little group of a complex null momenta is $\CC_*=U(1)\times \RR_+$ and representations are labelled by two helicities:
 \begin{enumerate}
     \item For the compact $U(1)$ subgroup $c=e^{i\theta}$ we have a discrete label $J_{1,3} \in \ZZ/2$ corresponding to the little group and helicity in (1,3) signature.
     \item While for the non-compact $\RR_+$ subgroup where $c$ is real we have a continuous imaginary label $J_{2,2} \in i\RR$ corresponding to the little group and helicity in (2,2) signature.
 \end{enumerate}
  So we write asymptotic particle states in complexified spacetime as
  \begin{equation}\label{eq: complexmomentumEigenstate}
    \ket{\lam, \lamt ; J_{1,3}, J_{2,2}}\,,
\end{equation}
and they transform under the complexified little group for $c=\abs{c} e^{i\theta}$ as
\begin{equation}\label{eq: complexLGhomogeneity}
    \boxed{\ket{c\,\lam, c^{-1}\lamt ; J_{1,3}, J_{2,2}}=(e^{i\theta})^{-2\,J_{1,3}}\abs{c}^{-2\,J_{2,2}} \ket{\,\lam, \lamt ; J_{1,3}, J_{2,2}}\,.}
\end{equation}

The complexified Lorentz group is $\slC \times \slC$ and the transformation of these states under this group is encoded in the spinors through
\begin{equation}\label{eq: GlobalLorentz}
    \ket{\ld{\alpha} , \ltd{\alpha};J_{1,3}, J_{2,2}}\rightarrow \ket{M_\alpha^{\,\,\beta}\ld{\beta} , \tilde{M}_{\dot{\alpha}}^{\,\,\dot{\beta}}\ltd{\beta}; J_{1,3}, J_{2,2} }\,,
\end{equation}
where $M_\alpha^{\,\,\beta}$ and $\tilde{M}_{\dot{\alpha}}^{\,\,\dot{\beta}}$ are independent $\slC$ matrices.

To restrict to $(1,3)$ signature we make the slice \eqref{eq: 13realitycondition} and this cuts the little group down to $c \in U(1)$ with states labelled by just $J_{1,3}$.

For split signature something more subtle happens, we enforce the reality condition \eqref{eq: 22realitycondition} and this time the $(2,2)$ restriction of the little group is $\RR_* =\RR_{+} \times \ZZ_2$ since we require that $c$ in \eqref{eq: complexLGhomogeneity} be real. The $\RR_{+}$ representation of the complexified little group survives and is still labelled by $J_{2,2}$ but now $U(1)$ degenerates into  $\ZZ_2$. This is the origin of the discrete helicity $s_J \in {0,1}$ labelling even/odd representations of $\ZZ_2$ in split signature.  The discrete helicity $s_J$ is directly related to $J_{1,3}$ by $ s_J := 2\,J_{1,3} \mod 2$) and so naturally corresponds to bosons and fermions in (2,2).

Returning now to complex momentum, we can also consider extending the complexified little group to include a complex boost symmetry. In complexified momentum space these boosts scale the momentum by a non-zero complex number, $b \in \CC_*$
\begin{equation}\label{eq: complexmomboost}
    p_{\alpha\dot{\alpha}}\rightarrow  b\, p_{\alpha\dot{\alpha}}\,.
\end{equation}
Hence we see that the extended little group of a complex null direction is $ \mathbb{C}_*\times  \mathbb{C}_*$, and corresponds to rescaling each spinor independently by any non-zero complex numbers $y,\tilde{y}$
\begin{equation}\label{eq: extendedLGy}
    \lamt\rightarrow y \lamt, \quad \lamt\rightarrow \tilde y \lamt\,. 
\end{equation}
This leads to asymptotic states defined on a complexified celestial torus $\CCP^{1} \times \CCP^{1}$, also discussed in \cite{Banerjee:2022wht}. Such states are covariant under the extended little group $\CC_* \times \CC_*$ and are labelled by its representations. As well as the two helicities $J_{1,3}, J_{2,2}$ we have two conformal dimensions associated to the boost transformation by $b=\abs{b}e^{i\phi}$
\begin{enumerate}
     \item For the non-compact $\RR_+$ subgroup we have a continuous parameter $\Delta \in 1+i\RR$ which is the usual conformal dimension in both $(1,3)$ and $(2,2)$ signature,
     \item While for the compact $U(1)$  subgroup we have a discrete parameter $\Delta_c \in \ZZ/2$. Note this is only present because the momenta is complex but as one might expect it descends to the $\ZZ_2$ symmetry in (2,2) signature which flips the sign of the momentum.
\end{enumerate}
The splitting of the extended little group into rotations and boosts is conceptually very clear, however we can obtain more compact formulae if we combine these transformations into the complex rescalings \eqref{eq: extendedLGy}. We define new weights as follows
\begin{equation}\label{eq: complexChiralWeights}
    \begin{split}
        H &:=\frac{1}{4}\big(\Delta+\Delta_c+J_{1,3}+J_{2,2}\big),\\
        \Htl &:=\frac{1}{4}\big(\Delta-\Delta_c+J_{1,3}-J_{2,2}\big),\\
        \Hb &:=\frac{1}{4}\big(\Delta+\Delta_c-(J_{1,3}+J_{2,2})\big),\\
        \bar{\Htl} &:=\frac{1}{4}\big(\Delta-\Delta_c-(J_{1,3}-J_{2,2})\big).
    \end{split}
\end{equation}
and write celestial states as
\begin{equation}
    \vert \lam,\lamt; \mathbf{H} \rangle.
\end{equation}
Now under the extended little group, for any $(y, \yt) \in\mathbb{C}_*\times \mathbb{C}_*$ we have the homogeneity property:
\begin{equation}\label{eq: homoCond}
    \boxed{\vert y\lam,\yt\lamt; \mathbf{H} \rangle = y^{-2H}(y^*)^{-2\Hb}\yt^{-2\Htl}(\yt^*)^{-2\bar{\Htl}}\vert \lam,\lamt; \mathbf{H} \rangle \,,}
\end{equation}
where $\mathbf{H}$ is shorthand for the four weights defined in \eqref{eq: complexChiralWeights} and are labels for representations of the little group $\mathbb{C}_*\times \mathbb{C}_*$ which acts independently on the $\lambda$ spinor and the `conjugate' spinor $\lambdat$. These celestial states transform as conformal primaries under $\slC \times \slC$.

The slices to Lorentzian signature require that $\yt=\yb$ and so we recover the extended little group $\CC_*$ and homogeneity law \eqref{eq: 13ELGhomo}. While in (2,2) we require that both $y,\yt$ are real and so we recover the homogeneity property \eqref{eq: 22ELGTransformation} for the extended little group $\RR_* \times \RR_*$ and states are labelled by weights $h,\hb, s_h, s_{\hb}$.

We can go further and apply the general logic set out in the prelude (Section \ref{sec: prelude}) to define a complexified chiral Mellin transform which integrates over the complex scales of the independent spinors with weights $\Hbf$. Similarly we can define a shadow transform exactly analogous to \eqref{eq: 13shadow} and which acts on the spinor $\lam$, as well as an independent dual shadow which acts on $\lamt$. Each of these is then immediately related by the commuting diagram \ref{fig: fouriershadow} to an independent bulk Fourier transform over either $\lam$ and $\lamt$. An appropriate slicing procedure to (1,3) and (2,2) signature would then give the transforms that we have already met in the main text, for example a real (2,2) slice of the dual shadow would give the dual light transform. We leave the further exploration of these ideas to future work.


\section{Inverse Chiral Mellin Transform in Homogeneous Coordinates}\label{sec: Mellininversion}

In this appendix we consider the (1,3) chiral Mellin transform in \eqref{eq: 13chiralMellin} and find its inverse. This can be written in terms of Fourier transforms by changing variables as $u = e^{\frac{x}{2}}e^{i\theta}$. This gives
\begin{equation}\label{eq: DerivChiralInverse}
    \begin{split}
        &\vert \lambda, \lambdat, h, \hb  \rangle= \frac{1}{2\pi}\int_{0}^{2\pi}d\theta\, \left(e^{i\theta}\right)^{2(h-\hb)} \int_{-\infty}^\infty dx\,
        e^{x\Delta} \vert e^{\frac{x}{2}}e^{i\theta}\lambda,e^{\frac{x}{2}}e^{-i\theta}\lambdat \rangle  \\
        =&\frac{1}{2\pi}\int_{0}^{2\pi}d\theta\, \left(e^{i\theta}\right)^{2(h-\hb)} \int_{-\infty}^\infty dx\, e^{ix\beta} \left(e^{xa} \vert e^{\frac{x}{2}}e^{i\theta}\lambda,e^{\frac{x}{2}}e^{-i\theta}\lambdat \rangle\right)\, ,
    \end{split}
\end{equation}
where in the last line we used $\Delta=a + i\beta$. This is just the composition of two Fourier transforms acting on $e^{xa} \vert e^{\frac{x}{2}}e^{i\theta}\lambda,e^{\frac{x}{2}}e^{-i\theta}\lambdat  \rangle$ - a continuous/non-compact one in $x$ and a discrete/compact one in $\theta$. The compact Fourier transform over a circle always converges, however the non-compact Fourier transform converges only as long as the function $e^{xa} \vert e^{\frac{x}{2}}e^{i\theta}\lambda,e^{\frac{x}{2}}e^{-i\theta}\lambdat  \rangle$ dies off sufficiently fast for large values of $x$ . This defines a strip of definition for allowed values of $a$ for the Mellin transform of a state. 

We can now invert \eqref{eq: DerivChiralInverse} using the standard inversion theorems for Fourier transforms to find
\begin{equation}
    e^{ra} \vert e^{\frac{r}{2}}e^{i\theta}\lambda,e^{\frac{r}{2}}e^{-i\theta}\lambdat  \rangle= \frac{1}{2\pi}  \sum_{2(h-\hb)\in \mathbb{Z}} \left(e^{-i\theta}\right)^{2(h-\hb)} \int_{-\infty}^\infty db\,e^{-ir\beta} \vert \lambda,\lambdat,h, \hb  \rangle .
\end{equation}
Thus, setting $r=0$, $\theta=0$ we recover our original function
\begin{equation}\label{eq: InverseChiralMellin}
        \vert \lambda,\lambdat \rangle = \frac{1}{2\pi}  \sum_{2(h-\hb)\in \mathbb{Z}}  \int_{-\infty}^\infty db\, \vert \lambda,\lambdat,h,\hb \rangle 
                    = \frac{1}{2\pi i}  \sum_{2(h-\hb)\in \mathbb{Z}}  \int_{a-i\infty}^{a+i\infty} d\Delta\, \vert \lambda,\lambdat,h,\hb  \rangle.
\end{equation}
Equation \eqref{eq: InverseChiralMellin} is the inversion formula in homogeneous coordinates. This formula, as expected, follows the general form of inverse transforms given in Section \ref{sec: prelude}: a sum over representations of $\mathbb{C}$. The choice of the value of $a \in \mathbb{R}$ is in principle free within the strip of definition, however the choice $a=1$ is selected  as giving a basis of conformal primary wavefunctions that are normaliseable with respect to the Klein Gordon product \cite{Pasterski:2017kqt}.

If we denote the chiral Mellin transform by $\cC$ then we can check directly that $\cC^{-1} \circ \cC = \text{Id}$,
\begin{equation}
    \begin{split}
    \mathcal{C}^{-1}\{\vert \lam, \lamt ; h, \hb  \rangle \}&:=\sum_{h-\hb\in \frac{\mathbb{Z}}{2}} \int_{a-i\infty}^{a+i\infty} \frac{d\Delta}{2\pi i} \vert \lam, \lamt ; h, \hb \rangle\\ 
    =&\sum_{h-\hb\in \frac{\mathbb{Z}}{2}} \int_{a-i\infty}^{a+i\infty} \frac{d\Delta}{2\pi i} \int_{0}^{2\pi}\frac{d\theta}{2\pi}\, \left(e^{i\theta}\right)^{2(h-\hb)}\int_{0}^{\infty}\! d\omega \, \omega^{\Delta-1} \vert \sqrt{\omega}e^{i\theta}\lam, \sqrt{\omega}e^{-i\theta}\lamt  \rangle \\
    =&\int_{0}^{2\pi}d\theta \delta(\theta)\int_0^{\infty}\! d\omega \delta(\omega-1) \vert \sqrt{\omega}\lam, \sqrt{\omega}\lamt  \rangle = \vert\lam,\lamt  \rangle .
    \end{split}
\end{equation}
Here we have used that
\begin{align}
    \int_{a-i\infty}^{a+i\infty} \frac{d\Delta}{2\pi i} \, \omega^{\Delta-1}& = \int_{-\infty}^{\infty} \frac{d\beta}{2\pi} \, e^{(a-1+i\beta )\text{ln}(\omega)} = e^{(a-1)\text{ln}(\omega)}\delta(\text{ln}(\omega)) =\delta(\text{ln}(\omega))\\
    =& \vert \partial_{\omega}\text{ln}(\omega) \vert^{-1}_{\omega=1} \delta(\omega -1) = \delta(\omega -1)\,,
\end{align}
and 
\begin{equation}
    \frac{1}{2\pi}\sum_{h-\hb\in \frac{\mathbb{Z}}{2}} \left(e^{i\theta}\right)^{2(h-\hb)} = \delta(\theta).
\end{equation}

We can also check $\cC \circ \cC^{-1} = \text{Id}$: 
\begin{equation}\label{checkinversionproperty}
    \begin{split}
    &\cC \circ \cC^{-1}\{\vert \lam, \lamt ;h,\hb\rangle \} =\frac{1}{2\pi i} \!\!\int_{\CC} d\ub\wedge du  u^{2k-1} \ub^{2\kb-1} \!\!\!\!\sum_{h-\hb\in \frac{\mathbb{Z}}{2}} \int_{a-i\infty}^{a+i\infty} \frac{d\Delta}{2\pi i} \vert u\lam, \ub\lamt ; h,\hb \rangle\\ 
    =&\sum_{h-\hb\in \frac{\mathbb{Z}}{2}} \int_{a-i\infty}^{a+i\infty} \frac{d\Delta}{2\pi i}\frac{1}{2\pi i} \!\!\int_{\CC} d\ub\wedge du  u^{2k-2h-1} \ub^{2\kb-2\hb-1}\,\vert \lam, \lamt ;  h, \hb \rangle\\
    =&\sum_{h-\hb\in \frac{\mathbb{Z}}{2}} \int_{-i\infty}^{i\infty} \frac{d\beta}{2\pi } \frac{1}{2\pi}\!\!\int_{0}^{2\pi}d\theta\, e^{2 i\theta(h-\hb-(k-\kb))}\int_{-\infty}^{\infty} dx e^{x(a'-a+i(\beta'-\beta))}\,\vert \lam, \lamt ;  h, \hb \rangle\\
    =&\sum_{h-\hb\in \frac{\mathbb{Z}}{2}} \int_{-i\infty}^{i\infty} \frac{d\beta}{2\pi }\delta_{h-\hb,k-\kb} 2\pi \delta(\beta'-\beta)\,\vert \lam, \lamt ;  h, \hb \rangle = \vert \lam, \lamt ;  k, \kb \rangle,
\end{split}
\end{equation}
where we have used a change of variable $u=e^{\frac{x}{2}} e^{i\theta}$. For the case in hand we have that both $h,\hb$ and $k,\kb$ are on the principal continuous series: $a'=a(=1)$, which guarantees that $\cC \circ \cC^{-1}$ is the identity. This condition can be relaxed if we admit generalised delta functions \cite{Donnay:2020guq}.

\section{Spinor Conventions}\label{app: conventions}
Here we state our conventions for spinor helicity, largely drawn from \cite{Henn:2014sjd}. 

The Pauli matrices are given by
\begin{equation}
    \sigma^1 = \begin{pmatrix}
                0 & 1 \\
                1 & 0 
                \end{pmatrix}\,,
    \quad 
    \sigma^2 = \begin{pmatrix}
                0 & -i \\
                i & 0 
                \end{pmatrix}\,,
    \quad 
    \sigma^3 = \begin{pmatrix}
                1 & 0 \\
                0 & -1 
                \end{pmatrix}\,, 
    \quad
    \tvec{\sigma}= (\sigma^1,\sigma^2,\sigma^3)\,,
\end{equation}
and we then define
\begin{equation}
    \begin{split}
         &(\sigma^\mu)_{\alpha \dot{\alpha}}=(\mathbbm{1},\tvec{\sigma}), \quad (\bar{\sigma}^\mu)^{\dot{\alpha} \alpha}=\ep^{\dot{\alpha}\dot{\beta}}\ep^{\alpha \beta}(\sigma^\mu)_{\beta \dot{\beta}}=(\mathbbm{1},-\tvec{\sigma}),\\
         &(\sigma_\mu)_{\alpha \dot{\alpha}}=(\mathbbm{1},-\tvec{\sigma}), \quad (\bar{\sigma}_\mu)^{\dot{\alpha} \alpha}=\ep^{\dot{\alpha}\dot{\beta}}\ep^{\alpha \beta}(\sigma_\mu)_{\beta \dot{\beta}}=(\mathbbm{1},\tvec{\sigma})\,.
    \end{split}
\end{equation}
For any given 4-vector $p^\mu$ we define
\begin{equation}
    p_{\alpha\dot{\alpha}}=(p_\mu\sigma^\mu)_{\alpha\dot{\alpha}}\,
\end{equation}
If $p^2=0$ we can decompose this matrix into spinors
\begin{equation}
    p_{\alpha\dot{\alpha}}=\lam_{\alpha}\lamt_{\dot{\alpha}}.
\end{equation}
The two-dimensional Levi-Civita symbol, used to raise and lower two-spinor indices, is defined by 
\begin{equation}
    \ep^{\alpha \beta}=\ep^{\dot{\alpha}\dot{\beta}}=   \begin{pmatrix}
                                                            0 & 1 \\
                                                            -1 & 0 
                                                        \end{pmatrix},
    \quad 
    \ep_{\alpha \beta}=\ep_{\dot{\alpha}\dot{\beta}}=   \begin{pmatrix}
                                                            0 & -1 \\
                                                            1 & 0 
                                                        \end{pmatrix}.
\end{equation}
 We raise and lower indices as follows
\begin{equation}
    \begin{split}
        \lam^\alpha = \ep^{\alpha \beta} \lam_\beta \quad  \lamt^{\dot{\alpha}} = \ep^{\dot{\alpha} \dot{\beta}} \lamt_{\dot{\beta}}\,.
    \end{split}
\end{equation}
Products of two different spinors $i$ and $j$ are given by
\begin{equation}
    \langle\lam_i\lam_j\rangle= \lam_i^\alpha\lam_{j\alpha}, \quad [\lamt_i\lamt_j]= \lamt_{i\dot{\alpha}}\lamt_{j}^{\dot{\alpha}}\,,
\end{equation}
and additionally if $p^\mu, q^\mu$ are massless we may write
\begin{equation}
    (p+q)^2=2p \cdot q = \langle pq\rangle [qp]\,.
\end{equation}
We also define the spinor derivatives
\begin{equation}
    \partial_{\alpha}\coloneqq\partdiv{}{\lam^\alpha}\,,\quad  \tilde\partial_{\dot{\alpha}}\coloneqq\partdiv{}{\lamt^{\dot{\alpha}}}\,,
\end{equation}
whose indices must be raised and lowered in the opposite way
\begin{equation}
\begin{split}
         &\partial^{\alpha}\coloneqq\partdiv{}{\lam^\alpha}= \ep^{ \beta\alpha}  \partial_{\beta}\\
         &\tilde\partial^{\alpha}\coloneqq\partdiv{}{\lamt^{\dot\alpha}}= \ep^{ \dot{\beta}\dot{\alpha}}  \tilde\partial_{\dot\beta}\,.
\end{split}
\end{equation}
\newpage

\bibliographystyle{utphys}
\bibliography{remainder}

\end{document}